\documentclass{aa}  
\usepackage[breaklinks=true,colorlinks,citecolor=blue]{hyperref}
\usepackage{graphicx}
\usepackage{enumerate}
\usepackage{epsfig}
\usepackage{color}
\usepackage{txfonts}
\usepackage{natbib}
\usepackage{multirow}
\usepackage{ulem} 
\usepackage{amssymb}
\usepackage{xcolor}

\defcitealias{KhoperskovHESTIA-1}{Paper I}
\defcitealias{KhoperskovHESTIA-2}{Paper II}
\defcitealias{KhoperskovHESTIA-3}{Paper III}

\definecolor{ao}{rgb}{0.0, 0.5, 0.0}

\newcommand{\kmps}{\rm km~s\ensuremath{^{-1} }\,}

\newcommand{\Msun}{M\ensuremath{_\odot}}

\newcommand{\VV}{V\ensuremath{_\phi}\,}

\newcommand{\aFe}{\ensuremath{\rm [\alpha/Fe]}\,}

\newcommand{\FeH}{\ensuremath{\rm [Fe/H]}\,}

\newcommand{\MgFe}{\ensuremath{\rm [Mg/Fe]}\,}

\newcommand{\ELz}{\ensuremath{\rm E-L_z}\,}

\begin{document} 

\title{The stellar halo in Local Group Hestia simulations III. \\ Chemical abundance relations for accreted and in situ stars}

\titlerunning{HESTIA simulations: chemo-kinematics of stellar halo}

\author{Sergey Khoperskov$^1$\thanks{E-mail: sergey.khoperskov@gmail.com}, Ivan Minchev$^1$, Noam Libeskind$^{1,2}$,  Vasily Belokurov$^{3,4}$, Matthias Steinmetz$^1$,  \\ Facundo A. Gomez$^{5,6}$, Robert J. J. Grand$^{7,8,9}$,   Yehuda Hoffman$^{10}$, Alexander Knebe$^{11,12,13}$,   \\ Jenny G. Sorce$^{14,15,1}$, Martin Spaare$^{16,1}$,  Elmo Tempel$^{17,18}$, Mark Vogelsberger$^{19}$}

\authorrunning{Khoperskov et al.}

\institute{$^1$ Leibniz-Institut für Astrophysik Potsdam (AIP), An der Sternwarte 16, 14482 Potsdam, Germany\\
        $^2$ University of Lyon, UCB Lyon 1, CNRS/IN2P3, IUF, IP2I Lyon, France \\
        $^3$ Institute of Astronomy, Madingley Road, Cambridge CB3 0HA, UK \\
        $^4$ Center for Computational Astrophysics, Flatiron Institute, 162 5th Avenue, New York, NY 10010, USA \\
        $^5$ Instituto de Investigación Multidisciplinar en Ciencia y Tecnología, Universidad de La Serena, Raúl Bitrán 1305, La Serena, Chile \\
        $^6$ Departamento de Astronomía, Universidad de La Serena, Av. Juan Cisternas 1200 Norte, La Serena, Chile \\
        $^7$ Max-Planck-Institut für Astrophysik, Karl-Schwarzschild-Str 1, D-85748 Garching, Germany \\
        $^8$ Instituto de Astrofísica de Canarias, Calle Váa Láctea s/n, E-38205 La Laguna, Tenerife, Spain\\
        $^{9}$ Departamento de Astrofísica, Universidad de La Laguna, Av. del Astrofísico Francisco Sánchez s/n, E-38206 La Laguna, Tenerife, Spain\\
        $^{10}$ Racah Institute of Physics, Hebrew University, Jerusalem 91904, Israel \\        
        $^{11}$ Departamento de Física Teórica, Módulo 15, Facultad de Ciencias, Universidad Autónoma de Madrid, E-28049 Madrid, Spain \\
        $^{12}$ Centro de Investigación Avanzada en Física Fundamental (CIAFF), Facultad de Ciencias, Universidad Autónoma de Madrid, E-28049 Madrid, Spain \\ 
        $^{13}$ International Centre for Radio Astronomy Research, University of Western Australia, 35 Stirling Highway, Crawley, Western Australia 6009, Australia \\
        $^{14}$ Universit\'e Paris-Saclay, CNRS, Institut d'Astrophysique Spatiale, 91405, Orsay, France \\
        $^{15}$ Univ. Lille, CNRS, Centrale Lille, UMR 9189 CRIStAL, F-59000 Lille, France\\
        $^{16}$ Institut für Physik und Astronomie, Universität Potsdam, Campus Golm, Haus 28, Karl-Liebknecht Straße 24-25, D-14476 Potsdam \\
        $^{17}$ Tartu Observatory, University of Tartu, Observatooriumi 1, 61602 Tõravere, Estonia \\
        $^{18}$ Estonian Academy of Sciences, Kohtu 6, 10130 Tallinn, Estonia \\
        $^{19}$ Department of Physics, Kavli Institute for Astrophysics and Space Research, Massachusetts Institute of Technology, Cambridge, MA 02139, USA
        }
        
\date{Received ; accepted }

\abstract{
Stellar chemical abundances and kinematics provide key information for recovering the assembly history of galaxies. In this work we explore the chemo-chrono-kinematics of accreted and in situ stellar populations, by analyzing six M31/Milky Way (MW) analogues from the HESTIA suite of cosmological hydrodynamics zoom-in simulations of the Local Group. We show that elemental abundances~(\FeH, \MgFe) of merger debris in the stellar haloes are chemically distinct from the survived dwarf galaxies, in that they are [$\alpha$/Fe]-enhanced and have lower metallicity in the same stellar mass range. Therefore, mergers debris have abundances expected for stars originating from dwarfs that had their star formation activity quenched at early times. Accreted stellar haloes, including individual debris, reveal \FeH and \MgFe gradients in the \ELz plane, with the most metal-rich, [$\alpha$/Fe]-poor stars, which have formed in the inner parts of the disrupted systems before the merger, contributing mainly to the central regions of the host galaxies. This results in negative metallicity gradients in the accreted components of stellar haloes at $\rm z=0$, seen also for the individual merger debris. We suggest, therefore, that abundance measurements of halo stars in the inner MW will allow constraining better the parameters, such as the mass and merger time, of MW's most massive merger Gaia-Sausage-Enceladus. The metallicity distribution functions (MDFs) of the individual debris show several peaks and the majority of debris have lower metallicity than the in situ stars in the prograde part of the \ELz space. At the same time, non-rotating and retrograde accreted populations are very similar to the in situ stars in terms of \FeH abundance. Prograde accreted stars show a prominent knee in the \FeH-\MgFe plane, reaching up to solar \MgFe, while retrograde stars typically contribute to the high-\MgFe sequence only. We find that the most metal-poor stars~($\FeH\lesssim-1$) of the HESTIA galaxies exhibit net rotation up to 80~\kmps, which is consistent with the Aurora population recently identified in the MW. At higher metallicities ($\FeH\approx -0.5\pm0.1$) we detect a sharp transition~(spin-up) from the turbulent phase to a regular disk-like rotation. Different merger debris appear similar in the \FeH-\MgFe plane, thus making it difficult to identify individual events. However, combining a set of abundances, and especially stellar age, makes it possible to distinguish between different debris.
}


\maketitle

\section{Introduction}
Although a substantial amount of stellar mass is formed in situ~\citep{2009ApJ...701.1765M,2010ApJ...725.1277M,2011ApJ...739...24B}, galaxies also grow via hierarchical clustering of smaller systems. Their stellar content at $\rm z=0$ represents all the systems accreted at different epochs~\citep{1991ApJ...379...52W,1993MNRAS.262..627L,2005Natur.435..629S,2010MNRAS.406.2267F,2012AnP...524..507F}. Earlier mergers, happening at high redshifts, appear to contribute to the diffuse inner stellar halo, while more recent ones can be seen as coherent streams, shells, and other unmixed structures~\citep{1994Natur.370..194I,1995ApJ...451..598J, 2005ApJ...635..931B,2006MNRAS.365..747A,2006ApJ...642L.137B,2010MNRAS.406..744C,2012ApJ...748L..24M,2007MNRAS.380...15F,2014ApJ...787...19M,2018ApJ...862..114S,2019ApJ...883L..32V,2019ApJ...886..109C}. In this scenario, stellar haloes are made of different stellar populations that spatially overlap with each other, thus making it hard to uncover their parental galaxy remnants. Using kinematic information does allow for the capture of some individual stellar substructures~\citep{1999Natur.402...53H,2001ApJ...557..137J,2004ApJ...610L..97H,2009MNRAS.399.1223S,2018MNRAS.478..611B}, however, since merging galaxies are affected by the dynamical friction and evolving gravitational field. As a result, different merger debris can be fully phase-mixed with each other~\citep{2005MNRAS.357L..35K,2008MNRAS.385..236V,2008MNRAS.385.1365L,2013MNRAS.436.3602G,2017MNRAS.464.2882A,2018MNRAS.473.1656T}. In this situation, the chemical abundances of stars are the only fossil records of the physical conditions of the star formation in the galaxies at different epochs. Therefore, an analysis of the chemical abundance patterns of resolved stellar populations in the halo with the phase-space data makes it possible to extract more precise information on  galactic building blocks~\citep{2006ApJ...646..886F,2021arXiv211002957C,2016ApJ...821....5D,2018AJ....156..179R,2020ARA&A..58..205H}.

Recent studies of the Milky Way~(MW) have undoubtedly shown that the inner stellar halo is dominated by accreted stars from a single dwarf galaxy~\citep[Gaia-Sausage-Enceladus, GSE,][]{2018MNRAS.478..611B,2018ApJ...863..113H,2018Natur.563...85H} accreted at $10-11$~Gyr ago along~\citep{2019NatAs...3..932G}, with dynamically heated in situ stars~\citep{2019A&A...632A...4D,2020MNRAS.494.3880B}. However, these two main building blocks are not the only components that contribute to the diffuse stellar halo. In particular, one more component was discovered soon after the GSE remnants, namely: Sequoia~\citep[][see also \cite{2019ApJ...874L..35M} and \cite{2020MNRAS.497.1236M}]{2019MNRAS.488.1235M}, which is characterised by a weak retrograde rotation while the GSE's remnant stars are on radial orbits. The sequoia remnant was discovered by using the globular clusters~(GCs) chemical abundances and kinematics and the same data sets were used to predict the presence of several other merger debris~\citep[see, e.g.][]{2019MNRAS.486.3180K,2020MNRAS.493..847F}. Although the presence of many other potential merger debris has been identified, the number of stars that can be associated with them is quite limited. In this context, it is worth mentioning Wukong~\citep{2020ApJ...901...48N}, Thamnos~\citep[][but see also \cite{2017A&A...598A..58H}]{2019A&A...631L...9K}, I'itol, and Arjuna~\citep{2020ApJ...901...48N}; however, the estimated masses of their parental galaxies are by a factor of $\approx 10$ smaller compared to the GSE progenitor~\citep[see Table 1 in][and references therein]{2022arXiv220409057N}.

\begin{figure*}[t]
\begin{center}
\includegraphics[width=1\hsize]{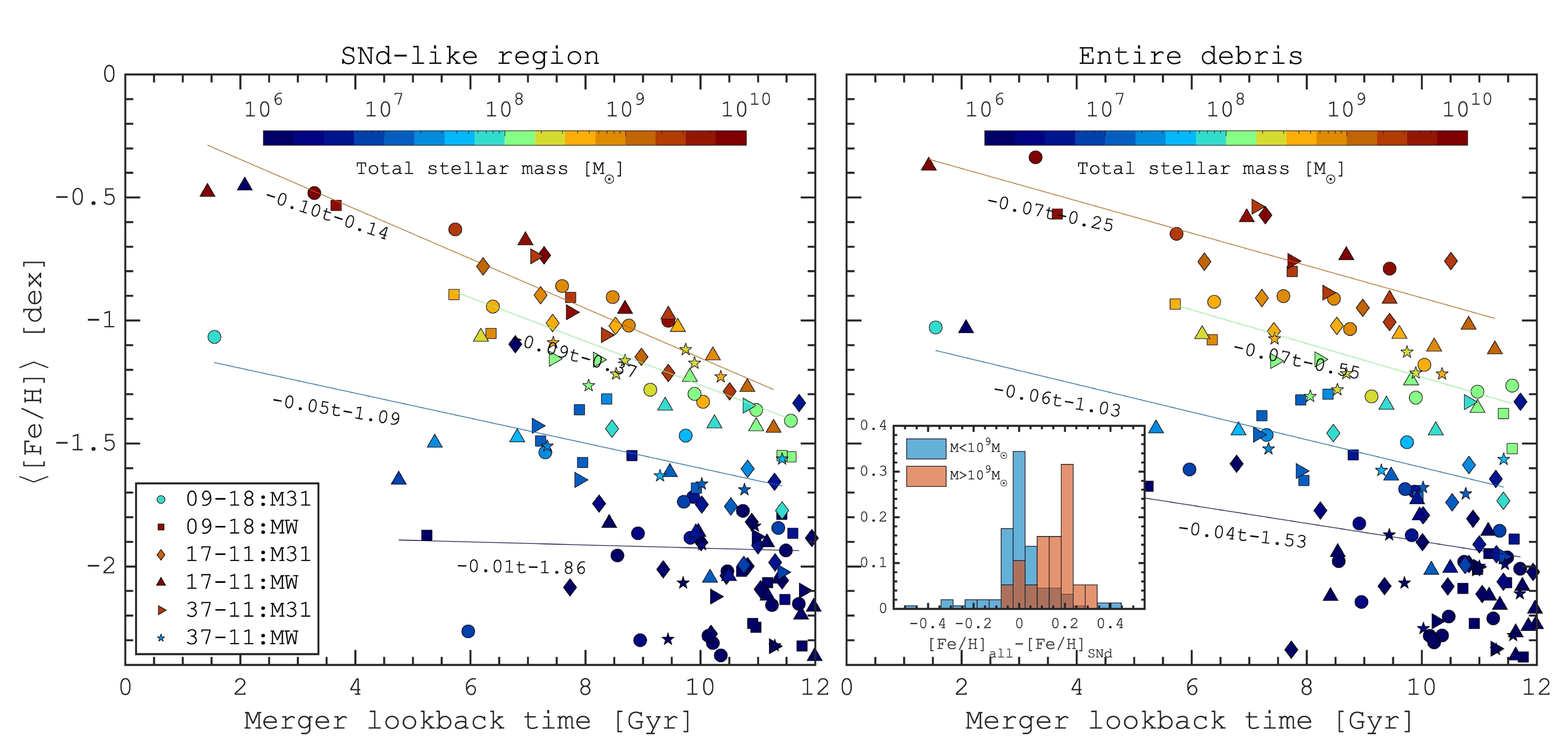}
\caption{Relation between mean metallicity and merger time for all the merger debris in the six HESTIA galaxies (indicated by different symbols in the bottom left of the left panel). The symbols are colour-coded according to the debris stellar mass. The left and right panels show the SNd-like and the entire population, respectively. In the right panel, the subplot depicts the difference between the mean metallicity of the entire debris and that of the SNd-like region~($(0.5-2) R_d$ and $|z|<10$~kpc, where $R_d$ is the disk scale length from \cite{2020MNRAS.498.2968L}). Two distributions are shown for the low-mass debris~($<10^9$~\Msun, blue) and the high-mass debris~($>10^9$~\Msun, red). The lines of different colours show the linear fits of the mean metallicity-and-merger time relation in several bins of the stellar mass.}
\label{fig3::feh_accr_time}
\end{center}
\end{figure*}

The above-mentioned substructures have mainly been identified by using combinations of the integrals of motions, orbital parameters and some chemical abundances. For instance, Wukong appears as a pair of overdensities in the \ELz coordinates with a cutoff at low \FeH~\citep{2020ApJ...901...48N}. At the same time, Wukong is not a prominent group of stars in the orbital parameters and \aFe-\FeH spaces. Thamnos's stars were selected using several circularity cuts in the retrograde region of the \ELz space where the debris is presumably made of two components with different \VV~\citep{2019A&A...631L...9K}. We note that, in contrast, for instance, to Wukong, there are no GCs found in the Thamnos selection regions. In the \aFe-\FeH space, while having slightly larger \MgFe, Thamnos substantially overlaps with the GSE and other substructures. Arjuna and I'itoi are a pair of substructures with the net retrograde rotation, again selected with sharp boundaries in \ELz coordinates, together with the Sequoia result in three peaks in the metallicity distribution function~\citep[MDF,][]{2020ApJ...901...48N}. Similarly to the previously reported structures, Arjuna and I'itoi, overlap in the \aFe-\FeH space and exhibit a broad range of orbital eccentricities $\approx 0.1-0.7$. 

The identification of prograde structures in the halo is even more difficult compared to the retrograde ones because of a strong contamination from the heated MW disc. Nevertheless, the list of potential mergers debris includes Helmi stream~\citep{1999Natur.402...53H}, Aleph~\citep{2020ApJ...901...48N}, Nyx~\citep{2020ApJ...903...25N}, Cetus~\citep{2009ApJ...700L..61N}, and Orphan/Chenab~\citep{2019MNRAS.485.4726K,2021ApJ...911..149L}. For instance, the Aleph substructure was identified as a sequence below the high-[$\alpha$/Fe] sequence, or as a continuation of the low-[$\alpha$/Fe] sequence ($\FeH \approx -0.5$), with a very low-eccentricity, typical for the disk stars. In the \ELz space, Aleph also looks like a continuation of the prograde disk populations. For a more comprehensive complete compilation of the halo substructures, we refer to recent works by \cite{2020ApJ...901...48N}, \cite{2022ApJ...926..107M}, and \cite{2022arXiv220410326M}.

In practice, a group of stars selected in the kinematical (chemical) space results in a broad distribution in chemical~(kinematical) coordinates. This problem was recently investigated, for instance, by~\cite{2022MNRAS.510.2407B}, who showed how different chemo-kinematic selections affect the resulting sample of potentially accreted stars in other coordinates~\citep[see also ][]{2021MNRAS.508.1489F,2022arXiv220404233H}. Another concern regarding the origin of the substructures identified in the MW halo, is that often a new group of stars is being 'discovered' because some of GCs (sometimes one or two) appearing in the same region of the \ELz. However, it is not evident that the GCs and the field stars of a dwarf galaxy, once it is undergoing accretion, experience the same orbital evolution~\citep[][]{2013MNRAS.433.1813S,Pagnini2022}, at least because of the dynamical friction affecting the motions of the globular clusters but with no impact on a low-density field stars~\citep{Pagnini2022}.

As mentioned above, a large number of substructures have been discovered in the halo of the MW and more features in the halo phase-space will be identified thanks to the publication of forthcoming Gaia data releases. Nowadays, there is a tendency to associate each halo substructure with an accretion of a dwarf galaxy. However, the link between a given substructure and the accreted dwarf galaxy is not so trivial. In particular, a set of difficulties of the halo substructures identification and their link to ancient merger events raises from the results of various types of simulations~(cosmological and tailored mergers of galaxies) which clearly show that: {\it i)} a single merger debris span over the large range in \ELz space~\citep{2013MNRAS.436.3602G,2017A&A...604A.106J,2019MNRAS.487L..72G,2019MNRAS.490L..32S}; {\it ii)} a single merger could result in a number of prominent overdensities in \ELz coordinates and other kinematic spaces~\citep{2017A&A...604A.106J,2019MNRAS.487L..72G,2020A&A...642L..18K,KhoperskovHESTIA-2}; {\it iii)} finally, both $E$ and $L_z$ are not conserved quantities in case of evolving and non-axisymmetric galaxy, thus the stars accreted $>8-10$ Gyr can change their orbital characteristics~(including actions) due to substantial mass growth of the galaxy~\citep{2021ApJ...920...10P,KhoperskovHESTIA-2} and/or close passages of massive satellites -- for instance, the LMC/SMC system~\citep{2021MNRAS.506.2677E,2021Natur.592..534C} or Sgr dwarf galaxy~\citep{2018MNRAS.481..286L} in the case of the MW.

\begin{figure*}[t]
\begin{center}
\includegraphics[width=1\hsize]{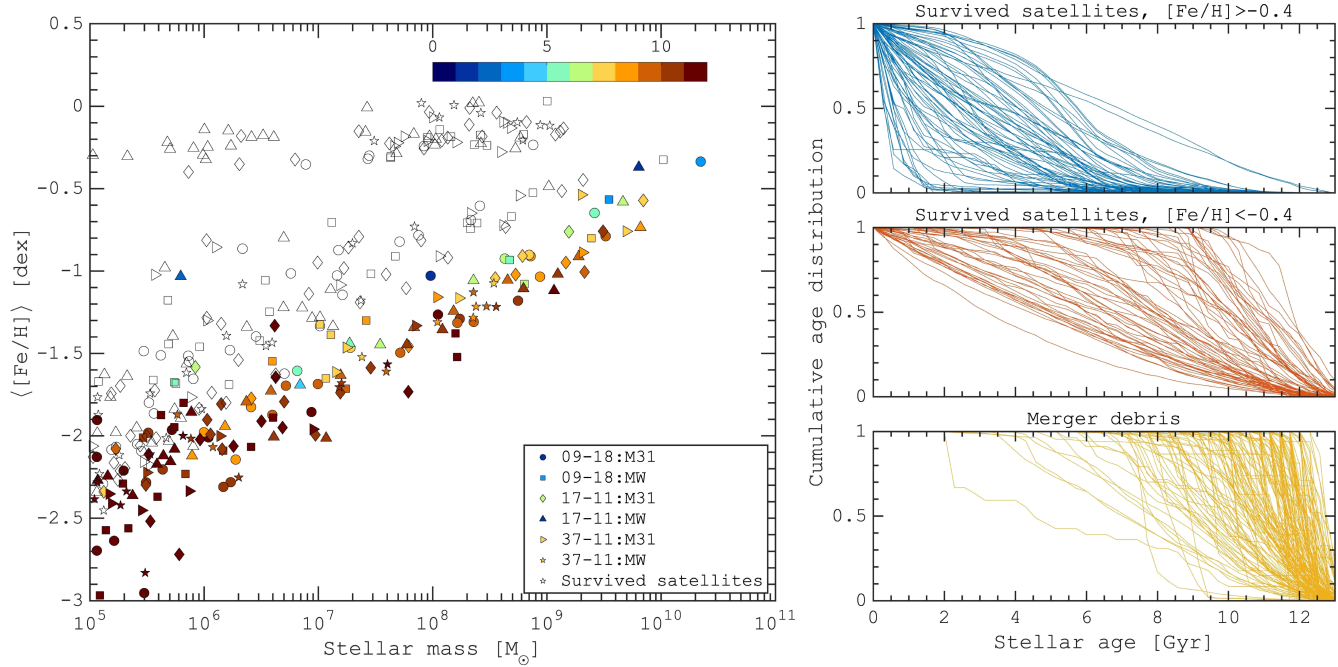}
\caption{Relation between stellar mass and mean stellar metallicity for all merged satellites~(coloured symbols) and surviving satellites~(empty symbols) for the six HESTIA M31/MW galaxies shown in the left panel. As in Fig.~\ref{fig3::feh_accr_time}, different galaxies are shown by different symbols, yet here they are colour-coded according to the merger lookback time, as indicated in the colour bar. The right panel shows the mass-weighted cumulative stellar age distribution for metal-rich satellites~(top), metal-poor satellites~(middle), and merger debris~(bottom). }\label{fig3::mass_metallicity}
\end{center}
\end{figure*}

In a series of works based on a new set of the HESTIA\footnote{https://hestia.aip.de} high-resolution cosmological simulations of Local Group~(LG) galaxies, we investigate the impact of the ancient mergers on the in situ stellar populations~\citep[][hereafter \citetalias{KhoperskovHESTIA-1}]{KhoperskovHESTIA-1} and phase-space evolution as well as the present-day structure of the merger debris~\citep[][hereafter \citetalias{KhoperskovHESTIA-2}]{KhoperskovHESTIA-2}. In this paper, we study the chemical abundance patterns as a function of the stellar ages and kinematics of both accreted and in situ stellar populations formed in six M31/MW analogues from the HESTIA simulations. The paper is structured as follows. The HESTIA simulations are briefly described in Section~\ref{sec3::all_model}. In Section~\ref{sec3::mergers_satellites} we analyse the relations between the mean stellar metallicity of the stellar debris and the mass and accretion time. We also compare the chemical composition of the merger debris and surviving satellites. In Section.~\ref{sec3::feh_in_Elz}, we focus on the metallicity variations of the individual debris in the \ELz space. In Section~\ref{sec3::afe_feh}, we analyse the properties of accreted and in situ stellar populations in the \MgFe-\FeH space as a function of stellar kinematics. In Section~\ref{sec3::more_elements}, we present the chemical abundances variations for a number of elements. Finally, in Section~\ref{sec3::concl}, we summarise our main results.

\section{Model}\label{sec3::all_model}

\subsection{HESTIA simulations}

In this study, we analyse the three highest resolution HESTIA simulations of the LG. Each simulation reproduces many LG properties, which have previously been described by \cite{2020MNRAS.498.2968L}. These include two massive disc galaxies resembling the MW and M31 analogues, as well as the population of orbiting smaller satellites at $\rm z=0$. 

The HESTIA simulations use the AREPO code~\citep{2005MNRAS.364.1105S, 2016MNRAS.455.1134P}, with gravitational forces computed using a hybrid TreePM technique, described in \cite{2005MNRAS.364.1105S}. The galaxy formation model used is from~\cite{2017MNRAS.467..179G}. HESTIA's cosmology is consistent with the best fit values~\citep{2014A&A...571A..16P}: $\sigma_8 = 0.83$ and $\rm H0 = 100~h\, km\, s^{-1} Mpc^{-1}$, where $h = 0.677$. Moreover, $\Omega_\Lambda = 0.682$, $\Omega_M = 0.270$, and $\Omega_b = 0.048$. We identified halos and subhaloes at each redshift by using the publicly available AHF\footnote{http://www.popia.ft.uam.es/AHF} halo finder~\citep{2009ApJS..182..608K}. We refer to the HESTIA simulations introductory paper~\citep{2020MNRAS.498.2968L} and \citetalias{KhoperskovHESTIA-1} for more details.

Throughout the paper, `in situ' refers to stars formed in the most massive M31 or MW galaxy progenitor, while `accreted' stars are formed in other galaxies and subsequently accreted onto the most massive one~\citep[see e.g.][]{2020MNRAS.497.4459F,2021MNRAS.503.5826A}. We define a merger event as the accretion of a dwarf galaxy that becomes gravitationally unbound from its own halo and gets bound to the main progenitor, according to the MergerTree tool from AHF. Therefore, all particles associated with a sub-halo right before this event are marked as accreted from a single merger and the last snapshot is used as the time of the merger. Smaller systems that are still bound at $\rm z=0$, while remaining inside the virial radius of the main progenitor, represent the population of surviving dwarf galaxies of M31/MW analogues.

\subsection{Orbital parameters calculation}

For each M31/MW simulated galaxy and every snapshot output, we defined a coordinate system $(x,y,z)$ centred at the position of the $10\%$ most bound in situ star particles and aligned with the principal axes of this in situ stellar component, such as the disk plane of the host galaxy located in the XY-plane and the z-axis along the direction of galactic rotation. Our study uses velocity in galactocentric cylindrical coordinates, tangential velocity, $V_{\phi}$, radial velocity, $V_r$, and velocity in the $z$ direction, $V_z$. We also made use of the integrals of motion, focusing on the angular momentum in the $z$ direction $L_z$, and the total orbital energy per mass $E$. To calculate the orbital eccentricity of star particles, we did not trace the motion of the particles across the snapshots, but instead integrated the orbits of star particles in a smooth potential. This allowed us to obtain the instantaneous values of the orbital parameters for each snapshot independently, which is not possible directly from the output data.

To compute the instantaneous orbital parameters of star particles, we used AGAMA~\citep{2019MNRAS.482.1525V} to model the smooth potential. In order to avoid the perturbation of orbits from massive satellites, we interpolated the galaxy potential using the particles associated with the host galaxy only. The potential due to dark matter and halo gas was represented by a symmetric expansion in spherical harmonics up to $l = 4$, while the potential of stars and the gaseous disk was approximated by an azimuthal harmonic expansion up to $m = 4$. Next, we integrated the orbits of star particles in this potential for $20$ Gyr. This long timescale was chosen to account for the many periods expected for halo particles with small orbital frequencies. The orbits were integrated using the eighth-order Runge-Kutta DOP853 integrator with an adaptive time step from AGAMA~\citep{2019MNRAS.482.1525V}.

\subsection{HESTIA: Merger histories}
The merger history of the HESTIA galaxies is explored in detail in \cite{KhoperskovHESTIA-1,KhoperskovHESTIA-2}~(see also \cite{Dupuy_etal} for an analysis of direction of accretion of satellites). The total number of mergers in M31/MW HESTIA galaxies varies within the range of $\approx 10-40$ for different galaxies; however, the number of massive mergers with $\mu_{*}>0.2$~(i.e., at least $1:5$ mergers) is only $1-4$~(see Fig. 4 in \cite{KhoperskovHESTIA-1}). For the analysis of individual mergers in this work, we selected the five most significant ones per galaxy, namely the mergers with the largest stellar mass relative to the host galaxy at the time of the merger. We refer to these mergers as `significant' because not all of them can be classified as `major' mergers. For most of the galaxies, the last significant merger has taken place $>8$~Gyr ago. This expectation is similar to what is considered for the MW~\citep[see][for review and references in the introduction]{2020ARA&A..58..205H}, with the single exception of the M31 analogue in the 09-18 simulation. In the analysis, we marked the five most significant mergers in each M31/MW galaxy as M1-M5, that is, from the earliest to the latest one. Their stellar masses lie in the range $5\times10^8 - 2\times 10^{10}$~\Msun, which corresponds to $0.05-0.96$ stellar mass relative to the host at the time of the merger. We note that some of these mergers are rather massive relative to our expectations regarding the merger history of the MW~\citep[see e.g.][]{2019MNRAS.490.3426D,2020MNRAS.498.2472K}; however, since the HESTIA simulations cover pairs of the M31/MW analogues, they are in agreement with the merger history of the M31 galaxy~\citep{2018NatAs...2..737D}.

\begin{figure*}[t]
\begin{center}
\includegraphics[width=1\hsize]{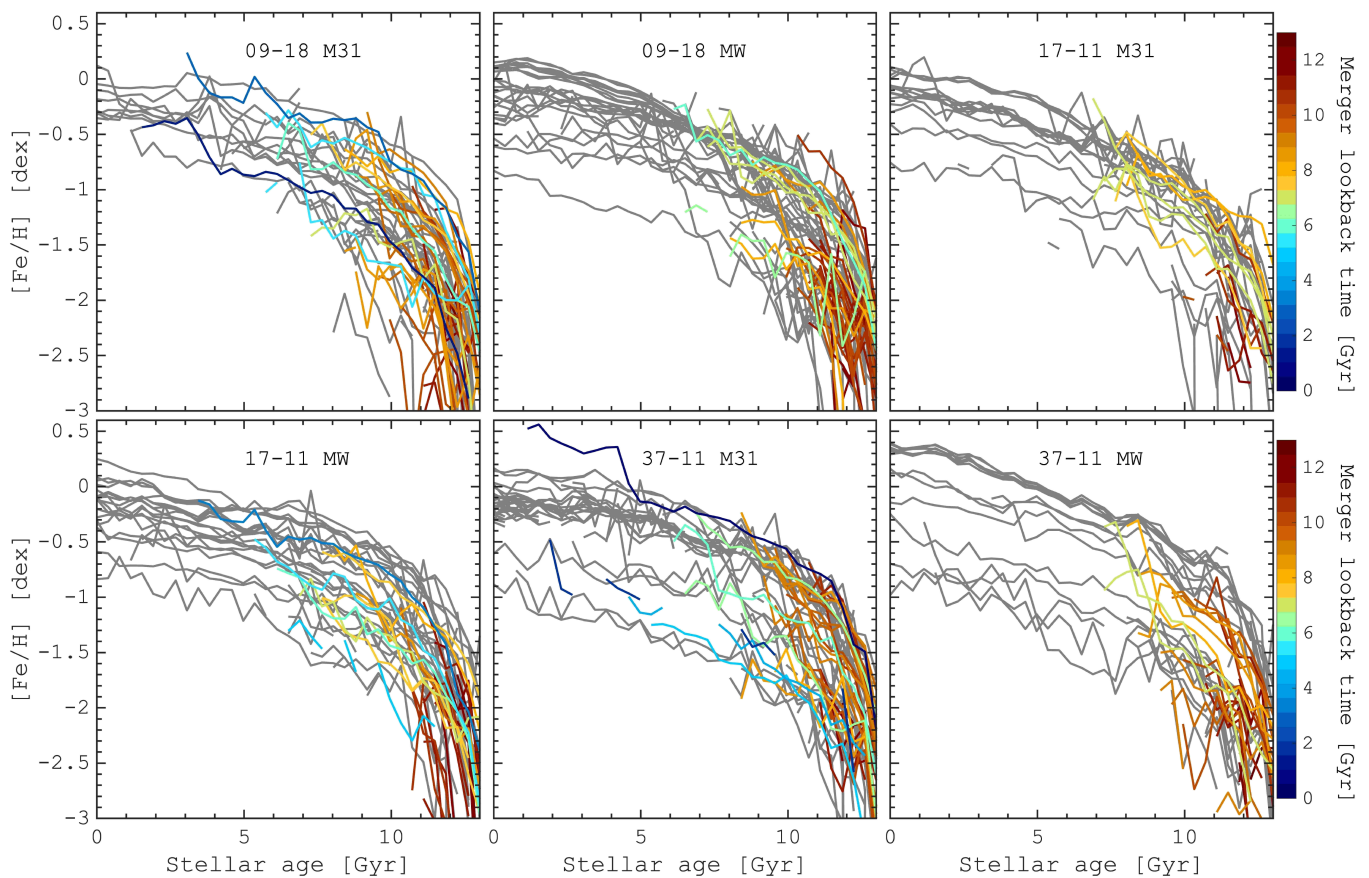}
\caption{Age-metallicity relation for the stellar merger debris~(coloured lines) and surviving satellites~(grey lines), for the six M31/MW HESTIA galaxies. The debris are colour-coded according to the merger lookback time.}\label{fig3::age_metallicity}
\end{center}
\end{figure*}

\section{Age-mass-metallicity relations for mergers debris and satellite galaxies}\label{sec3::mergers_satellites}
Chemical abundances of the merger debris are being used not only to identify substructures in the halo but also to recover the star formation history and the total mass of the parental dwarf galaxy~\citep[see e.g.][]{2018Natur.563...85H,2019MNRAS.487L..47V, 2019MNRAS.487.1462F,2019MNRAS.482.3426M,2021MNRAS.508.1489F,2021MNRAS.506.4321S}. These properties can be compared to the population of surviving dwarf galaxies orbiting around the MW. In some sense, both surviving satellites and those that have merged at different epochs are expected to experience similar evolutionary paths and to exhibit similarities in their chemical composition. However, in the MW, the elemental abundance measurements of stars in local dwarf galaxies show that these systems are chemically distinct from halo stars and often from other low-mass systems~\citep{2003AJ....125..684S,2009ARA&A..47..371T,2017ApJ...845..162H,2020MNRAS.497.4459F,2020ApJ...895...88N,2020A&A...641A.127R}. The observed differences suggest that the life paths of these systems are quite different; the evolution of merged galaxies stops once they merge and get buried in the host galaxy, while the surviving dwarfs experience a much longer evolution and enrichment history -- unless they have quenched their star formation.

\subsection{Metallicity-accretion time relation}
Before comparing the merger debris with the sample of dwarf galaxies in the HESTIA simulations, we look at the relation between the mean stellar metallicity of the merger debris as a function of the merger lookback time. In Fig.~\ref{fig3::feh_accr_time} we show the compilation of all the mergers in six M31/MW HESTIA galaxies colour-coded according to the total stellar mass of the debris. In the figure, we compare the relations based on the mean metallicity calculated for the solar neighbourhood~(SNd)-like region~(left)\footnote{We define the SNd-like region as a circular annulus $(0.5-2)R_d$ within $|z|<10$~kpc where $R_d$ is the radial scale length of the disk from~\cite{2020MNRAS.498.2968L}.} and the entire debris~(right).  First we can see a wide range of metallicities of the debris, from $\FeH\approx-2.5$ to $\FeH\approx-0.5$, where, as expected, lower-mass systems have on average lower mean metallicities. Also we find a clear trend where more massive debris have higher metallicity at the same accretion time and merger debris of similar masses have higher metallicity if they accreted later. These relations are somewhat expected because in order to form a more massive dwarf at the same time the SFR should be higher and thus, the chemical evolution needs to be faster and the metallicity reaches higher values on a shorter timescale. An interesting implication of the relations we find is that we can fit the mean metallicity as a function of the accretion time with a simple linear function in different mass bins~(see numbers in the figure). The fits we provide can be used to constrain the time of accretion of the merger debris. For instance, if we use the mass of the GSE of $10^{8.7}$~\Msun and the mean metallicity of $-1.18$~\citep{2022arXiv220409057N} then, following the HESTIA galaxies relation, we can estimate the accretion time of $\approx 11$~Gyr, which agrees well with other estimates~\citep{2018Natur.563...85H,2019A&A...632A...4D,2019NatAs...3..932G}.  As shown in Fig.~\ref{fig3::feh_accr_time}, the relation between the accretion time and mean metallicity of stellar debris, estimated in a SNd-like region, is slightly different from the one based on the entire debris. In this case, however, our estimate of $\approx 9$~Gyr is still consistent with the lower limit of  GSE accretion time estimates \citep{2019MNRAS.482.3426M,2020MNRAS.498.2472K, lu22}.

Another issue we address in Fig.~\ref{fig3::feh_accr_time} is the difference between the mean metallicity measured across the entire debris and in a local SNd-like region. As we show in the subpanel on the right, the difference usually does not exceed $0.2$~dex, however, it is slightly biased towards larger metallicities for the entire debris compared to the locally measured values. Although the difference we find is not significant for most of the mergers remnants, already at this point, we can see that the metallicity is not uniform across the debris, which we investigate in the subsequent section~(Sect.~\ref{sec3::feh_in_Elz}). Nevertheless, the  \FeH difference found for the most massive debris~(>$10^9$~\Msun) can reach up to $\approx 0.2$~dex. This mass range is close to the estimated mass of the GSE, and since its mass is based on the data biased towards the SNd, one could expect that the mean metallicity of the GSE debris is higher and, thus, its total stellar mass can be underestimated. Consequently, the GSE could have been accreted at a smaller redshift. Therefore, we suggest that in order to constrain better the GSE merger parameters, it is vital to analyse the abundances of accreted stars in the central parts of the MW~\citep{2020MNRAS.496.4964A,2021A&A...656A.156Q}, which will be soon accessible for the MOONS~\citep{2020Msngr.180...18G} and 4MOST~\citep{2019Msngr.175....3D} spectroscopic surveys.

\subsection{Mass-metallicity relation: Merger debris and surviving dwarf galaxies}
Dependence of the stellar metallicity on the mass is a fundamental relation reflecting the efficiency of the star formation and the physical conditions in galaxies~\citep{2003MNRAS.341...33K,2004ApJ...613..898T,2009ARA&A..47..371T,2013ApJ...779..102K}. In Fig.~\ref{fig3::mass_metallicity}, we show the relation between the mean metallicity of stars and the total stellar mass of merger debris~(coloured symbols) and surviving at $\rm z=0$ dwarf galaxies~(empty symbols), where the symbols correspond to the M31/MW HESTIA galaxies in three simulations (as noted in the legend). Once we combined the information about all the merger debris from all the galaxies in our sample, we could see a very narrow relation between their stellar mass and the mean metallicity. However, all the merger debris set up a 'metallicity floor' at a given stellar mass. A striking feature that can be seen in Fig.~\ref{fig3::mass_metallicity} is a split of the relation for the surviving satellites, where the most metal-rich dwarfs seen as a tight sequence with the metallicities from $\approx -0.4$ to $0$. The split of the relation is explained by the star formation history, which we show in the right panels of the figure and where we calculate a cumulative mass fraction as a function of the stellar age. As we can see, the metal-rich satellites~(top right) have extended star formation histories with a rather high intensity in a recent epoch~($<5$~Gyr). The star formation history for the low-metallicity satellites is qualitatively different because, in most cases, the active phase of the star formation stops at early times ($>5$~Gyr) or the star formation is quiescent on a very long time scale~(middle right). The SFH for the merger debris is even a more extreme case, whereby the star formation shuts down early on at the time of the merger. Therefore, the differences in the star formation histories aptly explain the mass-metallicity relation for the dwarf galaxies and the merger debris. In other words, the merger debris have the SFHs similar to the dwarf galaxies that quenched the star formation early on, and, as a result, their metallicities are more similar for the same stellar mass systems.

\begin{figure*}[t]
\begin{center}
\includegraphics[width=1\hsize]{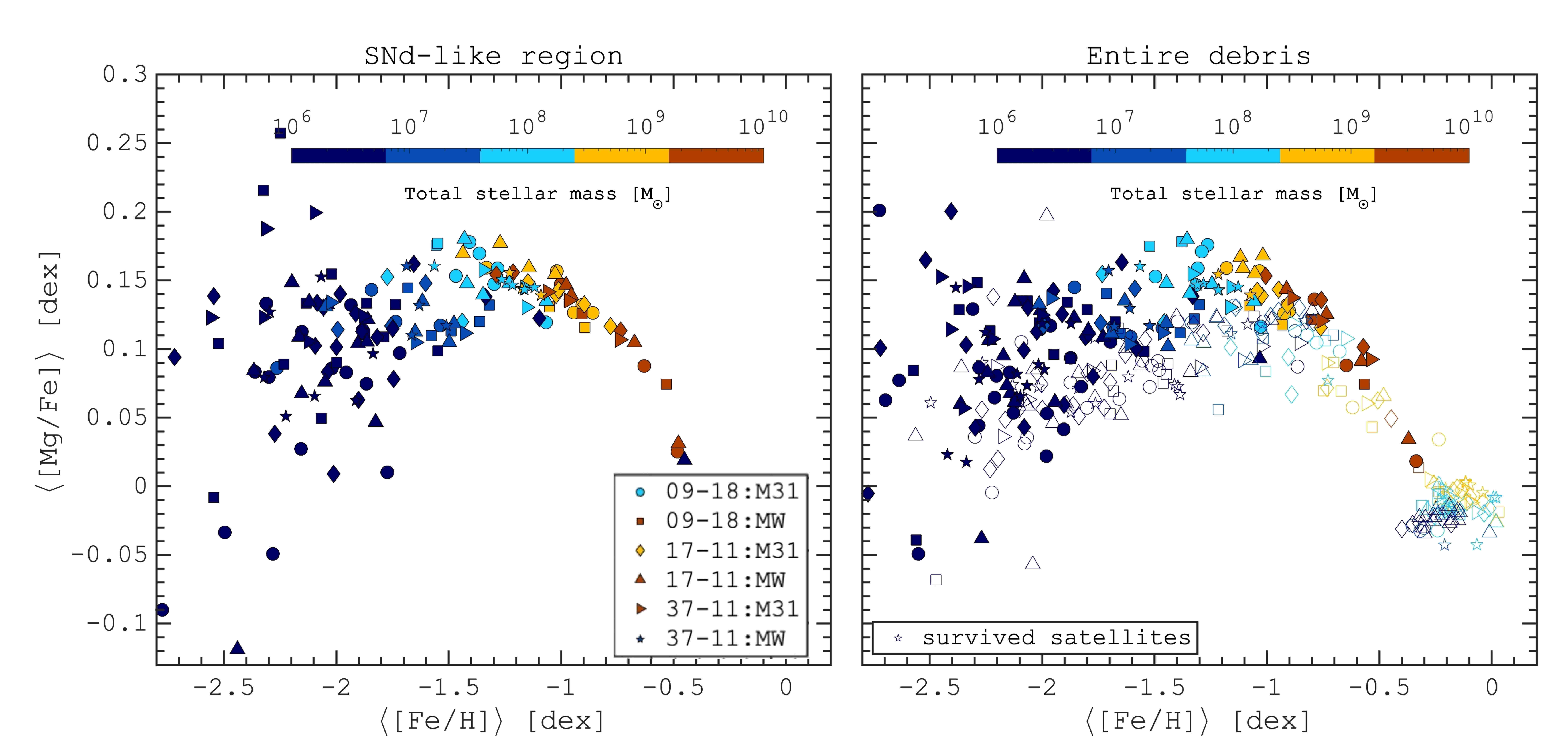}
\caption{Mean \MgFe abundance vs mean metallicity, <\MgFe>$-$<\FeH>, for the merger debris~(filled symbols) and surviving satellites ~(empty symbols). In both panels, the symbols are colour-coded according to the total stellar mass. The left panel, similarly to Fig.~\ref{fig3::feh_accr_time}, shows the relation for the SNd-like selection, while the right panel shows the values averaged over the entire system. }\label{fig3::MgFe_FeH_debris_satellites}
\end{center}
\end{figure*}

Next, we analysed the age-metallicity relation by comparing the merger debris with surviving satellites. In Fig.~\ref{fig3::age_metallicity}, we show the mean stellar metallicity as a function of the stellar age, where surviving satellites are shown by the grey lines and the merger debris are colour-coded according to the lookback time of the merger. This figure clearly explains the difference between the metallicity behaviour of dwarf galaxies and the merger debris. The enrichment of the merger debris stops much earlier compared to the dwarf galaxies, especially the ones that have extended chemical evolution. Meanwhile, at early times, the age-metallicity relations of the merger debris and surviving dwarfs are essentially the same; thus, they represent a similar type of high-redshift objects. If we turn this point around, the stellar population of dwarf galaxies with quenched star formation should represent the stellar populations of accreted systems buried in the halo. Therefore, we suggest that the analysis of old dwarf galaxies without recent star formation should offer a better idea about the chemical composition of the merger debris.

\subsection{\MgFe - \FeH relation for the merger debris and dwarf galaxies}

It is widely known that the $\aFe-\FeH$ relation provides a vital information about the star formation activity and the chemical enrichment variations~\citep[see][for the review]{2021A&ARv..29....5M}. In our analysis, we used \MgFe as a proxy of the [$\alpha$/Fe] elements. Similarly to Fig.~\ref{fig3::feh_accr_time}, in Fig.~\ref{fig3::MgFe_FeH_debris_satellites}, we show the \MgFe-\FeH plane for the mean abundances of the merger debris in the SNd-like region~(left) and the entire debris~(right). In the right panel, we also add the mean abundances of the surviving dwarf satellites, shown by the empty symbols. In both panels, the colour of the symbols indicate the total stellar mass. Although there is a large scatter of the mean abundances and significant overlap between the merger debris, it is evident that the dwarf galaxies are more metal-rich and have lower [$\alpha$/Fe] abundances. For the same stellar mass, disrupted galaxies, which are accreted early on, have little time to form their stellar content, which is likely not enough to enrich the ISM by the SNI and, thus, to reduce the [$\alpha$/Fe]-abundances of newly formed stars. On the other hand, dwarf galaxies have more time to evolve, and they have the same masses as the merged ones if the star formation activity is quiescent on a longer times scale. The differences in the star formation activity that lead to the observed \aFe relations are shown in Fig.~\ref{fig3::mass_metallicity}. Our results regarding the $\alpha-\FeH$ relation between the merger debris and the dwarf galaxies are in quantitative agreement with recent findings by \cite{2022arXiv220409057N}, who compared the merger debris abundances delivered by the H3 Survey with the MW dwarfs. 

\begin{figure*}[t]
\begin{center}
\includegraphics[width=1\hsize]{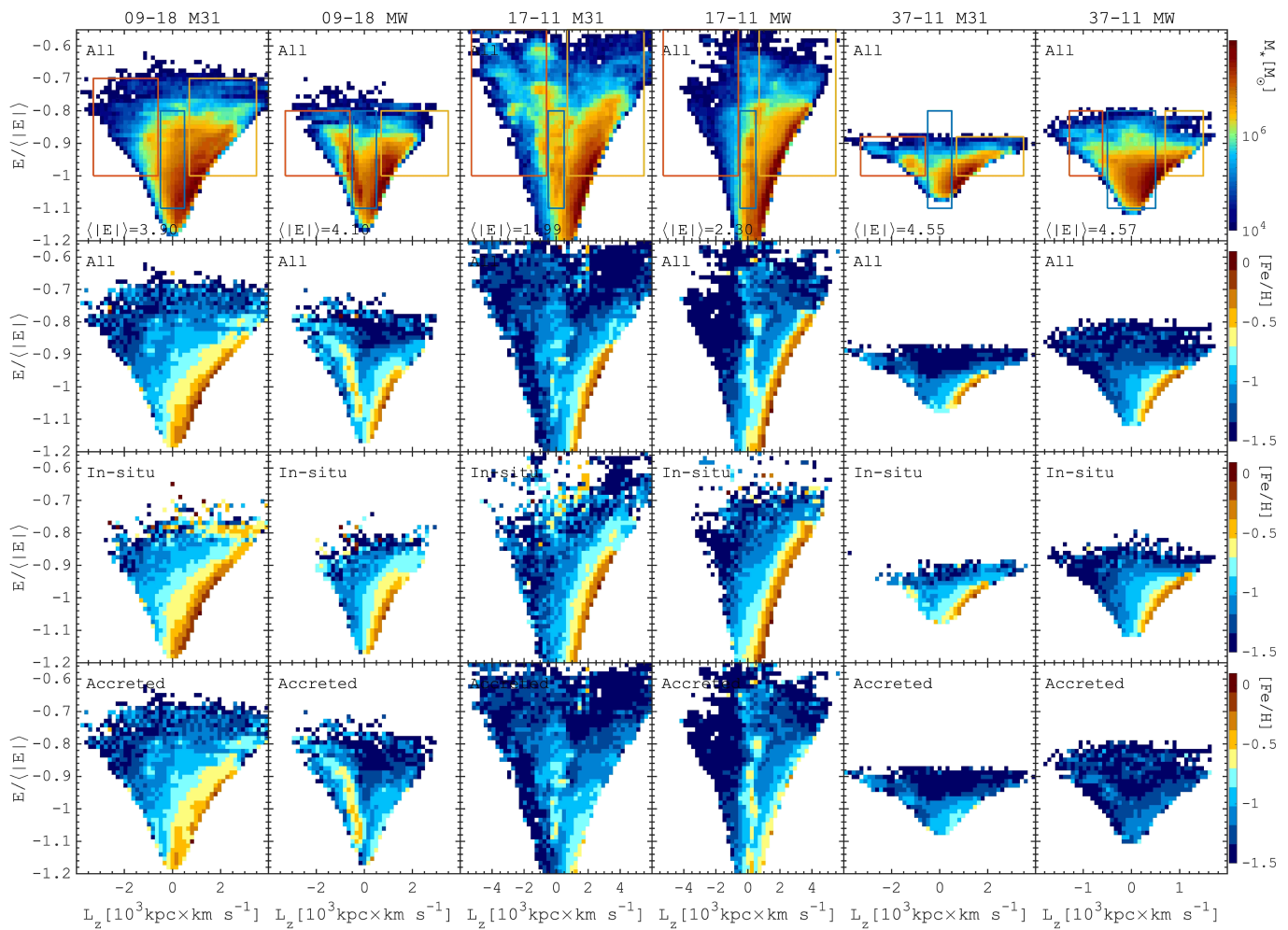}
\caption{Energy-angular momentum relation for the SNd-like region for all (top two rows), in situ (third row), and accreted (bottom row) populations. The first and second rows are coloured by the stellar number density and the mean stellar metallicity, respectively. In the bottom two rows, the colour corresponds to mean metallicity. The coloured boxes in the top row show the selection of retrograde~(red), non-rotating~(blue), and prograde~(yellow) stellar populations that are featured in Fig.~\ref{fig3::MDFs}. The energy is normalised to the mean energy of the corresponding sample for a given galaxy, marked at the bottom of each top panel and in units of $\rm 10^6\, kpc\times km^2s^{-2}$.}\label{fig3::ELz_selections}
\end{center}
\end{figure*}

\begin{figure*}[t]
\begin{center}
\includegraphics[width=1\hsize]{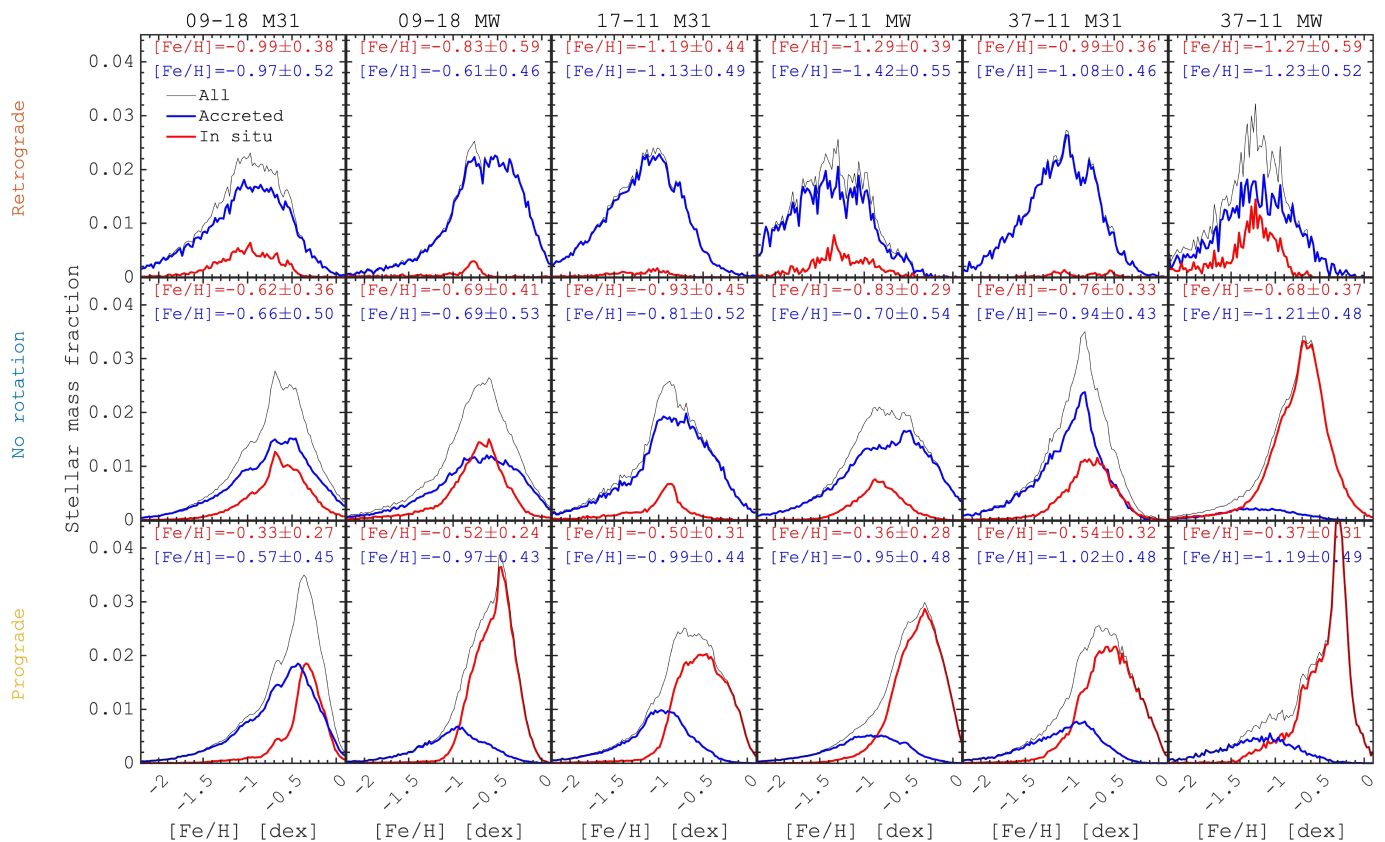}
\caption{Metallicty distribution functions (MDFs) for the retrograde~(top), non-rotating~(middle), and prograde~(bottom) stars selected in the \ELz coordinates in Fig.~\ref{fig3::ELz_selections}. All accreted and in situ populations are shown by the blue and red lines, respectively, while the black lines correspond to the overall distribution. The MDFs are normalised by the total mass of stars, including both accreted and in situ populations. Prograde in situ MDFs always peak at higher [Fe/H] compared to prograde accreted ones, while the two are mostly degenerate for the retrograde and non-rotating stars.
}\label{fig3::MDFs}
\end{center}
\end{figure*}

\begin{figure*}[t]
\begin{center}
\includegraphics[width=1\hsize]{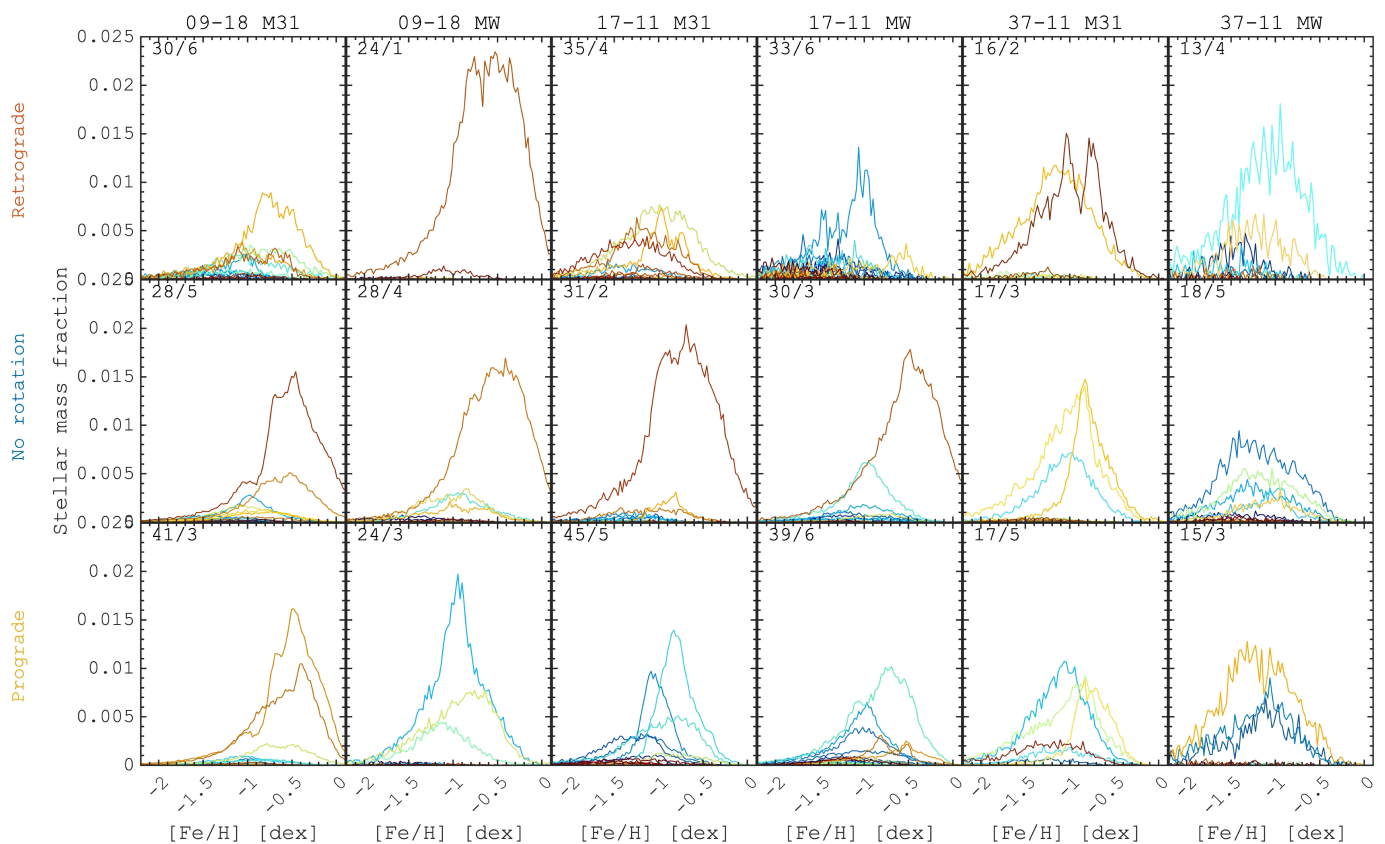}
\caption{MDFs for the retrograde~(top), non-rotating~(middle), and prograde~(bottom) stars but showing individual mergers in lines of different colours. The numbers in the top-left corner of each panel indicate the number of all mergers and the number of mergers, which contribute at least $5\%$ of the total mass of the accreted stellar populations in a given \ELz region. The MDFs are normalised by the total mass of accreted stellar populations. This figure shows that both retrograde and non-rotating accreted stars in the SNd-like region predominantly belong to just a few merger events.
}\label{fig3::MDFs_individual_mergers}
\end{center}
\end{figure*}

\begin{figure*}[t]
\begin{center}
\includegraphics[width=1\hsize]{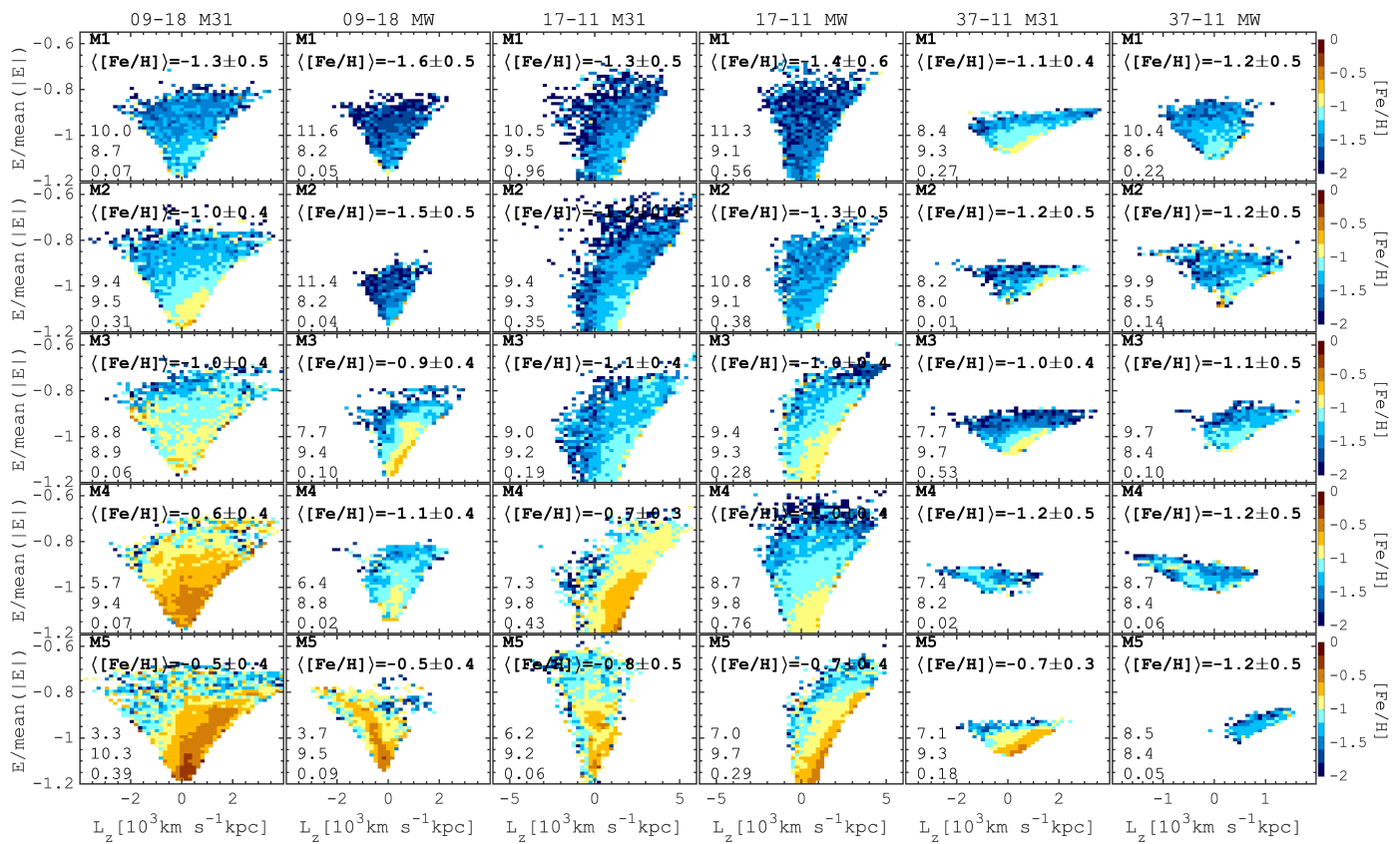}
\caption{Mean metallicity distribution for five of the most significant merger debris in \ELz coordinates. Different HESTA simulations are shown in different columns.  The merger accretion time~(Gyr), total stellar mass of the merger debris at the time of the merger ($\rm log_{10}(M_{*}/M_\odot)$), and the stellar mass ratio  ($\mu_{*}$) relative to the main M31/MW progenitor at the time of the merger are given in the bottom of each panel. The mean metallicity of debris associated with the given merger, together with its standard deviation, are shown in each panel. In most cases, individual merger debris show \FeH gradients, where the most metal-rich stars, formed in the centers of accreting galaxies, have the lowest energies. Meanwhile, low-\FeH stars formed in the outer regions of dwarf galaxies accreted first and, thus, have higher energy inside the host M31/MW galaxies.}\label{fig3::FeH_in_E_Lz_individual}
\end{center}
\end{figure*}

\section{Chemical abundance variations in \ELz}\label{sec3::feh_in_Elz}

\subsection{Metallicity distribution in \ELz for accreted and in situ stars}
In this section, we turn our attention from the mean chemical abundances of the merger debris to the metallicity and \MgFe variations of the accreted stars in several phase-space coordinates, which we also compare to the in situ stellar populations. As we already noted in the introduction, differences in metallicity in different regions of the \ELz space is often used to claim different dwarf galaxy progenitor for the MW halo substructures. In order to discuss the metallicity behaviour in the HESTIA galaxies similarly to the ones available in the MW, in Fig.~\ref{fig3::ELz_selections} we show the \ELz space for the SNd-like selection~(based on the circular annulus $(0.5-2)R_d$ within $|z|<10$~kpc). In the top row, we show the stellar density distribution, while the following rows present the mean stellar metallicity for all stars, both the in situ and accreted stellar populations. In the top panels, we can see a number of substructures in the \ELz space; however, as we showed in \citetalias{KhoperskovHESTIA-2}, not all these features correspond to individual merger events~\citep[see also][]{2017A&A...604A.106J,2019MNRAS.487L..72G}. In particular, the internal structure of a dwarf galaxy, prior to the merger, can result in the formation of different substructures once the galaxy is accreted~\citep[see, e.g.][]{2020A&A...642L..18K,2022arXiv220412187A}. The mean metallicity distribution is also not featureless. On top of the global \FeH diagonal gradient where the metallicity increases from larger energies to the lower and from negative $\rm L_z$ to the positive; there are several relatively high-metallicity features. These local metallicity peaks are the most prominent near $L_z \approx 0$ in both M31/MW 17-11 galaxies and in the retrograde tail of the MW analogue in the 09-18 simulation. The location of the high-\FeH spots suggest that they are linked to the accretion events. 

Two bottom rows in Fig.~\ref{fig3::ELz_selections} show the metallicity distributions for the in situ and accreted stars separately. It is clear that in most of the galaxies, the larger-scale diagonal gradient in \ELz~(see the second row in the figure) is mainly caused by the in situ stellar populations where the metallicity increases from the retrograde to prograde tails of the overall distribution. At the same time, the accreted stars display a diverse array of behaviours with regard to  the metallicity distribution, where the higher metallicity traces the location of the most massive, and, thus, more metal-rich debris. This debris is the most prominent on top of the overall distribution, especially if the accreted stars are not aligned with the host galaxy rotation. If the merger debris is prograde, then it is also difficult to distinguish the accreted stars from the in situ ones because the fraction of accreted stars in the prograde region is way smaller compared to the in situ component~(see more details in \citetalias{KhoperskovHESTIA-2}).

For a deeper understanding of the \FeH behaviour in the \ELz space, we analysed the metallicity distribution function~(MDF) of stars selected in three regions: retrograde, non-rotating, and prograde, which are marked by the red, blue and yellow rectangles, respectively, in the top row of Fig.~\ref{fig3::ELz_selections}. In Fig.~\ref{fig3::MDFs}, we present the \FeH distributions separating accreted~(blue lines) and in situ~(red lines) stars where the mean \FeH values are also marked in each panel. Interestingly, the MDFs are very similar in terms of the mean values in the retrograde part of the \ELz; however, the accreted stars dominate in this region. The mean \FeH is also very similar for the non-rotating stars selection~(second row), where (with a single exception: MW-37-11) the number of accreted stars is larger than the in situ ones. In this region, the accreted stars have a broader MDF. Therefore, the substantial overlap makes it very hard to disentangle between the accreted and in situ stars by using the metallicity information, only because here we show the stars with similar energy and angular momentum. The prograde region is the only one where we detected a prominent difference between the MDFs of accreted and in situ stars. Obviously, the in situ stars have a narrower distribution with a higher mean metallicity. Such behaviour is clear because at the time of accretion, all the merging dwarf galaxies are smaller than the main progenitor and, thus, they have lower metallicity~(see below for more details). Therefore, the HESTIA simulations, similarly to the case of the MW~\citep[see e.g.][]{2018Natur.563...85H,2019A&A...632A...4D}, suggest that the in situ stars which are heated up~(that is why they are in the same \ELz region) by the mergers are metal-rich compared to the merger debris.

\subsection{Origin of the \FeH variations in the \ELz plane for accreted stars and individual mergers}
In Fig.~\ref{fig3::MDFs} we can notice several peaks in the MDFs of distribution of the accreted stars, seen mainly in the retrograde part of the \ELz space. We recall that the presence of peaks in the MDFs is often used as an argument in favour of multiple distinct substructures in the MW halo; in other words, it is assumed that different peaks of the MDFs are associated with different progenitors~\citep[see e.g. ][]{2020ApJ...901...48N}. The relevance of such assumption is easy to check in the HESTIA simulations by comparing the MDFs of the individual merger debris in the same \ELz regions. In Fig.~\ref{fig3::MDFs_individual_mergers}, we show the contribution of all the merger debris to the MDFs of the accreted stars. The individual MDFs show that both retrograde and non-rotating accreted stars of the SNd-like region predominantly belong to just a few merger events. Meanwhile, in the prograde region~(bottom row), the number of significant mergers is way larger. This picture depends on the metallicity values, where at lower metallicities, the number of mergers increases, but their total mass fraction is rather low. Nevertheless, the MDFs are mainly shaped by just a few merger remnants, which, however, often show a complex behaviour. In some cases, for instance, M31 in 37-11, M31/MW in 09-18 simulations, the MDFs of the most significant mergers have a few prominent peaks, siggesting that some of the halo substructures in the MW may originate from a single progenitor, despite the fact that they have different metallicities and even linked to the peaks of the MDF.

The complex metallicity behaviour of the individual merger debris is also essential to explore in the \ELz coordinates.  In Fig.~\ref{fig3::FeH_in_E_Lz_individual}, we show the mean \FeH maps for five of the most significant merger debris for each M31/MW HESTIA galaxy~(see Fig.~4 in \citetalias{KhoperskovHESTIA-1}) by selecting stars in the radial range $(0.5-2)R_d$ and within $|z|<10$~kpc. The merger debris are sorted from the top to the bottom by decreasing the lookback time of accretion; in other words, the mergers accreted earlier are shown on the top. Before we discuss the distribution of the metallicity in the individual merger debris, we discuss a few general trends we observe in Fig.~\ref{fig3::FeH_in_E_Lz_individual}. In particular, the mean metallicity increases with the decrease of the lookback time of the merger, which suggests that a longer evolution of the dwarf galaxies prior to the merger allows them to be enriched to higher metallicities~(see Fig.~\ref{fig3::feh_accr_time}). However, the difference in the mean metallicity for different debris is substantial, especially in the case of the M31 galaxy in the 09-18 simulation, where the mergers occur on a long time scale~(left-most column). On the opposite, the difference in the metallicity of the merger debris is almost negligible in the case of the MW 37-11 galaxy~(right-most column), where all the significant mergers happen during $\approx 2$~Gyr.

The most striking feature of the distributions in Fig.~\ref{fig3::FeH_in_E_Lz_individual} is the metallicity gradient, optimally seen as a function of the total energy. Similarly to the total metallicity maps shown in Fig.~\ref{fig3::ELz_selections}, the mean stellar metallicity increases with lower energy, which suggests that more bound parts of the debris are more metal-rich. Since the most metal-rich stars of dwarf galaxies are located close to the center~(before the merger), where the star formation takes longer and possibly is more intense; thus, we can clearly see that the structure of the accreted galaxies is imprinted in the metallicity distribution of their stellar debris~\citep[see e.g.][]{2022arXiv220412187A}. The metallicity gradients are quite prominent in the debris of the galaxies already accreted $\approx 10$~Gyr ago, while for some earlier events, it is simply not enough time to create a gradient inside the dwarfs before they accreted. In all other cases, the gradients in the individual debris are quite prominent. This result again suggests that the metallicity variations are expected in the debris of massive dwarfs accreted onto the MW, with parameters similar to the GSE system.

\begin{figure*}[t]
\begin{center}
\includegraphics[width=1\hsize]{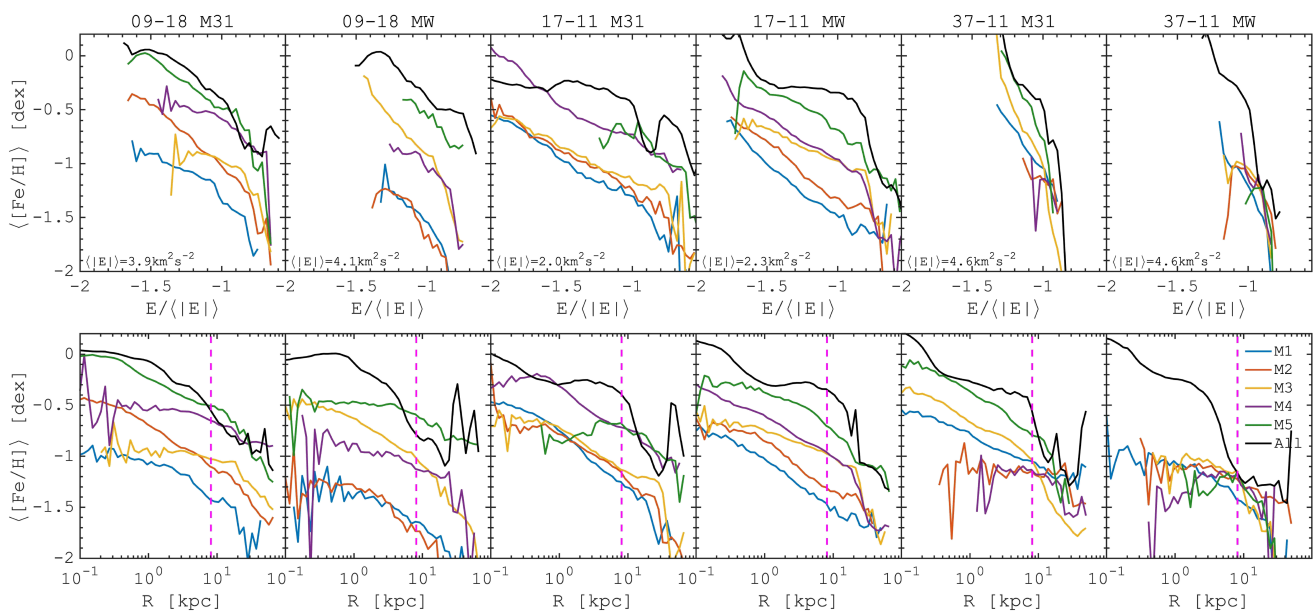}
\caption{Mean metallicity of the stellar debris for the five most significant mergers as a function of the total energy~(top row) and galactocentric distance~(bottom row). The chemical abundance variations within dwarf galaxies before the accretion result in the negative radial metallicity gradients in the main progenitors stellar halos once they are accreted.
}\label{fig3::FeH_E0_Rg}
\end{center}
\end{figure*}

Next, we elaborate more on possible observational consequences that can be used to interpret the \FeH variations in the \ELz space.  As we mentioned already, in Fig.~\ref{fig3::FeH_in_E_Lz_individual}, we  show the metallicity variations or gradients along with the total energy. To highlight this relation in Fig.~\ref{fig3::FeH_E0_Rg}~(top), we show the mean metallicity as a function of the total energy for the most significant merger debris~(coloured lines) and the total ones~(all the accreted stars in black). In all the HESTIA galaxies, we see a very prominent gradient of the metallicity and this gradient can be translated to the galactocentric distance because, on average, the lowest energy corresponds to the deepest parts of the potential well in the innermost parts of the host galaxy. 

In Fig.~\ref{fig3::FeH_E0_Rg}~(bottom), we present the mean metallicity as a function of the galactocentric distance, where we also draw a vertical line at the Solar radius of $8.2$~kpc. This figure demonstrates the metallicity gradients of the individual merger debris, together with the total accreted stars metallicity profiles, which show a clear decrease of the mean metallicity with radius~\citep[see also][]{2011MNRAS.416.2802F,2016MNRAS.459L..46M,2019MNRAS.485.2589M}. In some cases, we observe a flattening of the negative gradient, but some of the debris have a very steep profile. This is in agreement with some previous models suggesting that the metallicity of stellar haloes is flat along radius~\citep[see e.g.][]{2008MNRAS.391...14D,2010MNRAS.406..744C,2012MNRAS.419.2163G}, however, models that include the contribution of in situ star formation reveal stronger negative metallicity gradients~\citep{2011MNRAS.416.2802F, 2014MNRAS.439.3128T}. Our results suggest that once being accreted, more metal-rich parts of the disrupted galaxies are dragged towards the central parts of the host galaxy that creates the large-scale gradient~(see also Figs.~6 and 7 in \citetalias{KhoperskovHESTIA-2}). This effect is more evident for the most massive accretion events. Also, the metallicity gradient suggests that the metallicity of the massive debris taken in the SNd-like region is not meant to be representative of the mean metallicity of the debris~(see also Figs.~\ref{fig3::feh_accr_time} and ~\ref{fig3::MgFe_FeH_debris_satellites}), because the density of the debris increases towards the centre. Therefore, the measurements made in the MW should consider a possibility for the metallicity gradient of the stellar halo, even if it is dominated by a single debris~(e.g., GSE). At the same time, the mean metallicity of the debris in the SNd-like region cannot be used to precisely date the merger and estimate the total stellar mass of the parental galaxy. 

\section{In situ and accreted populations in the \aFe-\FeH space}\label{sec3::afe_feh}

\subsection{Chemical abundance space}

\begin{figure*}[t]
\begin{center}
\includegraphics[width=1\hsize]{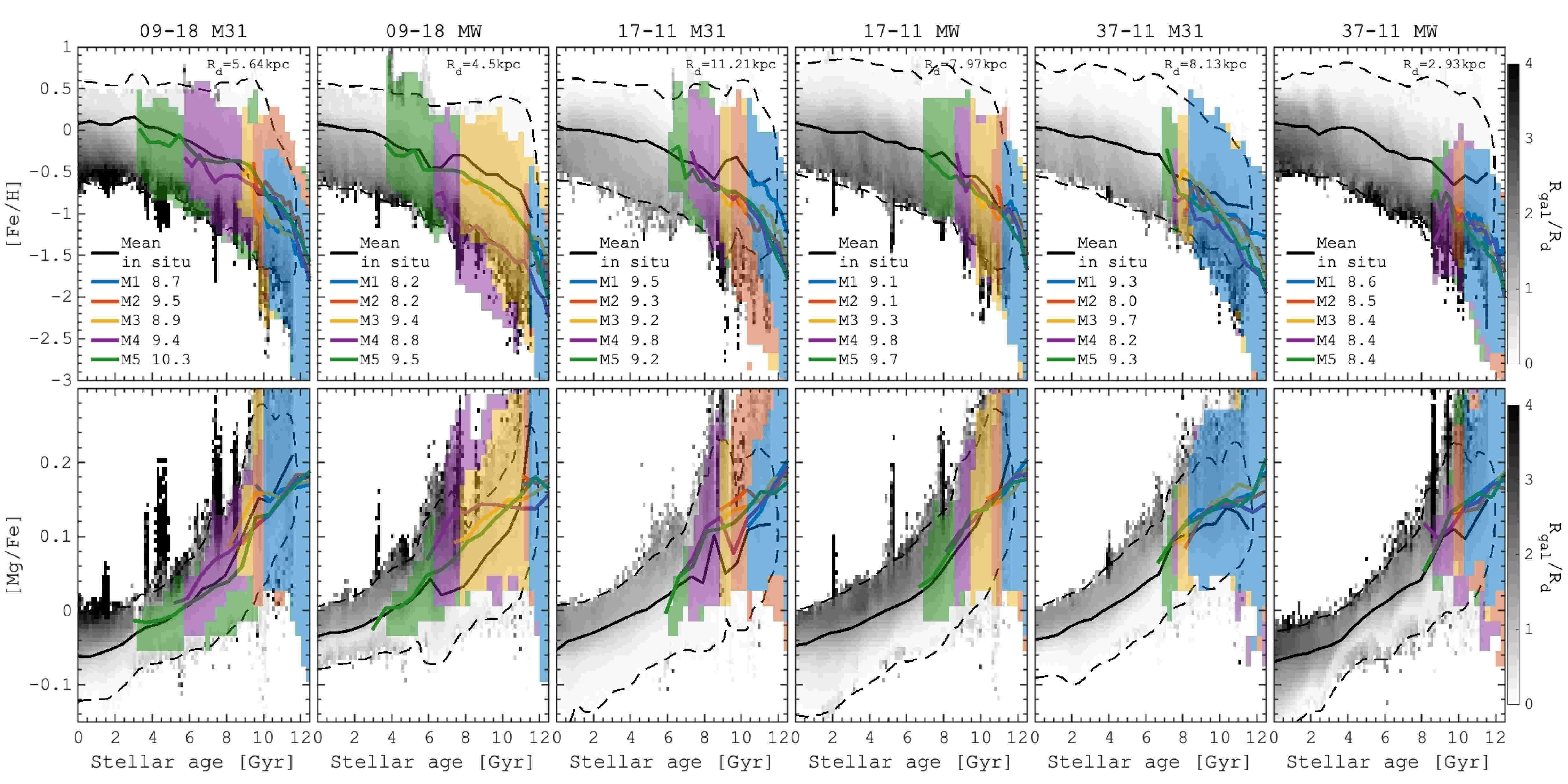}
\caption{Age-metallicity~(top row) and age-\MgFe (bottom row) relations for in situ stars~(grey shading) and the five most significant mergers~(coloured areas). The total stellar mass of the merger debris at the time of the merger ($\rm log_{10}(M_{*}/M_\odot)$) is mentioned in the legend in each top panel. Mean values are shown by the coloured solid curves~(except for the black one, which corresponds to the in situ stars). The grey-scale for the in situ stars represents the mean galactocentric distance in the units of the disc scale lengths from \cite{2020MNRAS.498.2968L}, as seen in the colour bars on the right, while the black dashed curves correspond to the $3\sigma$ around the mean. For the accreted populations, there is no shading and the boundaries correspond to $3\sigma$ around the mean.
}\label{fig3::age_met_with_insitu}
\end{center}
\end{figure*}

In this section, together with \FeH, we include the abundances of [$\alpha$/Fe] elements~(in particular \MgFe) and combine this information with the kinematics of stars to describe the properties of accreted and in situ populations.  In Fig.~\ref{fig3::age_met_with_insitu}, we present the age-\FeH and age-\MgFe relations for the in situ populations and five of the most significant merger debris. The in situ component is shown by the grey-scale distributions that are colour-coded by the mean galactocentric distance of stars at $\rm z=0$. The accreted stars are shown by the transparent filled areas where the mean age-abundance tracks are marked by the solid lines of the same colour. The yellow lines show the mean values for the in situ stars. The broad range of \FeH/\MgFe at a given age represents the fact that disc galaxies exhibit a gradient in metallicity with radius, which is often discussed in the context of inside-out galaxy formation~\citep{1976MNRAS.176...31L, 2010ApJ...710L.156R, 2012A&A...540A..56P, 2012MNRAS.426..690B}. In the age-\FeH relation, despite an overlap between the accreted and in situ stars of the same age, the accreted stars tend to have lower \FeH; in this sense, it is similar to the outer disc stellar populations. The offset between the mean trends of accreted and in situ stars at a given age does not exceed $0.2-0.3$~dex for the low-mass debris~(see Section~\ref{sec3::more_elements} for more details about the offset between the accreted and in situ populations). 

From Fig.~\ref{fig3::age_met_with_insitu} we can see that the maximum stellar metallicity of accreted dwarf galaxies reaches up the solar values at $\approx 10$~Gyr ago, while at the same time, the stars in the center of the main could have even higher values. Nevertheless, it is quite intriguing that the accreted stars could have a chemical composition of the Sun. Of course, the amount of such stars is quite small, but searching for them in observations is very important because it will allow us to for better constrains to be placed on the chemical evolution, mass growth, and star formation history of accreted galaxies. One unfortunate detail, however, is the large scatter of age-\FeH/\MgFe relations for the accreted populations, which also overlap with each other~\citep[see e.g.][]{2019MNRAS.482.3426M,2020MNRAS.495.2645B}, making it difficult to separate them when there are just a few abundances available.

\begin{figure*}[t]
\begin{center}
\includegraphics[width=1\hsize]{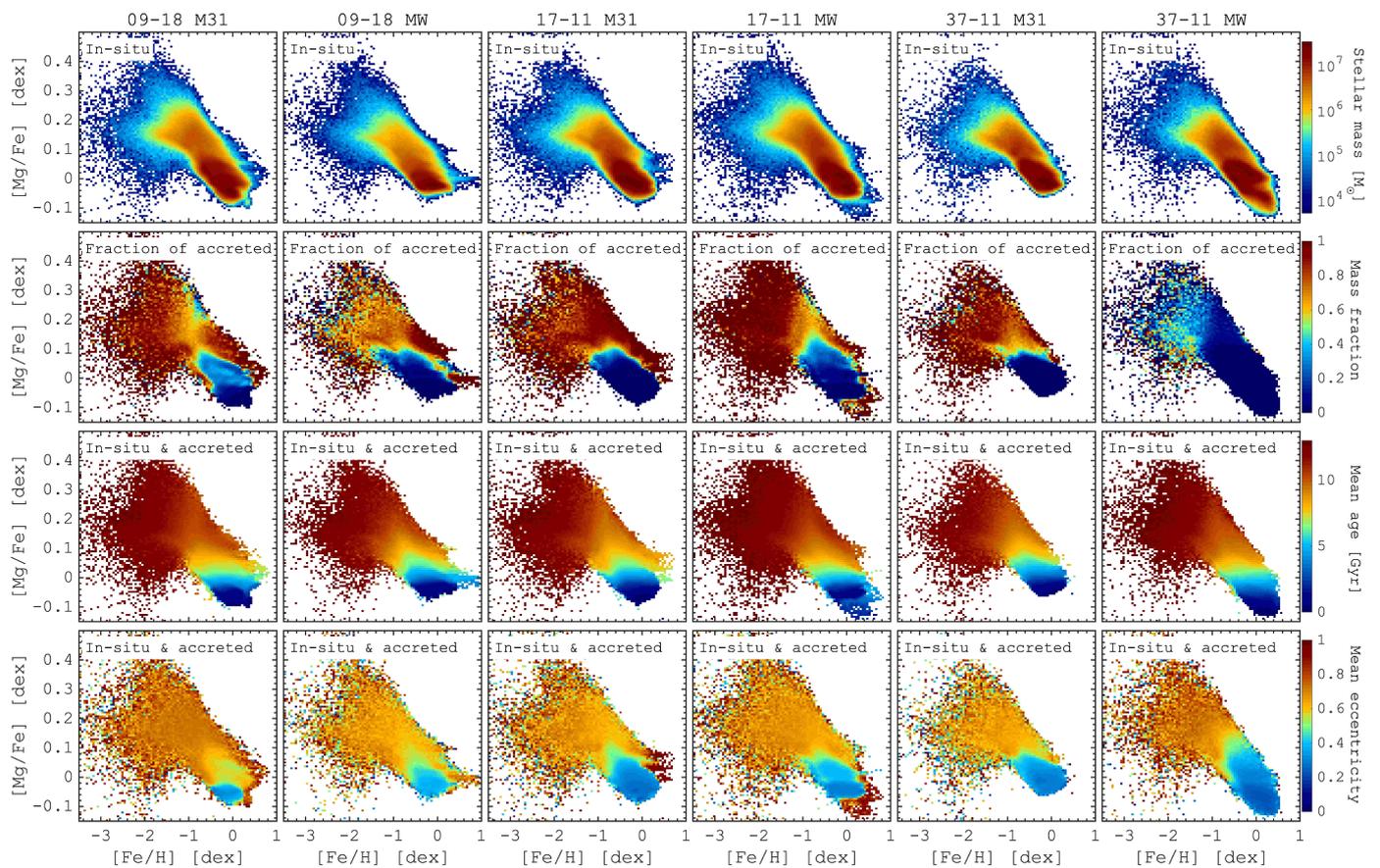}
\caption{\MgFe-\FeH relations for the six M31/MW HESTIA galaxies (different columns), including all stars. From top to bottom: Colour contours in different rows represent the stellar number density, the fraction of accreted stars, the mean stellar age, and the mean eccentricity.
}\label{fig3::mgfe_feh_small}
\end{center}
\end{figure*}

\begin{figure*}[t]
\begin{center}
\includegraphics[width=1\hsize]{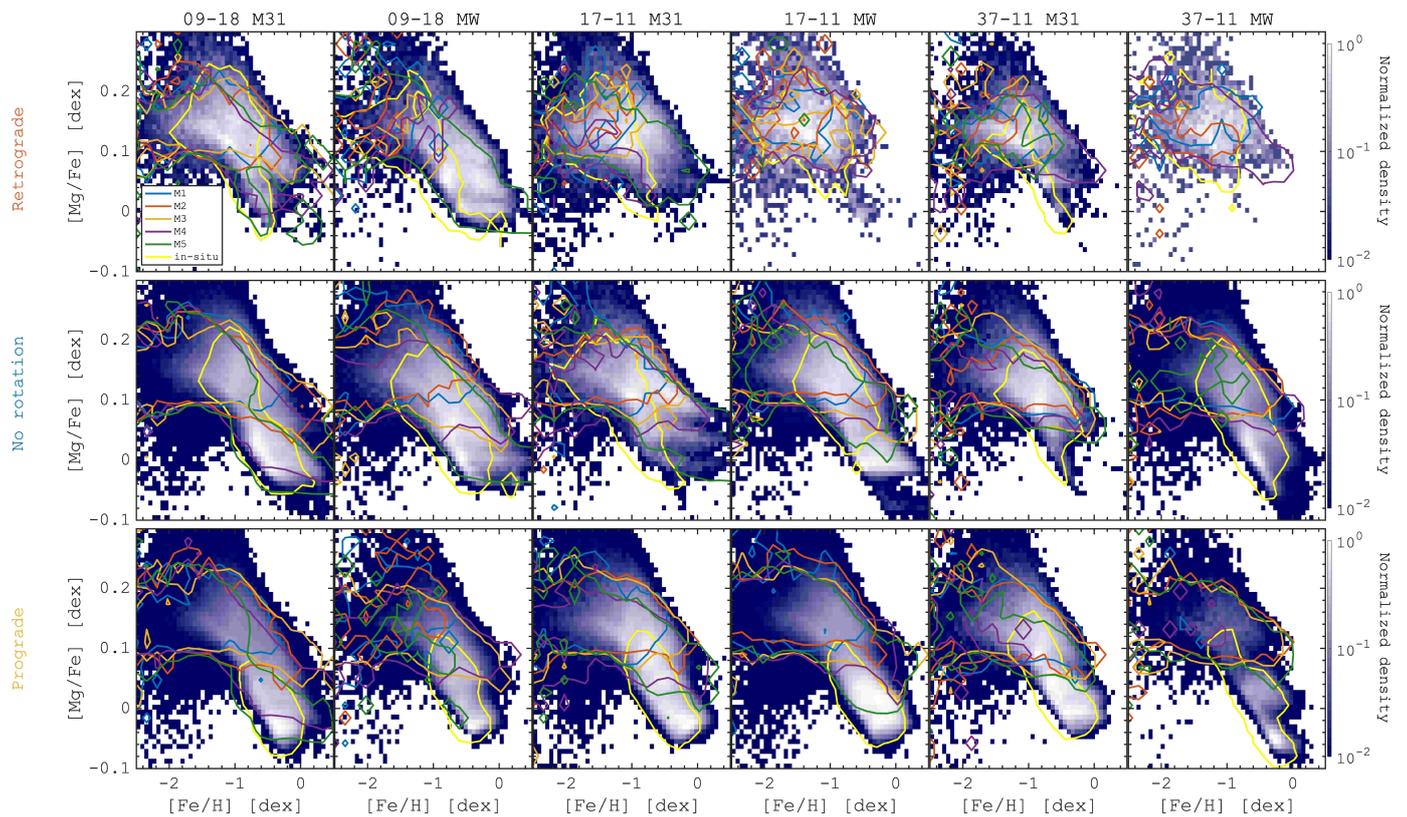}
\caption{\MgFe-\FeH relation for retrograde~(top), non-rotating~(middle), and prograde~(bottom) stars as selected in the blue, red, and orange rectangles in the top row of Fig.~\ref{fig3::ELz_selections}. The five most significant mergers are shown by colour-coded contours~($75\%$~density level), while the yellow contours correspond to the in situ stars. This figure shows that depending on the orbital properties, single merger debris can mimic chemically different substructures. }\label{fig3::mgfe_feh_from_ELz}
\end{center}
\end{figure*}

In Fig.~\ref{fig3::mgfe_feh_small} we show the widely explored in the literature \MgFe-\FeH relation for the simulated M31/MW HESTIA galaxies, selecting stars in the radial range $(0.5-2)R_d$ and $|z|<10$~kpc. In the top row, we plot the stellar mass distributions, which show several substructures that are similar to the ones recently explored, for instance, in cosmological simulations~\citep{2004ApJ...612..894B,2005ApJ...630..298B,2018MNRAS.474.3629G,2018MNRAS.477.5072M}. In particular, we can see a dichotomy of the \MgFe abundance at high metallicities~(\FeH>-0.5), similar to the MW high-[$\alpha/Fe$] and low-[$\alpha/Fe$] sequences. The exact mechanism of formation of the [$\alpha$/Fe] bimodality in the in situ stars is beyond the scope of the present work, however, some recent results and competitive scenarios can be found in \cite{2018MNRAS.474.3629G,2019MNRAS.484.3476C,2020MNRAS.491.5435B, 2021MNRAS.503.5868R, 2021MNRAS.501.5176K}. The most interesting in the context of the present study is the relation between the in situ and accreted stars. 

The second row of Fig.~\ref{fig3::mgfe_feh_small} shows the fraction of accreted stars in the \MgFe-\FeH plane, where the accreted stars occupy low-metallicity and high-[$\alpha$/Fe] regions for all galaxies. More quantitatively, the high-[$\alpha$/Fe] sequence is dominated by the accreted stars at \MgFe>0.1, with the single exception of the MW analogue in galaxy 37-11~(rightmost panels), where the accreted component contribution is minor compared to the in situ~(see also Fig.~4 in \citetalias{KhoperskovHESTIA-1}, which shows that this galaxy has a very quiet accretion history).

In some cases, however, we can see that the accreted stars can dominate in the super-solar metallicity regions, being likely formed in the cores of the dwarf galaxies, thus, they can have a longer enrichment history. This is suggested by the third row of Fig.~\ref{fig3::mgfe_feh_small}, where we show that the distribution of the mean stellar age and the high-\FeH knots of the accreted stars have ages comparable to the low-[$\alpha$/Fe] sequence of in situ stars. More generally, regions with a higher fraction of accreted stars obviously contain older stars, and the mean stellar age decreases towards smaller \MgFe values and higher metallicity. Nevertheless, the orbital parameters of the accreted and in situ stars are very much different in the high-\FeH regions, as illustrated in the bottom row of Fig.~\ref{fig3::mgfe_feh_small}, colour-coded by the mean stellar eccentricity. The regions with a dominant fraction of accreted stars show higher orbital eccentricity~($>0.5$). At the same time, the low-[$\alpha$/Fe] in situ populations have disk-like kinematics\footnote{For a detailed comparison of the orbital eccentricities of accreted and in situ stars we refer the reader to \citetalias{KhoperskovHESTIA-2}}. 

\begin{figure*}[t]
\begin{center}
\includegraphics[width=1\hsize]{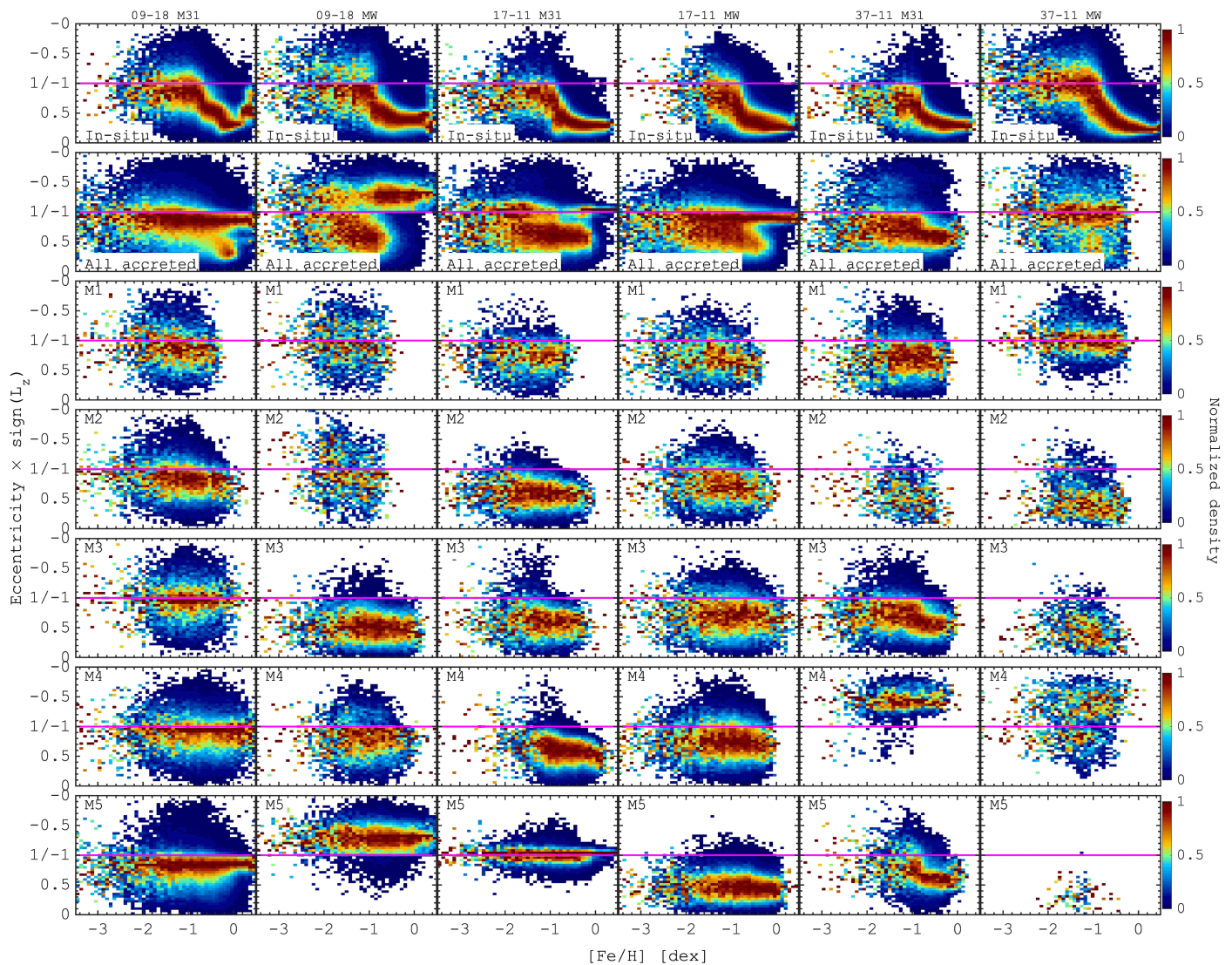}
\caption{Relation between orbital eccentricity and \FeH for the in situ~(first row), accreted~(second row), and the five most significant merger debris~(third through sixth rows). For improved visibility, the density maps are normalized in each metallicity bin. Stars with negative angular momentum have negative values for eccentricity. In other words, the stars above the pink line in each panel are counter-rotating. Although the eccentricity distribution of accreted stars shows a complex behaviour, the individual merger debris have nearly constant eccentricity as a function of \FeH. The features of the in situ stars are linked to the merger events that heat up the disc and, thus, increase the eccentricity of the pre-existing populations. }\label{fig3::feh_ecc_insitu_accr_individual}
\end{center}
\end{figure*}

As we have already shown, different merger debris overlap in the age-\FeH/\MgFe relations~(see Fig.~\ref{fig3::age_met_with_insitu}). Thus, similar behaviour is expected in the \MgFe-\FeH plane. In order to add more dimensions to purely chemical abundance relation in Fig.~\ref{fig3::mgfe_feh_from_ELz} we show the \MgFe-\FeH for the stars taken in three different regions of the \ELz diagram. We also display in  Fig.~\ref{fig3::ELz_selections} three rectangles show the retrograde~(blue), non-rotating~(red), and prograde~(yellow) regions. The contours in the panels of Fig.~\ref{fig3::mgfe_feh_from_ELz} show the distributions of stars from five of the most significant mergers~(coloured contours), together with all accreted stars~(colour map) and in situ stars~(yellow line). We can see that the overall relations are somewhat different depending on the orbital parameters of stars. For instance, the metallicity peak of a given merger debris is shifting towards higher values from the retrograde to prograde stars. This also affects the shape of the relation, where for the retrograde stars, we missed a low-[$\alpha$/Fe] sequence for the in situ stars suggesting that the high-[$\alpha$/Fe] stars, being formed earlier, are the most contaminated by the mergers~(see more details about the mergers impact on the in situ populations in \citetalias{KhoperskovHESTIA-1}). In most cases, a single merger debris have slightly different distributions in retrograde, non-rotating and prograde regions of \ELz. In particular, in the retrograde region the merger debris seems not to have a prominent knee, mostly occupying the high-[$\alpha$/Fe] sequence, while in the non-rotating and especially in the prograde part, the knee is quite prominent, at least for the most massive merger debris. The explanation of such a differentiation is based on the accreting galaxy structure we already discussed above. Most likely, the retrograde part of the \ELz space is occupied by older stars formed earlier from the outer parts of the dwarf galaxies before they merged into the host. In \citetalias{KhoperskovHESTIA-2}~(see Fig. 6), we showed that the retrograde parts of the \ELz are mainly made of stars accreted from the outer parts of the dwarf galaxies, which, in most cases, have a lower-\FeH and do not thus experience long chemical enrichment history compared to the prograde parts of the debris, which are mainly made of the central regions of dwarf galaxies where the SNI were able to pull down the [$\alpha$/Fe]-abundances to nearly solar values. Of course, there are some exceptions in the HESTIA galaxies sample, where the satellites are being accreted on the retrograde orbits~(see e.g. MW in the 09-18 simulation). However, the main chemo-kinematic trends for these merger debris remain essentially the same.

\begin{figure*}[t]
\begin{center}
\includegraphics[width=1\hsize]{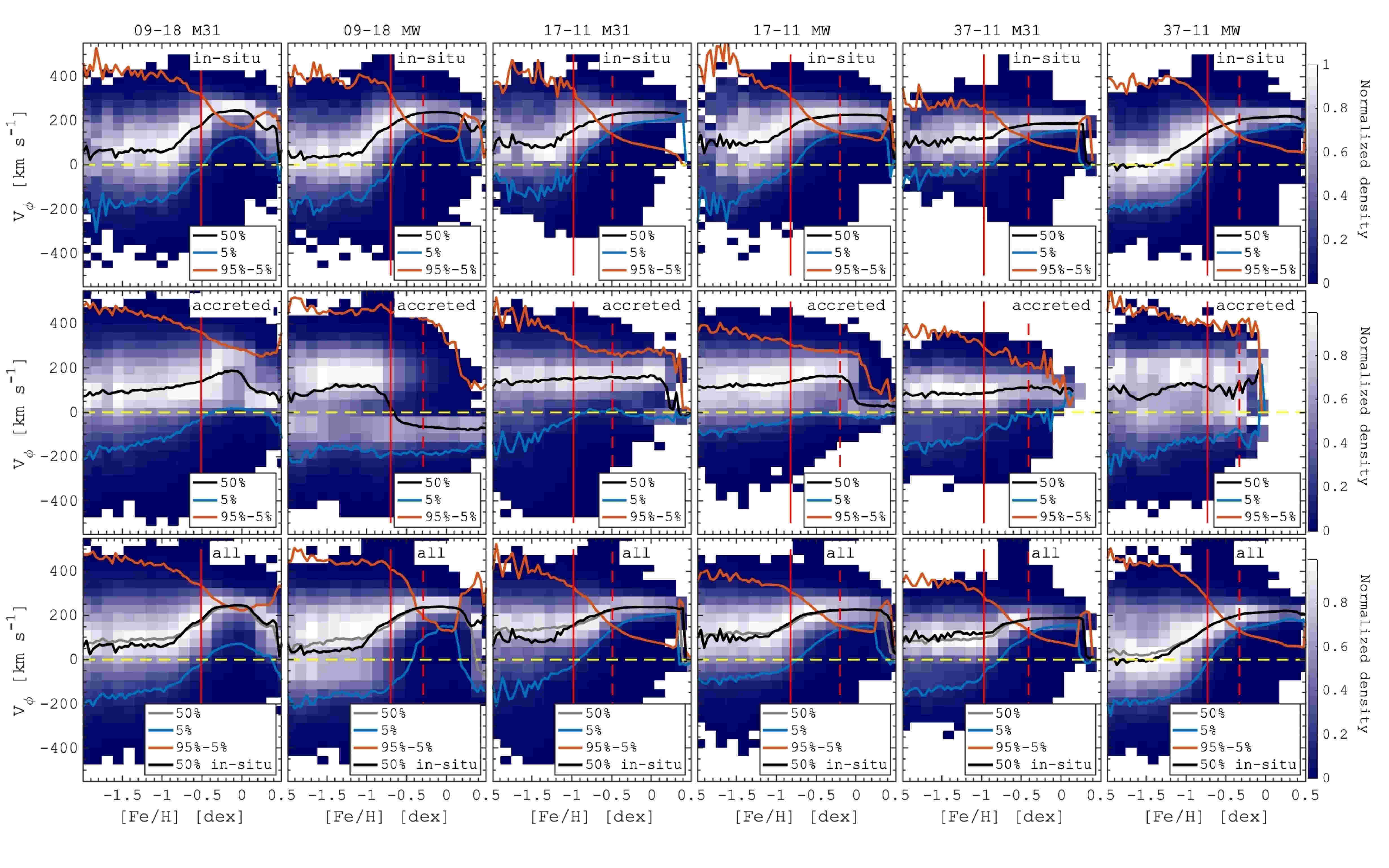}
\caption{Azimuthal velocity \VV as a function of metallicity: In situ~(top) and all accreted~(bottom) stellar populations. Black lines show the mean azimuthal velocity values, while blue and red show $5\%$ and $95-5\%$ statistics of the velocity distribution. The red vertical solid and dashed lines highlight the boundaries similar to the ones used to identify the Aurora and Splash populations in the MW~\citep[][]{2022arXiv220304980B}. The HESTIA galaxies show a behaviour similar to the MW with a little rotation of the low-\FeH~(<-1.2...-0.8) stars and a sharp spin-up of the net rotation for more metal-rich stars~($\FeH<-0.5$).}\label{fig3::vphi_feh}
\end{center}
\end{figure*}

\subsection{Coupling chemical abundances with orbital parameters of in situ and accreted populations}

In this section, we combine the orbital parameters and the kinematics of stellar populations and present their variations as a function of \FeH abundances. In Fig.~\ref{fig3::feh_ecc_insitu_accr_individual}, we show how the stellar eccentricity depends on the metallicity for the in situ, accreted and individual merger debris. In these figures, we extend the eccentricity scale to the negative values by multiplying it by the sign of the angular momentum. As a result, the top parts~(above the pink lines) in each panel show the relations between the stars on retrograde orbits. The in situ stellar populations~(top) show a very complex behaviour of the eccentricity as a function of \FeH. For the low-\FeH stars the eccentricity is very high, reaching up to the unity with a non-negligible fraction of the retrograde stars. For the higher-\FeH, the eccentricity drops down to $\approx 0.2$ in all the galaxies in our sample. Interestingly, in some cases, this transition is not smooth and we can see a few segments of nearly constant eccentricity in a range of \FeH. In \citetalias{KhoperskovHESTIA-1}, we showed that these sharp changes of the eccentricity as a function of the stellar age are linked to the merger events~(see Fig.~13 therein) where the oldest ones represent the stellar populations with kinematics similar to the Splash/Plume discovered in the MW~\citep{2019A&A...632A...4D,2020MNRAS.494.3880B}.

In Fig.~\ref{fig3::feh_ecc_insitu_accr_individual}, the accreted stars show even more complex behaviour compared to the in situ populations. We can see bimodal distributions and broadening and thickening of the distributions in some ranges of abundances. In most of the abundance ranges, the eccentricity of accreted stars is higher compared to the in situ stars, however, it is not always the case. In all the galaxies we can see the presence of rather regularly rotating accreted populations with the eccentricity down to $\approx 0.5$. However, the complexity of the relation breaks once we decompose the accreted stars onto the individual merger debris. Starting from the third rows in Figs.~\ref{fig3::feh_ecc_insitu_accr_individual} we present the eccentricity variations for the most significant debris. In this case, contrary to the entire sample of accreted stars, the individual debris are characterised by a nearly constant eccentricity along the \FeH. Most of the debris have a broad range of eccentricities with some contribution to the retrograde regions and there are few individual merger debris that are predominantly counterrotating~(e.g. M5 in MW 09-18 and M4 in M31 37-11). The results presented in Fig.~\ref{fig3::feh_ecc_insitu_accr_individual} suggests that although the accreted stars have a complex distribution of the orbital eccentricities, the individual merger debris tend to have roughly constant eccentricity along the \FeH. Therefore, once the scatter of about $0.1-0.2$ of the eccentricity is taken into account, this behaviour can be used to trace the stars associated with a single merger event. 

\begin{figure*}[t]
\begin{center}
\includegraphics[width=1\hsize]{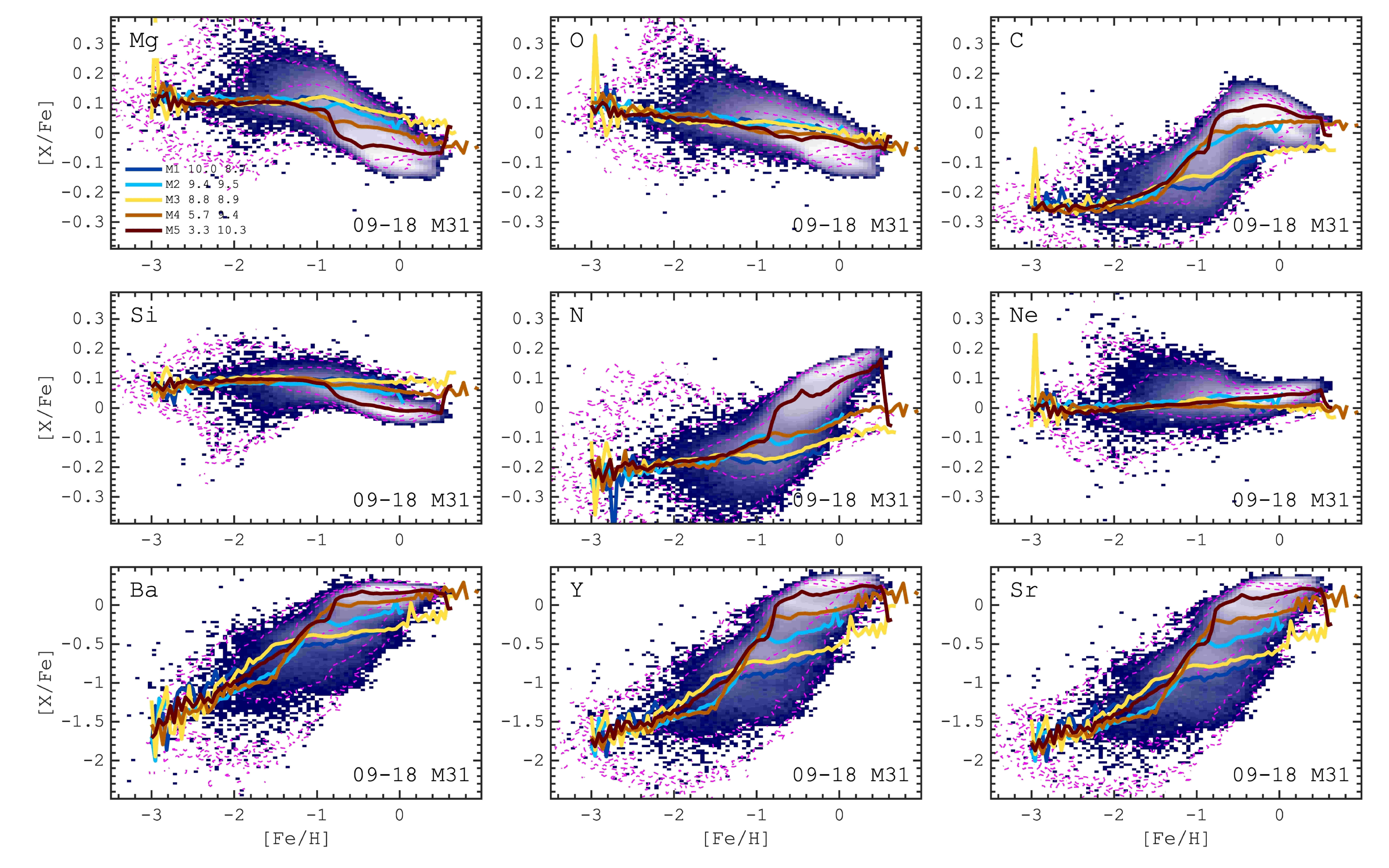}\\
\caption{Elemental abundance relations $\rm [X/Fe]-\FeH$ for the M31 galaxy analogue from the 09-18 HESTIA simulation. $\rm X$ is indicated in each panel. Background maps correspond to the in situ stars, while the magenta contours~($10^{-3}$, $10^{-2}$ and $10^{-1}$ levels of the normalised density distribution) show all accreted stars distribution. The evolutionary tracks corresponding to the five most significant mergers are shown by the coloured lines. The merger accretion time~(Gyr) and the total stellar mass of the merger debris at the time of the merger $\rm log_{10}(M_{*}/M_\odot)$ are shown in the top-left panel. The abundance patterns of both the accreted and in situ populations are very similar at very low metallicities~($\FeH\lesssim-1$), while at larger metallicities different merger debris start to be distinguishable from each other.}\label{fig3::XX_FeH}
\end{center}
\end{figure*}

Next, following a recent analysis of the APOGEE DR17 data by \cite{2022arXiv220304980B}, in Fig.~\ref{fig3::vphi_feh}, we show a variation of the azimuthal velocity components as a function of the stellar metallicity. Taking advantage of the simulations, we can separate the in situ and accreted populations in different rows. We recall that \cite{2022arXiv220304980B} found a population of chemically pre-selected in situ stars that show no~(or very little) net rotation at $\FeH<-1.3$, dubbed as Aurora. These stars have a high velocity dispersion and a large scatter in element abundances which, at such low metallicities, can be interpreted as the manifestation of the chaotic pre-disk stages of MW's evolution. At a higher metallicity~(between $\FeH = -1.3$ and $-0.9$), the rotational velocity of the MW stars increases with metallicity up to $\approx 150$~\kmps suggesting a rapid transformation~(``spin-up'') from the turbulent stages to a coherently rotating disc of the MW.

Figure~\ref{fig3::vphi_feh} suggests that all the HESTIA galaxies show similar behaviour of the azimuthal velocity as a function of \FeH. In particular, below $\FeH<-1.2$~(or $\FeH<-0.8$ depending on the galaxy), the azimuthal velocity profile is flat with a modest net rotation~($\approx 0-80$~\kmps). These populations, similar to the Aurora, have the highest rotational velocity dispersion. For more metal-rich~($\FeH \lesssim -0.5 \pm 0.1$) in situ stars, the velocity dispersion decreases sharply together with increasing median azimuthal velocity which is in good agreement with the spin-up of the MW~\citep[see also][]{2022arXiv220402989C}. 

Another piece of the analysis concerns the kinematics of the accreted stars. In the bottom row of Fig.~\ref{fig3::vphi_feh}, we show the azimuthal velocity as a function of the stellar metallicity for all accreted stellar populations, without dissecting them onto individual merger debris. A striking feature here is that even accreted stars in all the galaxies show some rotation with the mean rotational velocity slowly increasing with the metallicity. There is a single exception here, namely: MW in the 09-18 simulation, which experienced a retrograde merger $\approx 3.7$ Gyr ago~(see the M5 merger). Nevertheless, our small statistics (albeit biased towards the MW-like galaxies) suggest that the accretion of dwarf galaxies at early times happens predominantly with the same direction of rotation as the host disc, which is likely due to dynamical friction and a torque from the disc~\citep{2017MNRAS.472.3722G}.

\section{Multiple abundances}\label{sec3::more_elements}

Recent progress in the understanding of the MW stellar populations is driven by large spectroscopic surveys: RAVE~\citep{2020AJ....160...82S}, LAMOST~\citep{2015RAA....15.1095L}, APOGEE~\citep{2017AJ....154...94M}, GALAH~\citep{2015MNRAS.449.2604D}, and Gaia-ESO~\citep{2012Msngr.147...25G}. They deliver a number of chemical abundances with statistical errors that allow for studies of different components and stellar populations not only in the disk but, thanks to the mounting growth of the measurements,  in a less populated stellar halo as well. Concerning the stellar halo, using various combinations of the abundances~\citep[see, e.g.][]{2022MNRAS.510.2407B,2022arXiv220404233H,2022ApJ...938...21M,2022MNRAS.513.1557C}, it becomes possible to differentiate the halo substructures by their chemical composition, which can vary due to the initial conditions for the star formation, its efficiency, and the gas fraction. Meanwhile, cosmological and idealized tailored simulations have started to deliver reliable chemical abundances, not only for iron and some [$\alpha$/Fe]-elements. In this case, of course, the models are still limited by the implementation of the subgrid physics and adopted yields~\citep[see e.g.][]{2021MNRAS.508.3365B} which may not allow for reproducing observational data quantitatively. Nevertheless, the important information delivered by the simulations is the relative chemical abundance variation of different stellar populations. In particular, we suggest that even if the chemical abundances delivered by the simulations are not exactly on the same scale as the ones measured in the MW, we could still learn about the accreted systems once we compare their chemistry to the in situ stars.

In this section, we analyze a set of the chemical abundances delivered in the HESTIA simulations, which, for the reasons we stated above, may not perfectly match the MW data but still can guide further studies of the chemical abundance trends of accreted stellar populations in both simulations and observational data. First, in Fig.~\ref{fig3::XX_FeH}, we show the abundance ratio for $\rm X = Mg, O, C, Si, N, Ne, Ba, Y, Sr$ as a function of the iron abundance \FeH in a single galaxy M31 from 09-18 HESTIA simulation. We note that the stellar yields for the s-process elements include a contribution from AGB stars only~\citep{2015ApJS..219...40C,2016ApJ...825...26K}. Coloured lines correspond to the mean trends in $\rm [X/Fe] - \FeH$ for five of the most significant mergers, while the magenta contours highlight the distribution of all accreted stars and the background colour maps correspond to the distributions of the in situ stellar populations. 

We can see that the difference between the merger debris is negligible at very low metallicities, $\FeH \lesssim -2$. This is evidently because such low metallicities are typical for the stars formed very early on, $11-14$~Gyr ago~(see Figs.~\ref{fig3::age_metallicity} and \ref{fig3::age_met_with_insitu}) when the 'initial' conditions for the star formation are very similar inside progenitors of the dwarf galaxies, and the ISM is not enriched enough. At the same time, the main progenitor~(MW or M31 in our case) is also similar to other proto-dwarf galaxies and, thus, the first stellar populations have nearly identical abundance relations. At slightly higher (but still low metallicities) of $-2\lesssim \FeH \lesssim -1$, some elements~($\rm C, N, Ba, Y, Sr$) start to show a certain level of differentiation but taking into account observational errors in this metallicity range these tiny differences can not be distinguished by the large-scale surveys. The most promising metallicity region is $\FeH=-1...0$, where the chemical abundance tracks for individual debris show the largest separation. Obviously, the difference is larger between the mergers the most distant in time. In this case, more recently accreted galaxies, tend to have higher stellar mass; in addition, the highest metallicity stars have lower-[$\alpha$/Fe] abundances. 

\begin{figure*}[t]
\begin{center}
\includegraphics[width=1\hsize]{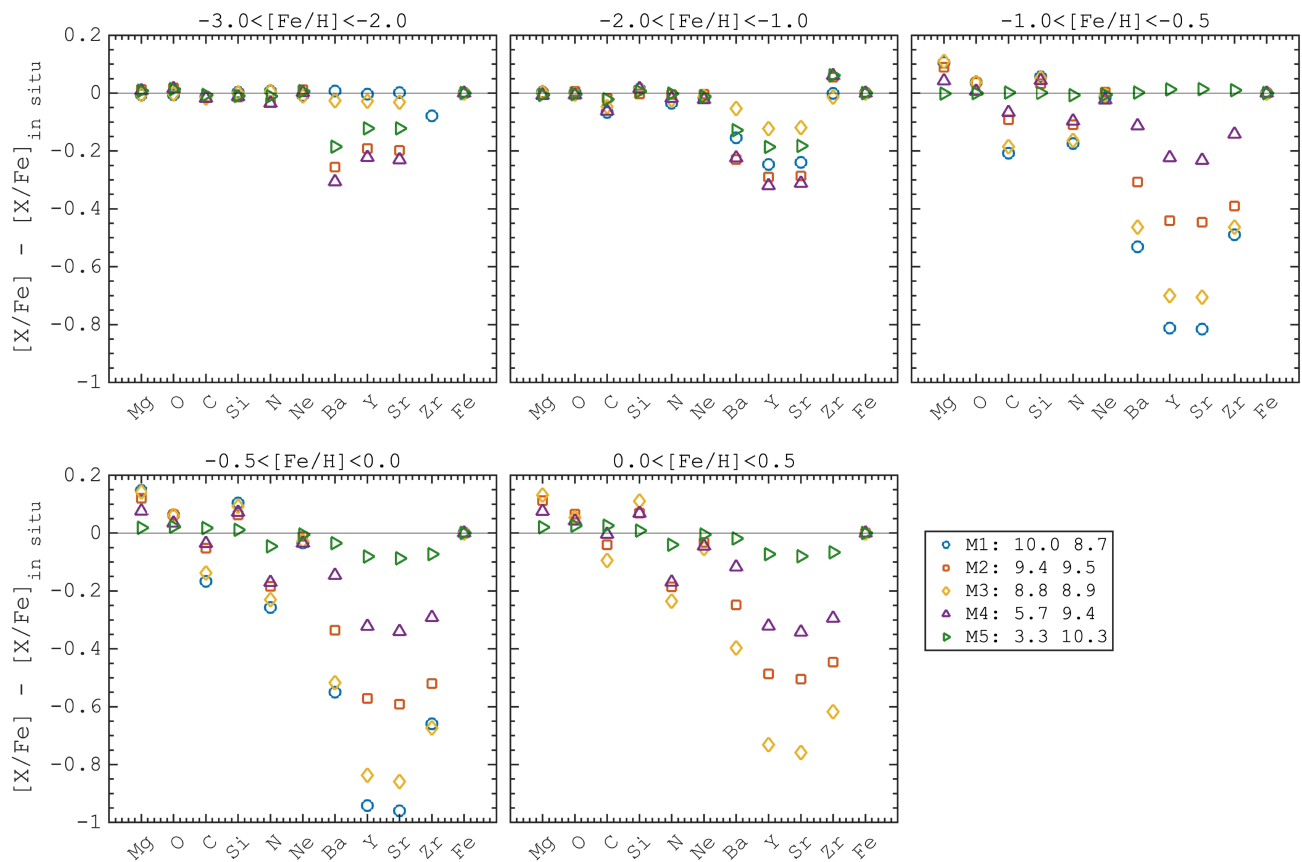}
\caption{Mean $\rm  [X/H]$ values for the five most significant merger debris relative to the mean values of the in situ populations as a function of metallicity in the M31 09-18 galaxy. The merger accretion time~(Gyr) and the total stellar mass of the merger debris at the time of the merger $\rm log_{10}(M_{*}/M_\odot)$ are shown in the legend. 
}\label{fig3::individual_elements_feh_age}
\end{center}
\end{figure*}

To better quantify the difference between the merger debris and the in situ stars, in Fig.~\ref{fig3::individual_elements_feh_age}, we present the difference between the mean individual merger debris abundance and the mean in situ values. We suggest that the residual abundances of the accreted stars relative to the in situ populations allow specific trends of the merger debris to be captured more accurately. The relative abundances~($\rm [X/Fe] - [X/Fe]_{in situ}$) are shown in several bins of the stellar metallicity \FeH. The idea here is to provide typical parameters space where the chemical abundance differences of different merger remnants and the in situ stars are maximal. In other words, we set the criteria that will be used to explore the MW stellar halo in the chemical abundance space with the aim of disentangling merger remnants and in situ stellar populations  -- and the ultimate goal of recovering the assembly history of the MW. From this figure, we can see that the individual merger debris are very much similar to each other at the very low metallicities, although they are slightly [$\alpha$/Fe]-enhanced relative to the in situ stars. This difference increases towards the higher-\FeH for the some [$\alpha$/Fe]-elements~($\rm Mg, O, Si$), however, it does not exceed $\approx 0.1$~dex. The next abundance value for the accreted stars that shows substantial deviation from the in situ population is carbon, where a significant gap is seen already at $-2\lesssim\FeH\lesssim-1$ that can be traced out to super-solar metallicities. Another group of elements showing the biggest difference, up to $1$~dex, is $\rm Ba, Y, Zr$. These abundances seem to be the most promising in the differentiation of the merger debris and the in situ populations.

\begin{figure*}[t]
\begin{center}
\includegraphics[width=1\hsize]{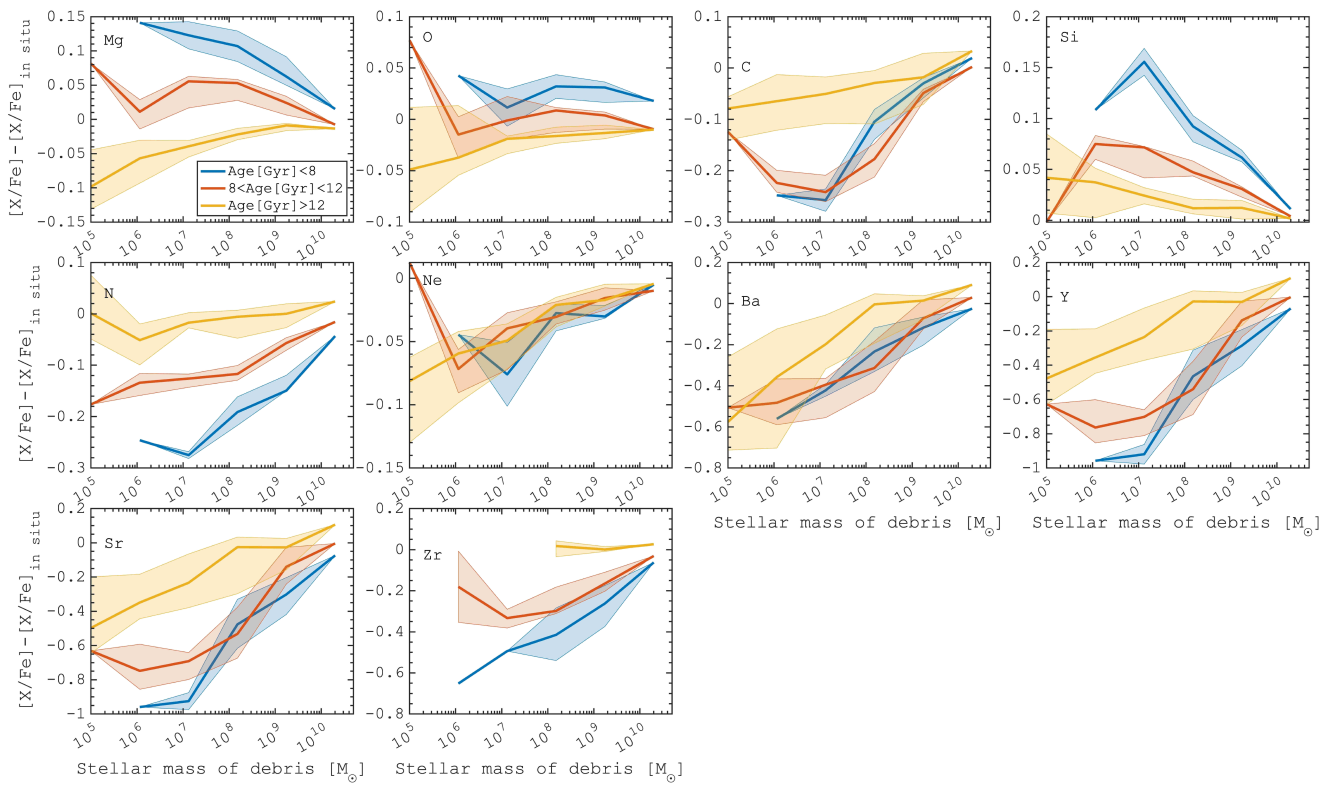}
\caption{Relative chemical abundance of accreted stars and in situ populations versus the total stellar mass of the merger debris. The statistics is based on all the merger debris from the six M31/MW HESTIA galaxies. Lines of different colour correspond to different stellar age bins~(see top-left panel), while the shaded area shows the $3\sigma$ level. The most striking feature of the relations is the dependence of the residual abundances on the stellar age.
}\label{fig3::accr_mass_abundances}
\end{center}
\end{figure*}

Finally, we summarise the trends presented for a single galaxy~(M31 from 09-18 simulation) in Figs.~\ref{fig3::XX_FeH} and ~\ref{fig3::individual_elements_feh_age} by combining all the merger debris we identified in all the HESTIA galaxies.  In Fig.~\ref{fig3::accr_mass_abundances}, we show the residual abundances~(relative to the in situ populations) of accreted stars as a function of the total stellar mass of the debris where the relations are shown for the stars in three age ranges: $<8$~Gyr~(blue), $8-12$~Gyr~(red) and $>12$~Gyr~(yellow). We already showed that separating the in situ and different accreted debris using chemical abundances poses a great challenge due to substantial overlap. Nevertheless, the broad statistics of various galaxies accreted onto the M31/MW allows us to find some systematic differences in the chemical abundances. The most striking feature of the relations we present here is the dependence of the residual abundances on the stellar age. Therefore, one of the key elements in the future studies of the MW assembly history is the age measurements for larger samples of stars. 

\section{Summary}\label{sec3::concl}

This paper, as part of a series based on the analysis of the HESTIA cosmological simulations of the Local Group, explores the chemical abundance variations among the accreted populations together with in situ stellar populations. Our main conclusions are as follows:

\begin{itemize}
    \item We show that the mean metallicity of the entire merger debris increases with the total stellar mass and decreases with the lookback time of accretion~($t_{accr}$). In particular, for the stellar masses larger than $10^6$~\Msun, the mean stellar metallicity can be fitted as $\FeH \propto -0.06 \times t_{accr}+C$ where $\rm C$ increases with the total stellar mass~(see Fig.~\ref{fig3::feh_accr_time}). We also find that the chemical abundances of stellar merger debris differ from the properties of the surviving dwarf galaxies. For the same stellar mass, the merger debris are more metal-poor and more [$\alpha$/Fe]-enhanced~(see, Figs.~\ref{fig3::mass_metallicity} and \ref{fig3::MgFe_FeH_debris_satellites}). In analysing the star formation histories of the satellites and merger debris, we show that the chemical abundances of the merger debris are the most similar to the dwarf galaxies that quench their star formation early on~(see Fig.~\ref{fig3::mass_metallicity}).
    
    \item By analysing the metallicity patterns in \ELz space, we find a prominent diagonal gradient of the mean stellar metallicity, which increases towards the prograde regularly rotating stellar populations~(see Fig.~\ref{fig3::ELz_selections}). The observed gradient  does not only result from the in situ stars' fraction variations, as the individual merger debris depict similar metallicity patterns~(see Fig.~\ref{fig3::feh_ecc_insitu_accr_individual}). More generally, the mean stellar metallicity of the individual debris increases towards the stars formed in the innermost parts of the disrupted dwarf galaxy. The latter results in the negative metallicity gradient as a function of the galactocentric distance for the massive~($M_{star}>10^8~\Msun$) merger debris~(see Fig.~\ref{fig3::FeH_E0_Rg}). This result also suggests that the metallicity of the debris measured at the Solar radius in the Milky Way may underestimate the mean metallicity of the debris and, thus, it may either underestimate the total mass or overestimate the accretion lookback time. 
    
    \item We find that the MDF of some individual merger debris reveal several peaks~(see Fig.~
    \ref{fig3::MDFs_individual_mergers}). This result is more prominent for the stars selected in the retrograde part of \ELz space. The complex composition of MDFs is likely linked to the differentiation of stellar populations accreted from a single dwarf galaxy, according to the initial kinematics inside the dwarf prior to the merger.  
    
    \item We show that, despite the substantial overlap with the in situ stars, the accreted populations dominate in the low-\FeH regions and in the high-[$\alpha$/Fe] sequence of the \MgFe-\FeH plane~(see Fig.~\ref{fig3::mgfe_feh_small}). The \MgFe-\FeH relations for the individual debris vary significantly depending on the mean angular momentum. We find that the retrograde stars of the merger debris usually do not show a knee while the prograde populations of the same debris show a more advanced chemical enrichment with prominent 'knees'~(see Fig.~\ref{fig3::mgfe_feh_from_ELz}). Therefore, depending on the orbital properties, individual merger debris can mimic chemically different substructures. 
    
    \item By coupling the chemical abundances with the kinematics of stars we found that the eccentricity of stars from the individual merger debris while having a large scatter in a given chemical abundance range, while remaining nearly constant along \FeH and \MgFe. At the same time, when we consider the accreted halo made up of all the merger debris, the eccentricity distributions are the complex functions of the \FeH and \MgFe; for instance, there are regions of abundances with a bimodal distribution of the eccentricity. In situ stars also show a non-featureless distribution of the eccentricity as a function of the chemical composition of stars~(see Fig.~\ref{fig3::feh_ecc_insitu_accr_individual}). In this case, the variations of the eccentricity are aptly explained by the impact of the mergers~(see also \citetalias{KhoperskovHESTIA-2}), where the more metal-poor stars formed before the most massive accretion events were heated up by the mergers and, thus, have higher mean eccentricity.
    
    \item In agreement with a recent analysis of the APOGEE data~\citep{2022arXiv220304980B}, the most metal-poor in situ stars~($\FeH\lesssim-1$) in the HESTIA galaxies have a constant azimuthal velocity with a net rotation of $0-80$~\kmps~(for different models), which reproduces the Aurora stellar population of the MW well. At higher metallicities~(up to $\FeH \approx -0.5\pm 0.1$), similar to the case of the MW, we detected the rapid spin-up, where the rotational velocity sharply increases up to $150-200$~\kmps. The latter is a manifestation of the formation of rotating stellar components in the HESTIA galaxies early on~(see also \citetalias{KhoperskovHESTIA-1}). 
    
    \item In the \FeH-\MgFe plane, different merger debris show a similar behaviour, thus making it difficult to capture various individual events. Nevertheless, we show that combining a set of the abundances allows us to capture specific patterns corresponding to different merger debris~(see Figs.~\ref{fig3::XX_FeH} and ~\ref{fig3::individual_elements_feh_age}), thus making it possible to uncover the assembly history of the MW, with the data being delivered by current and forthcoming spectroscopic surveys. According to the HESTIA simulations, a larger differentiation of stellar debris appears for metallicities $\FeH \approx -1$ and $\FeH \approx 0$.

    \item The HESTIA simulations predict systematic variations of the residual abundances of accreted stars~(relative to the mean abundances of the in situ stars) from the total stellar mass of the debris~(see  Fig.~\ref{fig3::accr_mass_abundances}). The relations show different behaviour depending on the age of the stellar populations. This makes the determination of the ages vital for larger samples of the MW stellar populations, and once they are used together with the chemical abundances, it will allow for better constraints to be placed on the merger history of the MW galaxy. 
    
\end{itemize}

The chemical composition of accreted and in situ stars in the MW reveals distinct characteristics that provide clues about their origins. Accreted stars often exhibit different elemental abundance patterns compared to in situ stars, indicating their diverse stellar birthplaces. These differences in chemical composition arise from the accretion of dwarf galaxies or satellite systems with distinct enrichment histories and enrichment mechanisms. In analyzing the suite of the HESTIA simulations, we highlight the importance of chemical abundances in tracing accreted stars. This, together with growing samples of stars with precise ages, will allow us to trace the history of Galactic growth, along with the various sources that have contributed to the MW stellar halo.

\begin{acknowledgements}
We thank the anonymous referee for their valuable comments. SK acknowledgements the HESTIA collaboration for providing access to the simulations. SK also thanks Paola Di Matteo and Misha Haywood for useful comments. FAG acknowledges support from ANID FONDECYT Regular 1211370 and from the ANID BASAL project FB210003. FAG, acknowledge funding from the Max Planck Society through a Partner Group grant. 
JS acknowledges support from the French Agence Nationale de la Recherche for the LOCALIZATION project under grant agreements ANR-21-CE31-0019.
AK is supported by the Ministerio de Ciencia e Innovaci\'{o}n (MICINN) under research grant PID2021-122603NB-C21 and further thanks Verve for a storm in heaven. YH has been partially supported by the Israel Science Foundation grant ISF 1358/18.
ET acknowledges support by ETAg grant PRG1006 and by EU through the ERDF CoE TK133.

\newline
{\it Software:} \texttt{IPython} \citep{2007CSE.....9c..21P}, \texttt{Astropy} \citep{2013A&A...558A..33A, 2018AJ....156..123A}, \texttt{NumPy} \citep{2011CSE....13b..22V}, \texttt{SciPy} \citep{2020SciPy-NMeth}, \texttt{AGAMA} \citep{2019MNRAS.482.1525V}, \texttt{Matplotlib} \citep{2007CSE.....9...90H}, \texttt{Pandas} \citep{mckinney-proc-scipy-2010}, TOPCAT~\citep{2005ASPC..347...29T}.
\end{acknowledgements}

\bibliographystyle{aa}
\bibliography{references-1}

\begin{thebibliography}{143}
\expandafter\ifx\csname natexlab\endcsname\relax\def\natexlab#1{#1}\fi

\bibitem[{{Abadi} {et~al.}(2006){Abadi}, {Navarro}, \&
  {Steinmetz}}]{2006MNRAS.365..747A}
{Abadi}, M.~G., {Navarro}, J.~F., \& {Steinmetz}, M. 2006, \mnras, 365, 747

\bibitem[{{Agertz} {et~al.}(2021){Agertz}, {Renaud}, {Feltzing}, {Read},
  {Ryde}, {Andersson}, {Rey}, {Bensby}, \& {Feuillet}}]{2021MNRAS.503.5826A}
{Agertz}, O., {Renaud}, F., {Feltzing}, S., {et~al.} 2021, \mnras, 503, 5826

\bibitem[{{Amarante} {et~al.}(2022){Amarante}, {Debattista}, {Beraldo e Silva},
  {Laporte}, \& {Deg}}]{2022arXiv220412187A}
{Amarante}, J. A.~S., {Debattista}, V.~P., {Beraldo e Silva}, L., {Laporte}, C.
  F.~P., \& {Deg}, N. 2022, \apj, 937, 12

\bibitem[{{Amorisco}(2017)}]{2017MNRAS.464.2882A}
{Amorisco}, N.~C. 2017, \mnras, 464, 2882

\bibitem[{{Arentsen} {et~al.}(2020){Arentsen}, {Starkenburg}, {Martin},
  {Aguado}, {Zucker}, {Allende Prieto}, {Hill}, {Venn}, {Carlberg},
  {Gonz{\'a}lez Hern{\'a}ndez}, {Mashonkina}, {Navarro}, {S{\'a}nchez-Janssen},
  {Schultheis}, {Thomas}, {Youakim}, {Lewis}, {Simpson}, {Wan}, {Cohen},
  {Geisler}, \& {O'Connell}}]{2020MNRAS.496.4964A}
{Arentsen}, A., {Starkenburg}, E., {Martin}, N.~F., {et~al.} 2020, \mnras, 496,
  4964

\bibitem[{{Astropy Collaboration} {et~al.}(2018){Astropy Collaboration},
  {Price-Whelan}, {Sip{\H{o}}cz}, {G{\"u}nther}, {Lim}, {Crawford}, {Conseil},
  {Shupe}, {Craig}, {Dencheva}, {Ginsburg}, {Vand erPlas}, {Bradley},
  {P{\'e}rez-Su{\'a}rez}, {de Val-Borro}, {Aldcroft}, {Cruz}, {Robitaille},
  {Tollerud}, {Ardelean}, {Babej}, {Bach}, {Bachetti}, {Bakanov}, {Bamford},
  {Barentsen}, {Barmby}, {Baumbach}, {Berry}, {Biscani}, {Boquien}, {Bostroem},
  {Bouma}, {Brammer}, {Bray}, {Breytenbach}, {Buddelmeijer}, {Burke},
  {Calderone}, {Cano Rodr{\'\i}guez}, {Cara}, {Cardoso}, {Cheedella}, {Copin},
  {Corrales}, {Crichton}, {D'Avella}, {Deil}, {Depagne}, {Dietrich}, {Donath},
  {Droettboom}, {Earl}, {Erben}, {Fabbro}, {Ferreira}, {Finethy}, {Fox},
  {Garrison}, {Gibbons}, {Goldstein}, {Gommers}, {Greco}, {Greenfield},
  {Groener}, {Grollier}, {Hagen}, {Hirst}, {Homeier}, {Horton}, {Hosseinzadeh},
  {Hu}, {Hunkeler}, {Ivezi{\'c}}, {Jain}, {Jenness}, {Kanarek}, {Kendrew},
  {Kern}, {Kerzendorf}, {Khvalko}, {King}, {Kirkby}, {Kulkarni}, {Kumar},
  {Lee}, {Lenz}, {Littlefair}, {Ma}, {Macleod}, {Mastropietro}, {McCully},
  {Montagnac}, {Morris}, {Mueller}, {Mumford}, {Muna}, {Murphy}, {Nelson},
  {Nguyen}, {Ninan}, {N{\"o}the}, {Ogaz}, {Oh}, {Parejko}, {Parley}, {Pascual},
  {Patil}, {Patil}, {Plunkett}, {Prochaska}, {Rastogi}, {Reddy Janga},
  {Sabater}, {Sakurikar}, {Seifert}, {Sherbert}, {Sherwood-Taylor}, {Shih},
  {Sick}, {Silbiger}, {Singanamalla}, {Singer}, {Sladen}, {Sooley},
  {Sornarajah}, {Streicher}, {Teuben}, {Thomas}, {Tremblay}, {Turner},
  {Terr{\'o}n}, {van Kerkwijk}, {de la Vega}, {Watkins}, {Weaver}, {Whitmore},
  {Woillez}, {Zabalza}, \& {Astropy Contributors}}]{2018AJ....156..123A}
{Astropy Collaboration}, {Price-Whelan}, A.~M., {Sip{\H{o}}cz}, B.~M., {et~al.}
  2018, \aj, 156, 123

\bibitem[{{Astropy Collaboration} {et~al.}(2013){Astropy Collaboration},
  {Robitaille}, {Tollerud}, {Greenfield}, {Droettboom}, {Bray}, {Aldcroft},
  {Davis}, {Ginsburg}, {Price-Whelan}, {Kerzendorf}, {Conley}, {Crighton},
  {Barbary}, {Muna}, {Ferguson}, {Grollier}, {Parikh}, {Nair}, {Unther},
  {Deil}, {Woillez}, {Conseil}, {Kramer}, {Turner}, {Singer}, {Fox}, {Weaver},
  {Zabalza}, {Edwards}, {Azalee Bostroem}, {Burke}, {Casey}, {Crawford},
  {Dencheva}, {Ely}, {Jenness}, {Labrie}, {Lim}, {Pierfederici}, {Pontzen},
  {Ptak}, {Refsdal}, {Servillat}, \& {Streicher}}]{2013A&A...558A..33A}
{Astropy Collaboration}, {Robitaille}, T.~P., {Tollerud}, E.~J., {et~al.} 2013,
  \aap, 558, A33

\bibitem[{{Belokurov} {et~al.}(2018){Belokurov}, {Erkal}, {Evans}, {Koposov},
  \& {Deason}}]{2018MNRAS.478..611B}
{Belokurov}, V., {Erkal}, D., {Evans}, N.~W., {Koposov}, S.~E., \& {Deason},
  A.~J. 2018, \mnras, 478, 611

\bibitem[{{Belokurov} \& {Kravtsov}(2022)}]{2022arXiv220304980B}
{Belokurov}, V. \& {Kravtsov}, A. 2022, \mnras, 514, 689

\bibitem[{{Belokurov} {et~al.}(2020){Belokurov}, {Sanders}, {Fattahi}, {Smith},
  {Deason}, {Evans}, \& {Grand}}]{2020MNRAS.494.3880B}
{Belokurov}, V., {Sanders}, J.~L., {Fattahi}, A., {et~al.} 2020, \mnras, 494,
  3880

\bibitem[{{Belokurov} {et~al.}(2006){Belokurov}, {Zucker}, {Evans}, {Gilmore},
  {Vidrih}, {Bramich}, {Newberg}, {Wyse}, {Irwin}, {Fellhauer}, {Hewett},
  {Walton}, {Wilkinson}, {Cole}, {Yanny}, {Rockosi}, {Beers}, {Bell},
  {Brinkmann}, {Ivezi{\'c}}, \& {Lupton}}]{2006ApJ...642L.137B}
{Belokurov}, V., {Zucker}, D.~B., {Evans}, N.~W., {et~al.} 2006, \apjl, 642,
  L137

\bibitem[{{Brammer} {et~al.}(2011){Brammer}, {Whitaker}, {van Dokkum},
  {Marchesini}, {Franx}, {Kriek}, {Labb{\'e}}, {Lee}, {Muzzin}, {Quadri},
  {Rudnick}, \& {Williams}}]{2011ApJ...739...24B}
{Brammer}, G.~B., {Whitaker}, K.~E., {van Dokkum}, P.~G., {et~al.} 2011, \apj,
  739, 24

\bibitem[{{Brook} {et~al.}(2005){Brook}, {Gibson}, {Martel}, \&
  {Kawata}}]{2005ApJ...630..298B}
{Brook}, C.~B., {Gibson}, B.~K., {Martel}, H., \& {Kawata}, D. 2005, \apj, 630,
  298

\bibitem[{{Brook} {et~al.}(2004){Brook}, {Kawata}, {Gibson}, \&
  {Freeman}}]{2004ApJ...612..894B}
{Brook}, C.~B., {Kawata}, D., {Gibson}, B.~K., \& {Freeman}, K.~C. 2004, \apj,
  612, 894

\bibitem[{{Brook} {et~al.}(2020){Brook}, {Kawata}, {Gibson}, {Gallart}, \&
  {Vicente}}]{2020MNRAS.495.2645B}
{Brook}, C.~B., {Kawata}, D., {Gibson}, B.~K., {Gallart}, C., \& {Vicente}, A.
  2020, \mnras, 495, 2645

\bibitem[{{Brook} {et~al.}(2012){Brook}, {Stinson}, {Gibson}, {Kawata},
  {House}, {Miranda}, {Macci{\`o}}, {Pilkington}, {Ro{\v{s}}kar}, {Wadsley}, \&
  {Quinn}}]{2012MNRAS.426..690B}
{Brook}, C.~B., {Stinson}, G.~S., {Gibson}, B.~K., {et~al.} 2012, \mnras, 426,
  690

\bibitem[{{Buck}(2020)}]{2020MNRAS.491.5435B}
{Buck}, T. 2020, \mnras, 491, 5435

\bibitem[{{Buck} {et~al.}(2021){Buck}, {Rybizki}, {Buder}, {Obreja},
  {Macci{\`o}}, {Pfrommer}, {Steinmetz}, \& {Ness}}]{2021MNRAS.508.3365B}
{Buck}, T., {Rybizki}, J., {Buder}, S., {et~al.} 2021, \mnras, 508, 3365

\bibitem[{{Buder} {et~al.}(2022){Buder}, {Lind}, {Ness}, {Feuillet}, {Horta},
  {Monty}, {Buck}, {Nordlander}, {Bland-Hawthorn}, {Casey}, {de Silva},
  {D'Orazi}, {Freeman}, {Hayden}, {Kos}, {Martell}, {Lewis}, {Lin},
  {Schlesinger}, {Sharma}, {Simpson}, {Stello}, {Zucker}, {Zwitter},
  {Ciuc{\u{a}}}, {Horner}, {Kobayashi}, {Ting}, {Wyse}, \&
  {Wyse}}]{2022MNRAS.510.2407B}
{Buder}, S., {Lind}, K., {Ness}, M.~K., {et~al.} 2022, \mnras, 510, 2407

\bibitem[{{Bullock} \& {Johnston}(2005)}]{2005ApJ...635..931B}
{Bullock}, J.~S. \& {Johnston}, K.~V. 2005, \apj, 635, 931

\bibitem[{{Carlin} {et~al.}(2019){Carlin}, {Garling}, {Peter}, {Crnojevi{\'c}},
  {Forbes}, {Hargis}, {Mutlu-Pakdil}, {Pucha}, {Romanowsky}, {Sand},
  {Spekkens}, {Strader}, \& {Willman}}]{2019ApJ...886..109C}
{Carlin}, J.~L., {Garling}, C.~T., {Peter}, A. H.~G., {et~al.} 2019, \apj, 886,
  109

\bibitem[{{Carrillo} {et~al.}(2022){Carrillo}, {Hawkins}, {Jofr{\'e}}, {de
  Brito Silva}, {Das}, \& {Lucey}}]{2022MNRAS.513.1557C}
{Carrillo}, A., {Hawkins}, K., {Jofr{\'e}}, P., {et~al.} 2022, \mnras, 513,
  1557

\bibitem[{{Clarke} {et~al.}(2019){Clarke}, {Debattista}, {Nidever}, {Loebman},
  {Simons}, {Kassin}, {Du}, {Ness}, {Fisher}, {Quinn}, {Wadsley}, {Freeman}, \&
  {Popescu}}]{2019MNRAS.484.3476C}
{Clarke}, A.~J., {Debattista}, V.~P., {Nidever}, D.~L., {et~al.} 2019, \mnras,
  484, 3476

\bibitem[{{Conroy} {et~al.}(2021){Conroy}, {Naidu}, {Garavito-Camargo},
  {Besla}, {Zaritsky}, {Bonaca}, \& {Johnson}}]{2021Natur.592..534C}
{Conroy}, C., {Naidu}, R.~P., {Garavito-Camargo}, N., {et~al.} 2021, \nat, 592,
  534

\bibitem[{{Conroy} {et~al.}(2022){Conroy}, {Weinberg}, {Naidu}, {Buck},
  {Johnson}, {Cargile}, {Bonaca}, {Caldwell}, {Chandra}, {Han}, {Johnson},
  {Speagle}, {Ting}, {Woody}, \& {Zaritsky}}]{2022arXiv220402989C}
{Conroy}, C., {Weinberg}, D.~H., {Naidu}, R.~P., {et~al.} 2022, arXiv e-prints,
  arXiv:2204.02989

\bibitem[{{Cooper} {et~al.}(2010){Cooper}, {Cole}, {Frenk}, {White}, {Helly},
  {Benson}, {De Lucia}, {Helmi}, {Jenkins}, {Navarro}, {Springel}, \&
  {Wang}}]{2010MNRAS.406..744C}
{Cooper}, A.~P., {Cole}, S., {Frenk}, C.~S., {et~al.} 2010, \mnras, 406, 744

\bibitem[{{Cristallo} {et~al.}(2015){Cristallo}, {Straniero}, {Piersanti}, \&
  {Gobrecht}}]{2015ApJS..219...40C}
{Cristallo}, S., {Straniero}, O., {Piersanti}, L., \& {Gobrecht}, D. 2015,
  \apjs, 219, 40

\bibitem[{{Cunningham} {et~al.}(2022){Cunningham}, {Sanderson}, {Johnston},
  {Panithanpaisal}, {Ness}, {Wetzel}, {Loebman}, {Escala}, {Horta}, \&
  {Faucher-Gigu{\`e}re}}]{2021arXiv211002957C}
{Cunningham}, E.~C., {Sanderson}, R.~E., {Johnston}, K.~V., {et~al.} 2022,
  \apj, 934, 172

\bibitem[{{de Jong} {et~al.}(2019){de Jong}, {Agertz}, {Berbel}, {Aird},
  {Alexander}, {Amarsi}, {Anders}, {Andrae}, {Ansarinejad}, {Ansorge},
  {Antilogus}, {Anwand -Heerwart}, {Arentsen}, {Arnadottir}, {Asplund},
  {Auger}, {Azais}, {Baade}, {Baker}, {Baker}, {Balbinot}, {Baldry}, {Banerji},
  {Barden}, {Barklem}, {Barth{\'e}l{\'e}my-Mazot}, {Battistini}, {Bauer},
  {Bell}, {Bellido-Tirado}, {Bellstedt}, {Belokurov}, {Bensby}, {Bergemann},
  {Bestenlehner}, {Bielby}, {Bilicki}, {Blake}, {Bland-Hawthorn}, {Boeche},
  {Boland}, {Boller}, {Bongard}, {Bongiorno}, {Bonifacio}, {Boudon}, {Brooks},
  {Brown}, {Brown}, {Br{\"u}ggen}, {Brynnel}, {Brzeski}, {Buchert},
  {Buschkamp}, {Caffau}, {Caillier}, {Carrick}, {Casagrande}, {Case}, {Casey},
  {Cesarini}, {Cescutti}, {Chapuis}, {Chiappini}, {Childress}, {Christlieb},
  {Church}, {Cioni}, {Cluver}, {Colless}, {Collett}, {Comparat}, {Cooper},
  {Couch}, {Courbin}, {Croom}, {Croton}, {Daguis{\'e}}, {Dalton}, {Davies},
  {Davis}, {de Laverny}, {Deason}, {Dionies}, {Disseau}, {Doel}, {D{\"o}scher},
  {Driver}, {Dwelly}, {Eckert}, {Edge}, {Edvardsson}, {Youssoufi}, {Elhaddad},
  {Enke}, {Erfanianfar}, {Farrell}, {Fechner}, {Feiz}, {Feltzing}, {Ferreras},
  {Feuerstein}, {Feuillet}, {Finoguenov}, {Ford}, {Fotopoulou}, {Fouesneau},
  {Frenk}, {Frey}, {Gaessler}, {Geier}, {Fusillo}, {Gerhard}, {Giannantonio},
  {Giannone}, {Gibson}, {Gillingham}, {Gonz{\'a}lez-Fern{\'a}ndez},
  {Gonzalez-Solares}, {Gottloeber}, {Gould}, {Grebel}, {Gueguen}, {Guiglion},
  {Haehnelt}, {Hahn}, {Hansen}, {Hartman}, {Hauptner}, {Hawkins}, {Haynes},
  {Haynes}, {Heiter}, {Helmi}, {Aguayo}, {Hewett}, {Hinton}, {Hobbs}, {Hoenig},
  {Hofman}, {Hook}, {Hopgood}, {Hopkins}, {Hourihane}, {Howes}, {Howlett},
  {Huet}, {Irwin}, {Iwert}, {Jablonka}, {Jahn}, {Jahnke}, {Jarno}, {Jin},
  {Jofre}, {Johl}, {Jones}, {J{\"o}nsson}, {Jordan}, {Karovicova}, {Khalatyan},
  {Kelz}, {Kennicutt}, {King}, {Kitaura}, {Klar}, {Klauser}, {Kneib}, {Koch},
  {Koposov}, {Kordopatis}, {Korn}, {Kosmalski}, {Kotak}, {Kovalev}, {Kreckel},
  {Kripak}, {Krumpe}, {Kuijken}, {Kunder}, {Kushniruk}, {Lam}, {Lamer},
  {Laurent}, {Lawrence}, {Lehmitz}, {Lemasle}, {Lewis}, {Li}, {Lidman}, {Lind},
  {Liske}, {Lizon}, {Loveday}, {Ludwig}, {McDermid}, {Maguire}, {Mainieri},
  {Mali}, {Mandel}, {Mandel}, {Mannering}, {Martell}, {Martinez Delgado},
  {Matijevic}, {McGregor}, {McMahon}, {McMillan}, {Mena}, {Merloni}, {Meyer},
  {Michel}, {Micheva}, {Migniau}, {Minchev}, {Monari}, {Muller}, {Murphy},
  {Muthukrishna}, {Nandra}, {Navarro}, {Ness}, {Nichani}, {Nichol}, {Nicklas},
  {Niederhofer}, {Norberg}, {Obreschkow}, {Oliver}, {Owers}, {Pai},
  {Pankratow}, {Parkinson}, {Paschke}, {Paterson}, {Pecontal}, {Parry},
  {Phillips}, {Pillepich}, {Pinard}, {Pirard}, {Piskunov}, {Plank},
  {Pl{\"u}schke}, {Pons}, {Popesso}, {Power}, {Pragt}, {Pramskiy}, {Pryer},
  {Quattri}, {Queiroz}, {Quirrenbach}, {Rahurkar}, {Raichoor}, {Ramstedt},
  {Rau}, {Recio-Blanco}, {Reiss}, {Renaud}, {Revaz}, {Rhode}, {Richard},
  {Richter}, {Rix}, {Robotham}, {Roelfsema}, {Romaniello}, {Rosario},
  {Rothmaier}, {Roukema}, {Ruchti}, {Rupprecht}, {Rybizki}, {Ryde}, {Saar},
  {Sadler}, {Sahl{\'e}n}, {Salvato}, {Sassolas}, {Saunders}, {Saviauk},
  {Sbordone}, {Schmidt}, {Schnurr}, {Scholz}, {Schwope}, {Seifert}, {Shanks},
  {Sheinis}, {Sivov}, {Sk{\'u}lad{\'o}ttir}, {Smartt}, {Smedley}, {Smith},
  {Smith}, {Sorce}, {Spitler}, {Starkenburg}, {Steinmetz}, {Stilz}, {Storm},
  {Sullivan}, {Sutherland}, {Swann}, {Tamone}, {Taylor}, {Teillon}, {Tempel},
  {ter Horst}, {Thi}, {Tolstoy}, {Trager}, {Traven}, {Tremblay}, {Tresse},
  {Valentini}, {van de Weygaert}, {van den Ancker}, {Veljanoski}, {Venkatesan},
  {Wagner}, {Wagner}, {Walcher}, {Waller}, {Walton}, {Wang}, {Winkler},
  {Wisotzki}, {Worley}, {Worseck}, {Xiang}, {Xu}, {Yong}, {Zhao}, {Zheng},
  {Zscheyge}, \& {Zucker}}]{2019Msngr.175....3D}
{de Jong}, R.~S., {Agertz}, O., {Berbel}, A.~A., {et~al.} 2019, The Messenger,
  175, 3

\bibitem[{{De Lucia} \& {Helmi}(2008)}]{2008MNRAS.391...14D}
{De Lucia}, G. \& {Helmi}, A. 2008, \mnras, 391, 14

\bibitem[{{De Silva} {et~al.}(2015){De Silva}, {Freeman}, {Bland-Hawthorn},
  {Martell}, {de Boer}, {Asplund}, {Keller}, {Sharma}, {Zucker}, {Zwitter},
  {Anguiano}, {Bacigalupo}, {Bayliss}, {Beavis}, {Bergemann}, {Campbell},
  {Cannon}, {Carollo}, {Casagrande}, {Casey}, {Da Costa}, {D'Orazi}, {Dotter},
  {Duong}, {Heger}, {Ireland}, {Kafle}, {Kos}, {Lattanzio}, {Lewis}, {Lin},
  {Lind}, {Munari}, {Nataf}, {O'Toole}, {Parker}, {Reid}, {Schlesinger},
  {Sheinis}, {Simpson}, {Stello}, {Ting}, {Traven}, {Watson}, {Wittenmyer},
  {Yong}, \& {{\v{Z}}erjal}}]{2015MNRAS.449.2604D}
{De Silva}, G.~M., {Freeman}, K.~C., {Bland-Hawthorn}, J., {et~al.} 2015,
  \mnras, 449, 2604

\bibitem[{{Deason} {et~al.}(2019){Deason}, {Belokurov}, \&
  {Sanders}}]{2019MNRAS.490.3426D}
{Deason}, A.~J., {Belokurov}, V., \& {Sanders}, J.~L. 2019, \mnras, 490, 3426

\bibitem[{{Deason} {et~al.}(2016){Deason}, {Mao}, \&
  {Wechsler}}]{2016ApJ...821....5D}
{Deason}, A.~J., {Mao}, Y.-Y., \& {Wechsler}, R.~H. 2016, \apj, 821, 5

\bibitem[{{Di Matteo} {et~al.}(2019){Di Matteo}, {Haywood}, {Lehnert}, {Katz},
  {Khoperskov}, {Snaith}, {G{\'o}mez}, \& {Robichon}}]{2019A&A...632A...4D}
{Di Matteo}, P., {Haywood}, M., {Lehnert}, M.~D., {et~al.} 2019, \aap, 632, A4

\bibitem[{{D'Souza} \& {Bell}(2018)}]{2018NatAs...2..737D}
{D'Souza}, R. \& {Bell}, E.~F. 2018, Nature Astronomy, 2, 737

\bibitem[{{Dupuy} {et~al.}(2022){Dupuy}, {Libeskind}, {Hoffman}, {Courtois},
  {Gottl{\"o}ber}, {Grand}, {Knebe}, {Sorce}, {Tempel}, {Tully},
  {Vogelsberger}, \& {Wang}}]{Dupuy_etal}
{Dupuy}, A., {Libeskind}, N.~I., {Hoffman}, Y., {et~al.} 2022, \mnras, 516,
  4576

\bibitem[{{Erkal} {et~al.}(2021){Erkal}, {Deason}, {Belokurov}, {Xue},
  {Koposov}, {Bird}, {Liu}, {Simion}, {Yang}, {Zhang}, \&
  {Zhao}}]{2021MNRAS.506.2677E}
{Erkal}, D., {Deason}, A.~J., {Belokurov}, V., {et~al.} 2021, \mnras, 506, 2677

\bibitem[{{Fakhouri} {et~al.}(2010){Fakhouri}, {Ma}, \&
  {Boylan-Kolchin}}]{2010MNRAS.406.2267F}
{Fakhouri}, O., {Ma}, C.-P., \& {Boylan-Kolchin}, M. 2010, \mnras, 406, 2267

\bibitem[{{Fardal} {et~al.}(2007){Fardal}, {Guhathakurta}, {Babul}, \&
  {McConnachie}}]{2007MNRAS.380...15F}
{Fardal}, M.~A., {Guhathakurta}, P., {Babul}, A., \& {McConnachie}, A.~W. 2007,
  \mnras, 380, 15

\bibitem[{{Fattahi} {et~al.}(2020){Fattahi}, {Deason}, {Frenk}, {Simpson},
  {G{\'o}mez}, {Grand}, {Monachesi}, {Marinacci}, \&
  {Pakmor}}]{2020MNRAS.497.4459F}
{Fattahi}, A., {Deason}, A.~J., {Frenk}, C.~S., {et~al.} 2020, \mnras, 497,
  4459

\bibitem[{{Fern{\'a}ndez-Alvar} {et~al.}(2019){Fern{\'a}ndez-Alvar},
  {Fern{\'a}ndez-Trincado}, {Moreno}, {Schuster}, {Carigi}, {Recio-Blanco},
  {Beers}, {Chiappini}, {Anders}, {Santiago}, {Queiroz}, {P{\'e}rez-Villegas},
  {Zamora}, {Garc{\'\i}a-Hern{\'a}ndez}, \&
  {Ortigoza-Urdaneta}}]{2019MNRAS.487.1462F}
{Fern{\'a}ndez-Alvar}, E., {Fern{\'a}ndez-Trincado}, J.~G., {Moreno}, E.,
  {et~al.} 2019, \mnras, 487, 1462

\bibitem[{{Feuillet} {et~al.}(2021){Feuillet}, {Sahlholdt}, {Feltzing}, \&
  {Casagrande}}]{2021MNRAS.508.1489F}
{Feuillet}, D.~K., {Sahlholdt}, C.~L., {Feltzing}, S., \& {Casagrande}, L.
  2021, \mnras, 508, 1489

\bibitem[{{Font} {et~al.}(2006){Font}, {Johnston}, {Bullock}, \&
  {Robertson}}]{2006ApJ...646..886F}
{Font}, A.~S., {Johnston}, K.~V., {Bullock}, J.~S., \& {Robertson}, B.~E. 2006,
  \apj, 646, 886

\bibitem[{{Font} {et~al.}(2011){Font}, {McCarthy}, {Crain}, {Theuns}, {Schaye},
  {Wiersma}, \& {Dalla Vecchia}}]{2011MNRAS.416.2802F}
{Font}, A.~S., {McCarthy}, I.~G., {Crain}, R.~A., {et~al.} 2011, \mnras, 416,
  2802

\bibitem[{{Forbes}(2020)}]{2020MNRAS.493..847F}
{Forbes}, D.~A. 2020, \mnras, 493, 847

\bibitem[{{Frenk} \& {White}(2012)}]{2012AnP...524..507F}
{Frenk}, C.~S. \& {White}, S.~D.~M. 2012, Annalen der Physik, 524, 507

\bibitem[{{Gallart} {et~al.}(2019){Gallart}, {Bernard}, {Brook}, {Ruiz-Lara},
  {Cassisi}, {Hill}, \& {Monelli}}]{2019NatAs...3..932G}
{Gallart}, C., {Bernard}, E.~J., {Brook}, C.~B., {et~al.} 2019, Nature
  Astronomy, 3, 932

\bibitem[{{Gilmore} {et~al.}(2012){Gilmore}, {Randich}, {Asplund}, {Binney},
  {Bonifacio}, {Drew}, {Feltzing}, {Ferguson}, {Jeffries}, {Micela},
  {Negueruela}, {Prusti}, {Rix}, {Vallenari}, {Alfaro}, {Allende-Prieto},
  {Babusiaux}, {Bensby}, {Blomme}, {Bragaglia}, {Flaccomio}, {Fran{\c{c}}ois},
  {Irwin}, {Koposov}, {Korn}, {Lanzafame}, {Pancino}, {Paunzen},
  {Recio-Blanco}, {Sacco}, {Smiljanic}, {Van Eck}, {Walton}, {Aden}, {Aerts},
  {Affer}, {Alcala}, {Altavilla}, {Alves}, {Antoja}, {Arenou}, {Argiroffi},
  {Asensio Ramos}, {Bailer-Jones}, {Balaguer-Nunez}, {Bayo}, {Barbuy},
  {Barisevicius}, {Barrado y Navascues}, {Battistini}, {Bellas Velidis},
  {Bellazzini}, {Belokurov}, {Bergemann}, {Bertelli}, {Biazzo}, {Bienayme},
  {Bland-Hawthorn}, {Boeche}, {Bonito}, {Boudreault}, {Bouvier}, {Brandao},
  {Brown}, {de Bruijne}, {Burleigh}, {Caballero}, {Caffau}, {Calura},
  {Capuzzo-Dolcetta}, {Caramazza}, {Carraro}, {Casagrande}, {Casewell},
  {Chapman}, {Chiappini}, {Chorniy}, {Christlieb}, {Cignoni}, {Cocozza},
  {Colless}, {Collet}, {Collins}, {Correnti}, {Covino}, {Crnojevic}, {Cropper},
  {Cunha}, {Damiani}, {David}, {Delgado}, {Duffau}, {Edvardsson}, {Eldridge},
  {Enke}, {Eriksson}, {Evans}, {Eyer}, {Famaey}, {Fellhauer}, {Ferreras},
  {Figueras}, {Fiorentino}, {Flynn}, {Folha}, {Franciosini}, {Frasca},
  {Freeman}, {Fremat}, {Friel}, {Gaensicke}, {Gameiro}, {Garzon}, {Geier},
  {Geisler}, {Gerhard}, {Gibson}, {Gomboc}, {Gomez}, {Gonzalez-Fernandez},
  {Gonzalez Hernandez}, {Gosset}, {Grebel}, {Greimel}, {Groenewegen},
  {Grundahl}, {Guarcello}, {Gustafsson}, {Hadrava}, {Hatzidimitriou}, {Hambly},
  {Hammersley}, {Hansen}, {Haywood}, {Heber}, {Heiter}, {Held}, {Helmi},
  {Hensler}, {Herrero}, {Hill}, {Hodgkin}, {Huelamo}, {Huxor}, {Ibata},
  {Jackson}, {de Jong}, {Jonker}, {Jordan}, {Jordi}, {Jorissen}, {Katz},
  {Kawata}, {Keller}, {Kharchenko}, {Klement}, {Klutsch}, {Knude}, {Koch},
  {Kochukhov}, {Kontizas}, {Koubsky}, {Lallement}, {de Laverny}, {van Leeuwen},
  {Lemasle}, {Lewis}, {Lind}, {Lindstrom}, {Lobel}, {Lopez Santiago}, {Lucas},
  {Ludwig}, {Lueftinger}, {Magrini}, {Maiz Apellaniz}, {Maldonado}, {Marconi},
  {Marino}, {Martayan}, {Martinez-Valpuesta}, {Matijevic}, {McMahon},
  {Messina}, {Meyer}, {Miglio}, {Mikolaitis}, {Minchev}, {Minniti}, {Moitinho},
  {Momany}, {Monaco}, {Montalto}, {Monteiro}, {Monier}, {Montes}, {Mora},
  {Moraux}, {Morel}, {Mowlavi}, {Mucciarelli}, {Munari}, {Napiwotzki},
  {Nardetto}, {Naylor}, {Naze}, {Nelemans}, {Okamoto}, {Ortolani}, {Pace},
  {Palla}, {Palous}, {Parker}, {Penarrubia}, {Pillitteri}, {Piotto}, {Posbic},
  {Prisinzano}, {Puzeras}, {Quirrenbach}, {Ragaini}, {Read}, {Read}, {Reyle},
  {De Ridder}, {Robichon}, {Robin}, {Roeser}, {Romano}, {Royer}, {Ruchti},
  {Ruzicka}, {Ryan}, {Ryde}, {Santos}, {Sanz Forcada}, {Sarro Baro},
  {Sbordone}, {Schilbach}, {Schmeja}, {Schnurr}, {Schoenrich}, {Scholz},
  {Seabroke}, {Sharma}, {De Silva}, {Smith}, {Solano}, {Sordo}, {Soubiran},
  {Sousa}, {Spagna}, {Steffen}, {Steinmetz}, {Stelzer}, {Stempels},
  {Tabernero}, {Tautvaisiene}, {Thevenin}, {Torra}, {Tosi}, {Tolstoy}, {Turon},
  {Walker}, {Wambsganss}, {Worley}, {Venn}, {Vink}, {Wyse}, {Zaggia},
  {Zeilinger}, {Zoccali}, {Zorec}, {Zucker}, {Zwitter}, \& {Gaia-ESO Survey
  Team}}]{2012Msngr.147...25G}
{Gilmore}, G., {Randich}, S., {Asplund}, M., {et~al.} 2012, The Messenger, 147,
  25

\bibitem[{{G{\'o}mez} {et~al.}(2017){G{\'o}mez}, {Grand}, {Monachesi}, {White},
  {Bustamante}, {Marinacci}, {Pakmor}, {Simpson}, {Springel}, \&
  {Frenk}}]{2017MNRAS.472.3722G}
{G{\'o}mez}, F.~A., {Grand}, R. J.~J., {Monachesi}, A., {et~al.} 2017, \mnras,
  472, 3722

\bibitem[{{G{\'o}mez} {et~al.}(2013){G{\'o}mez}, {Helmi}, {Cooper}, {Frenk},
  {Navarro}, \& {White}}]{2013MNRAS.436.3602G}
{G{\'o}mez}, F.~A., {Helmi}, A., {Cooper}, A.~P., {et~al.} 2013, \mnras, 436,
  3602

\bibitem[{{G{\'o}mez} {et~al.}(2012){G{\'o}mez}, {Minchev}, {Villalobos},
  {O'Shea}, \& {Williams}}]{2012MNRAS.419.2163G}
{G{\'o}mez}, F.~A., {Minchev}, I., {Villalobos}, {\'A}., {O'Shea}, B.~W., \&
  {Williams}, M. E.~K. 2012, \mnras, 419, 2163

\bibitem[{{Gonzalez} {et~al.}(2020){Gonzalez}, {Mucciarelli}, {Origlia},
  {Schultheis}, {Caffau}, {Di Matteo}, {Randich}, {Recio-Blanco}, {Zoccali},
  {Bonifacio}, {Dalessandro}, {Schiavon}, {Pancino}, {Taylor}, {Valenti},
  {Rojas-Arriagada}, {Sacco}, {Biazzo}, {Bellazzini}, {Cioni}, {Clementini},
  {Contreras Ramos}, {de Laverny}, {Evans}, {Haywood}, {Hill}, {Ibata},
  {Lucatello}, {Magrini}, {Martin}, {Nisini}, {Sanna}, {Cirasuolo}, {Maiolino},
  {Afonso}, {Lilly}, {Flores}, {Oliva}, {Paltani}, \&
  {Vanzi}}]{2020Msngr.180...18G}
{Gonzalez}, O.~A., {Mucciarelli}, A., {Origlia}, L., {et~al.} 2020, The
  Messenger, 180, 18

\bibitem[{{Grand} {et~al.}(2018){Grand}, {Bustamante}, {G{\'o}mez}, {Kawata},
  {Marinacci}, {Pakmor}, {Rix}, {Simpson}, {Sparre}, \&
  {Springel}}]{2018MNRAS.474.3629G}
{Grand}, R. J.~J., {Bustamante}, S., {G{\'o}mez}, F.~A., {et~al.} 2018, \mnras,
  474, 3629

\bibitem[{{Grand} {et~al.}(2019){Grand}, {Deason}, {White}, {Simpson},
  {G{\'o}mez}, {Marinacci}, \& {Pakmor}}]{2019MNRAS.487L..72G}
{Grand}, R. J.~J., {Deason}, A.~J., {White}, S. D.~M., {et~al.} 2019, \mnras,
  487, L72

\bibitem[{{Grand} {et~al.}(2017){Grand}, {G{\'o}mez}, {Marinacci}, {Pakmor},
  {Springel}, {Campbell}, {Frenk}, {Jenkins}, \& {White}}]{2017MNRAS.467..179G}
{Grand}, R. J.~J., {G{\'o}mez}, F.~A., {Marinacci}, F., {et~al.} 2017, \mnras,
  467, 179

\bibitem[{{Hasselquist} {et~al.}(2017){Hasselquist}, {Shetrone}, {Smith},
  {Holtzman}, {McWilliam}, {Fern{\'a}ndez-Trincado}, {Beers}, {Majewski},
  {Nidever}, {Tang}, {Tissera}, {Fern{\'a}ndez Alvar}, {Allende Prieto},
  {Almeida}, {Anguiano}, {Battaglia}, {Carigi}, {Delgado Inglada},
  {Frinchaboy}, {Garc{\'\i}a-Hern{\'a}ndez}, {Geisler}, {Minniti}, {Placco},
  {Schultheis}, {Sobeck}, \& {Villanova}}]{2017ApJ...845..162H}
{Hasselquist}, S., {Shetrone}, M., {Smith}, V., {et~al.} 2017, \apj, 845, 162

\bibitem[{{Haywood} {et~al.}(2018){Haywood}, {Di Matteo}, {Lehnert}, {Snaith},
  {Khoperskov}, \& {G{\'o}mez}}]{2018ApJ...863..113H}
{Haywood}, M., {Di Matteo}, P., {Lehnert}, M.~D., {et~al.} 2018, \apj, 863, 113

\bibitem[{{Helmi}(2004)}]{2004ApJ...610L..97H}
{Helmi}, A. 2004, \apjl, 610, L97

\bibitem[{{Helmi}(2020)}]{2020ARA&A..58..205H}
{Helmi}, A. 2020, \araa, 58, 205

\bibitem[{{Helmi} {et~al.}(2018){Helmi}, {Babusiaux}, {Koppelman}, {Massari},
  {Veljanoski}, \& {Brown}}]{2018Natur.563...85H}
{Helmi}, A., {Babusiaux}, C., {Koppelman}, H.~H., {et~al.} 2018, \nat, 563, 85

\bibitem[{{Helmi} {et~al.}(2017){Helmi}, {Veljanoski}, {Breddels}, {Tian}, \&
  {Sales}}]{2017A&A...598A..58H}
{Helmi}, A., {Veljanoski}, J., {Breddels}, M.~A., {Tian}, H., \& {Sales}, L.~V.
  2017, \aap, 598, A58

\bibitem[{{Helmi} {et~al.}(1999){Helmi}, {White}, {de Zeeuw}, \&
  {Zhao}}]{1999Natur.402...53H}
{Helmi}, A., {White}, S. D.~M., {de Zeeuw}, P.~T., \& {Zhao}, H. 1999, \nat,
  402, 53

\bibitem[{{Horta} {et~al.}(2023){Horta}, {Schiavon}, {Mackereth}, {Weinberg},
  {Hasselquist}, {Feuillet}, {O'Connell}, {Anguiano}, {Allende-Prieto},
  {Beaton}, {Bizyaev}, {Cunha}, {Geisler}, {Garc{\'\i}a-Hern{\'a}ndez},
  {Holtzman}, {J{\"o}nsson}, {Lane}, {Majewski}, {M{\'e}sz{\'a}ros}, {Minniti},
  {Nitschelm}, {Shetrone}, {Smith}, \& {Zasowski}}]{2022arXiv220404233H}
{Horta}, D., {Schiavon}, R.~P., {Mackereth}, J.~T., {et~al.} 2023, \mnras, 520,
  5671

\bibitem[{{Hunter}(2007)}]{2007CSE.....9...90H}
{Hunter}, J.~D. 2007, Computing in Science and Engineering, 9, 90

\bibitem[{{Ibata} {et~al.}(1994){Ibata}, {Gilmore}, \&
  {Irwin}}]{1994Natur.370..194I}
{Ibata}, R.~A., {Gilmore}, G., \& {Irwin}, M.~J. 1994, \nat, 370, 194

\bibitem[{{Jean-Baptiste} {et~al.}(2017){Jean-Baptiste}, {Di Matteo},
  {Haywood}, {G{\'o}mez}, {Montuori}, {Combes}, \&
  {Semelin}}]{2017A&A...604A.106J}
{Jean-Baptiste}, I., {Di Matteo}, P., {Haywood}, M., {et~al.} 2017, \aap, 604,
  A106

\bibitem[{{Johnston} {et~al.}(2001){Johnston}, {Sackett}, \&
  {Bullock}}]{2001ApJ...557..137J}
{Johnston}, K.~V., {Sackett}, P.~D., \& {Bullock}, J.~S. 2001, \apj, 557, 137

\bibitem[{{Johnston} {et~al.}(1995){Johnston}, {Spergel}, \&
  {Hernquist}}]{1995ApJ...451..598J}
{Johnston}, K.~V., {Spergel}, D.~N., \& {Hernquist}, L. 1995, \apj, 451, 598

\bibitem[{{Karakas} \& {Lugaro}(2016)}]{2016ApJ...825...26K}
{Karakas}, A.~I. \& {Lugaro}, M. 2016, \apj, 825, 26

\bibitem[{{Kauffmann} {et~al.}(2003){Kauffmann}, {Heckman}, {White}, {Charlot},
  {Tremonti}, {Brinchmann}, {Bruzual}, {Peng}, {Seibert}, {Bernardi},
  {Blanton}, {Brinkmann}, {Castander}, {Cs{\'a}bai}, {Fukugita}, {Ivezic},
  {Munn}, {Nichol}, {Padmanabhan}, {Thakar}, {Weinberg}, \&
  {York}}]{2003MNRAS.341...33K}
{Kauffmann}, G., {Heckman}, T.~M., {White}, S. D.~M., {et~al.} 2003, \mnras,
  341, 33

\bibitem[{{Khoperskov} {et~al.}(2021){Khoperskov}, {Haywood}, {Snaith}, {Di
  Matteo}, {Lehnert}, {Vasiliev}, {Naroenkov}, \&
  {Berczik}}]{2021MNRAS.501.5176K}
{Khoperskov}, S., {Haywood}, M., {Snaith}, O., {et~al.} 2021, \mnras, 501, 5176

\bibitem[{{Khoperskov} {et~al.}(2022{\natexlab{a}}){Khoperskov}, {Minchev},
  {Libeskind}, {Haywood}, {Di Matteo}, {Belokurov}, {Steinmetz}, {Gomez},
  {Grand}, {Knebe}, {Sorce}, {Sparre}, {Tempel}, \&
  {Vogelsberger}}]{KhoperskovHESTIA-1}
{Khoperskov}, S., {Minchev}, I., {Libeskind}, N., {et~al.} 2022{\natexlab{a}},
  arXiv e-prints, arXiv:2206.04521

\bibitem[{{Khoperskov} {et~al.}(2022{\natexlab{b}}){Khoperskov}, {Minchev},
  {Libeskind}, {Haywood}, {Di Matteo}, {Belokurov}, {Steinmetz}, {Gomez},
  {Grand}, {Knebe}, {Sorce}, {Sparre}, {Tempel}, \&
  {Vogelsberger}}]{KhoperskovHESTIA-2}
{Khoperskov}, S., {Minchev}, I., {Libeskind}, N., {et~al.} 2022{\natexlab{b}},
  arXiv e-prints, arXiv:2206.04522

\bibitem[{{Kirby} {et~al.}(2013){Kirby}, {Cohen}, {Guhathakurta}, {Cheng},
  {Bullock}, \& {Gallazzi}}]{2013ApJ...779..102K}
{Kirby}, E.~N., {Cohen}, J.~G., {Guhathakurta}, P., {et~al.} 2013, \apj, 779,
  102

\bibitem[{{Knebe} {et~al.}(2005){Knebe}, {Gill}, {Kawata}, \&
  {Gibson}}]{2005MNRAS.357L..35K}
{Knebe}, A., {Gill}, S. P.~D., {Kawata}, D., \& {Gibson}, B.~K. 2005, \mnras,
  357, L35

\bibitem[{{Knollmann} \& {Knebe}(2009)}]{2009ApJS..182..608K}
{Knollmann}, S.~R. \& {Knebe}, A. 2009, \apjs, 182, 608

\bibitem[{{Koposov} {et~al.}(2019){Koposov}, {Belokurov}, {Li}, {Mateu},
  {Erkal}, {Grillmair}, {Hendel}, {Price-Whelan}, {Laporte}, {Hawkins}, {Sohn},
  {del Pino}, {Evans}, {Slater}, {Kallivayalil}, {Navarro}, \& {Orphan Aspen
  Treasury Collaboration}}]{2019MNRAS.485.4726K}
{Koposov}, S.~E., {Belokurov}, V., {Li}, T.~S., {et~al.} 2019, \mnras, 485,
  4726

\bibitem[{{Koppelman} {et~al.}(2020){Koppelman}, {Bos}, \&
  {Helmi}}]{2020A&A...642L..18K}
{Koppelman}, H.~H., {Bos}, R. O.~Y., \& {Helmi}, A. 2020, \aap, 642, L18

\bibitem[{{Koppelman} {et~al.}(2019){Koppelman}, {Helmi}, {Massari},
  {Price-Whelan}, \& {Starkenburg}}]{2019A&A...631L...9K}
{Koppelman}, H.~H., {Helmi}, A., {Massari}, D., {Price-Whelan}, A.~M., \&
  {Starkenburg}, T.~K. 2019, \aap, 631, L9

\bibitem[{{Kruijssen} {et~al.}(2020){Kruijssen}, {Pfeffer}, {Chevance},
  {Bonaca}, {Trujillo-Gomez}, {Bastian}, {Reina-Campos}, {Crain}, \&
  {Hughes}}]{2020MNRAS.498.2472K}
{Kruijssen}, J.~M.~D., {Pfeffer}, J.~L., {Chevance}, M., {et~al.} 2020, \mnras,
  498, 2472

\bibitem[{{Kruijssen} {et~al.}(2019){Kruijssen}, {Pfeffer}, {Reina-Campos},
  {Crain}, \& {Bastian}}]{2019MNRAS.486.3180K}
{Kruijssen}, J.~M.~D., {Pfeffer}, J.~L., {Reina-Campos}, M., {Crain}, R.~A., \&
  {Bastian}, N. 2019, \mnras, 486, 3180

\bibitem[{{Lacey} \& {Cole}(1993)}]{1993MNRAS.262..627L}
{Lacey}, C. \& {Cole}, S. 1993, \mnras, 262, 627

\bibitem[{{Laporte} {et~al.}(2018){Laporte}, {Johnston}, {G{\'o}mez},
  {Garavito-Camargo}, \& {Besla}}]{2018MNRAS.481..286L}
{Laporte}, C. F.~P., {Johnston}, K.~V., {G{\'o}mez}, F.~A., {Garavito-Camargo},
  N., \& {Besla}, G. 2018, \mnras, 481, 286

\bibitem[{{Larson}(1976)}]{1976MNRAS.176...31L}
{Larson}, R.~B. 1976, \mnras, 176, 31

\bibitem[{{Li} {et~al.}(2021){Li}, {Koposov}, {Erkal}, {Ji}, {Shipp}, {Pace},
  {Hilmi}, {Kuehn}, {Lewis}, {Mackey}, {Simpson}, {Wan}, {Zucker},
  {Bland-Hawthorn}, {Cullinane}, {Da Costa}, {Drlica-Wagner}, {Hattori},
  {Martell}, {Sharma}, \& {S5 Collaboration}}]{2021ApJ...911..149L}
{Li}, T.~S., {Koposov}, S.~E., {Erkal}, D., {et~al.} 2021, \apj, 911, 149

\bibitem[{{Li} \& {Helmi}(2008)}]{2008MNRAS.385.1365L}
{Li}, Y.-S. \& {Helmi}, A. 2008, \mnras, 385, 1365

\bibitem[{{Libeskind} {et~al.}(2020){Libeskind}, {Carlesi}, {Grand},
  {Khalatyan}, {Knebe}, {Pakmor}, {Pilipenko}, {Pawlowski}, {Sparre}, {Tempel},
  {Wang}, {Courtois}, {Gottl{\"o}ber}, {Hoffman}, {Minchev}, {Pfrommer},
  {Sorce}, {Springel}, {Steinmetz}, {Tully}, {Vogelsberger}, \&
  {Yepes}}]{2020MNRAS.498.2968L}
{Libeskind}, N.~I., {Carlesi}, E., {Grand}, R. J.~J., {et~al.} 2020, \mnras,
  498, 2968

\bibitem[{{Lu} {et~al.}(2022){Lu}, {Minchev}, {Buck}, {Khoperskov},
  {Steinmetz}, {Libeskind}, {Cescutti}, \& {Freeman}}]{lu22}
{Lu}, {Minchev}, I., {Buck}, T., {et~al.} 2022, arXiv e-prints,
  arXiv:2212.04515

\bibitem[{{Luo} {et~al.}(2015){Luo}, {Zhao}, {Zhao}, {Deng}, {Liu}, {Jing},
  {Wang}, {Zhang}, {Shi}, {Cui}, {Chu}, {Li}, {Bai}, {Wu}, {Cai}, {Cao}, {Cao},
  {Carlin}, {Chen}, {Chen}, {Chen}, {Chen}, {Chen}, {Chen}, {Chen},
  {Christlieb}, {Chu}, {Cui}, {Dong}, {Du}, {Fan}, {Feng}, {Fu}, {Gao}, {Gong},
  {Gu}, {Guo}, {Han}, {He}, {Hou}, {Hou}, {Hou}, {Hu}, {Hu}, {Hu}, {Huo},
  {Jia}, {Jiang}, {Jiang}, {Jiang}, {Jin}, {Kong}, {Kong}, {Lei}, {Li}, {Li},
  {Li}, {Li}, {Li}, {Li}, {Li}, {Li}, {Li}, {Li}, {Li}, {Li}, {Liang}, {Lin},
  {Liu}, {Liu}, {Liu}, {Liu}, {Lu}, {Luo}, {Mao}, {Newberg}, {Ni}, {Qi}, {Qi},
  {Shen}, {Shi}, {Song}, {Song}, {Su}, {Su}, {Tang}, {Tao}, {Tian}, {Wang},
  {Wang}, {Wang}, {Wang}, {Wang}, {Wang}, {Wang}, {Wang}, {Wang}, {Wang},
  {Wang}, {Wang}, {Wang}, {Wang}, {Wang}, {Wang}, {Wang}, {Wang}, {Wang},
  {Wang}, {Wei}, {Wei}, {Wu}, {Wu}, {Wu}, {Wu}, {Xing}, {Xu}, {Xu}, {Xu},
  {Yan}, {Yang}, {Yang}, {Yang}, {Yang}, {Yao}, {Yu}, {Yuan}, {Yuan}, {Yuan},
  {Yuan}, {Zhai}, {Zhang}, {Zhang}, {Zhang}, {Zhang}, {Zhang}, {Zhang},
  {Zhang}, {Zhang}, {Zhao}, {Zhou}, {Zhou}, {Zhu}, {Zhu}, {Zou}, \&
  {Zuo}}]{2015RAA....15.1095L}
{Luo}, A.~L., {Zhao}, Y.-H., {Zhao}, G., {et~al.} 2015, Research in Astronomy
  and Astrophysics, 15, 1095

\bibitem[{{Mackereth} {et~al.}(2018){Mackereth}, {Crain}, {Schiavon}, {Schaye},
  {Theuns}, \& {Schaller}}]{2018MNRAS.477.5072M}
{Mackereth}, J.~T., {Crain}, R.~A., {Schiavon}, R.~P., {et~al.} 2018, \mnras,
  477, 5072

\bibitem[{{Mackereth} {et~al.}(2019){Mackereth}, {Schiavon}, {Pfeffer},
  {Hayes}, {Bovy}, {Anguiano}, {Allende Prieto}, {Hasselquist}, {Holtzman},
  {Johnson}, {Majewski}, {O'Connell}, {Shetrone}, {Tissera}, \&
  {Fern{\'a}ndez-Trincado}}]{2019MNRAS.482.3426M}
{Mackereth}, J.~T., {Schiavon}, R.~P., {Pfeffer}, J., {et~al.} 2019, \mnras,
  482, 3426

\bibitem[{{Majewski} {et~al.}(2017){Majewski}, {Schiavon}, {Frinchaboy},
  {Allende Prieto}, {Barkhouser}, {Bizyaev}, {Blank}, {Brunner}, {Burton},
  {Carrera}, {Chojnowski}, {Cunha}, {Epstein}, {Fitzgerald}, {Garc{\'\i}a
  P{\'e}rez}, {Hearty}, {Henderson}, {Holtzman}, {Johnson}, {Lam}, {Lawler},
  {Maseman}, {M{\'e}sz{\'a}ros}, {Nelson}, {Nguyen}, {Nidever}, {Pinsonneault},
  {Shetrone}, {Smee}, {Smith}, {Stolberg}, {Skrutskie}, {Walker}, {Wilson},
  {Zasowski}, {Anders}, {Basu}, {Beland}, {Blanton}, {Bovy}, {Brownstein},
  {Carlberg}, {Chaplin}, {Chiappini}, {Eisenstein}, {Elsworth}, {Feuillet},
  {Fleming}, {Galbraith-Frew}, {Garc{\'\i}a}, {Garc{\'\i}a-Hern{\'a}ndez},
  {Gillespie}, {Girardi}, {Gunn}, {Hasselquist}, {Hayden}, {Hekker}, {Ivans},
  {Kinemuchi}, {Klaene}, {Mahadevan}, {Mathur}, {Mosser}, {Muna}, {Munn},
  {Nichol}, {O'Connell}, {Parejko}, {Robin}, {Rocha-Pinto}, {Schultheis},
  {Serenelli}, {Shane}, {Silva Aguirre}, {Sobeck}, {Thompson}, {Troup},
  {Weinberg}, \& {Zamora}}]{2017AJ....154...94M}
{Majewski}, S.~R., {Schiavon}, R.~P., {Frinchaboy}, P.~M., {et~al.} 2017, \aj,
  154, 94

\bibitem[{{Malhan} {et~al.}(2022){Malhan}, {Ibata}, {Sharma}, {Famaey},
  {Bellazzini}, {Carlberg}, {D'Souza}, {Yuan}, {Martin}, \&
  {Thomas}}]{2022ApJ...926..107M}
{Malhan}, K., {Ibata}, R.~A., {Sharma}, S., {et~al.} 2022, \apj, 926, 107

\bibitem[{{Marchesini} {et~al.}(2009){Marchesini}, {van Dokkum}, {F{\"o}rster
  Schreiber}, {Franx}, {Labb{\'e}}, \& {Wuyts}}]{2009ApJ...701.1765M}
{Marchesini}, D., {van Dokkum}, P.~G., {F{\"o}rster Schreiber}, N.~M., {et~al.}
  2009, \apj, 701, 1765

\bibitem[{{Marchesini} {et~al.}(2010){Marchesini}, {Whitaker}, {Brammer}, {van
  Dokkum}, {Labb{\'e}}, {Muzzin}, {Quadri}, {Kriek}, {Lee}, {Rudnick}, {Franx},
  {Illingworth}, \& {Wake}}]{2010ApJ...725.1277M}
{Marchesini}, D., {Whitaker}, K.~E., {Brammer}, G., {et~al.} 2010, \apj, 725,
  1277

\bibitem[{{Martin} {et~al.}(2014){Martin}, {Ibata}, {Rich}, {Collins},
  {Fardal}, {Irwin}, {Lewis}, {McConnachie}, {Babul}, {Bate}, {Chapman},
  {Conn}, {Crnojevi{\'c}}, {Ferguson}, {Mackey}, {Navarro}, {Pe{\~n}arrubia},
  {Tanvir}, \& {Valls-Gabaud}}]{2014ApJ...787...19M}
{Martin}, N.~F., {Ibata}, R.~A., {Rich}, R.~M., {et~al.} 2014, \apj, 787, 19

\bibitem[{{Mart{\'\i}nez-Delgado} {et~al.}(2012){Mart{\'\i}nez-Delgado},
  {Romanowsky}, {Gabany}, {Annibali}, {Arnold}, {Fliri}, {Zibetti}, {van der
  Marel}, {Rix}, {Chonis}, {Carballo-Bello}, {Aloisi}, {Macci{\`o}},
  {Gallego-Laborda}, {Brodie}, \& {Merrifield}}]{2012ApJ...748L..24M}
{Mart{\'\i}nez-Delgado}, D., {Romanowsky}, A.~J., {Gabany}, R.~J., {et~al.}
  2012, \apjl, 748, L24

\bibitem[{{Mateu}(2023)}]{2022arXiv220410326M}
{Mateu}, C. 2023, \mnras, 520, 5225

\bibitem[{{Matsuno} {et~al.}(2019){Matsuno}, {Aoki}, \&
  {Suda}}]{2019ApJ...874L..35M}
{Matsuno}, T., {Aoki}, W., \& {Suda}, T. 2019, \apjl, 874, L35

\bibitem[{{Matteucci}(2021)}]{2021A&ARv..29....5M}
{Matteucci}, F. 2021, \aapr, 29, 5

\bibitem[{{Monachesi} {et~al.}(2016){Monachesi}, {G{\'o}mez}, {Grand},
  {Kauffmann}, {Marinacci}, {Pakmor}, {Springel}, \&
  {Frenk}}]{2016MNRAS.459L..46M}
{Monachesi}, A., {G{\'o}mez}, F.~A., {Grand}, R. J.~J., {et~al.} 2016, \mnras,
  459, L46

\bibitem[{{Monachesi} {et~al.}(2019){Monachesi}, {G{\'o}mez}, {Grand},
  {Simpson}, {Kauffmann}, {Bustamante}, {Marinacci}, {Pakmor}, {Springel},
  {Frenk}, {White}, \& {Tissera}}]{2019MNRAS.485.2589M}
{Monachesi}, A., {G{\'o}mez}, F.~A., {Grand}, R. J.~J., {et~al.} 2019, \mnras,
  485, 2589

\bibitem[{{Monty} {et~al.}(2020){Monty}, {Venn}, {Lane}, {Lokhorst}, \&
  {Yong}}]{2020MNRAS.497.1236M}
{Monty}, S., {Venn}, K.~A., {Lane}, J. M.~M., {Lokhorst}, D., \& {Yong}, D.
  2020, \mnras, 497, 1236

\bibitem[{{Myeong} {et~al.}(2022){Myeong}, {Belokurov}, {Aguado}, {Evans},
  {Caldwell}, \& {Bradley}}]{2022ApJ...938...21M}
{Myeong}, G.~C., {Belokurov}, V., {Aguado}, D.~S., {et~al.} 2022, \apj, 938, 21

\bibitem[{{Myeong} {et~al.}(2019){Myeong}, {Vasiliev}, {Iorio}, {Evans}, \&
  {Belokurov}}]{2019MNRAS.488.1235M}
{Myeong}, G.~C., {Vasiliev}, E., {Iorio}, G., {Evans}, N.~W., \& {Belokurov},
  V. 2019, \mnras, 488, 1235

\bibitem[{{Naidu} {et~al.}(2020){Naidu}, {Conroy}, {Bonaca}, {Johnson}, {Ting},
  {Caldwell}, {Zaritsky}, \& {Cargile}}]{2020ApJ...901...48N}
{Naidu}, R.~P., {Conroy}, C., {Bonaca}, A., {et~al.} 2020, \apj, 901, 48

\bibitem[{{Naidu} {et~al.}(2022){Naidu}, {Conroy}, {Bonaca}, {Zaritsky},
  {Ting}, {Caldwell}, {Cargile}, {Speagle}, {Chandra}, {Johnson}, {Woody}, \&
  {Han}}]{2022arXiv220409057N}
{Naidu}, R.~P., {Conroy}, C., {Bonaca}, A., {et~al.} 2022, arXiv e-prints,
  arXiv:2204.09057

\bibitem[{{Necib} {et~al.}(2020){Necib}, {Ostdiek}, {Lisanti}, {Cohen},
  {Freytsis}, \& {Garrison-Kimmel}}]{2020ApJ...903...25N}
{Necib}, L., {Ostdiek}, B., {Lisanti}, M., {et~al.} 2020, \apj, 903, 25

\bibitem[{{Newberg} {et~al.}(2009){Newberg}, {Yanny}, \&
  {Willett}}]{2009ApJ...700L..61N}
{Newberg}, H.~J., {Yanny}, B., \& {Willett}, B.~A. 2009, \apjl, 700, L61

\bibitem[{{Nidever} {et~al.}(2020){Nidever}, {Hasselquist}, {Hayes}, {Hawkins},
  {Povick}, {Majewski}, {Smith}, {Anguiano}, {Stringfellow}, {Sobeck}, {Cunha},
  {Beers}, {Bestenlehner}, {Cohen}, {Garcia-Hernandez}, {J{\"o}nsson},
  {Nitschelm}, {Shetrone}, {Lacerna}, {Allende Prieto}, {Beaton}, {Dell'Agli},
  {Fern{\'a}ndez-Trincado}, {Feuillet}, {Gallart}, {Hearty}, {Holtzman},
  {Manchado}, {Mu{\~n}oz}, {O'Connell}, \& {Rosado}}]{2020ApJ...895...88N}
{Nidever}, D.~L., {Hasselquist}, S., {Hayes}, C.~R., {et~al.} 2020, \apj, 895,
  88

\bibitem[{{Pagnini} {et~al.}(2023){Pagnini}, {Di Matteo}, {Khoperskov},
  {Mastrobuono-Battisti}, {Haywood}, {Renaud}, \& {Combes}}]{Pagnini2022}
{Pagnini}, G., {Di Matteo}, P., {Khoperskov}, S., {et~al.} 2023, \aap, 673, A86

\bibitem[{{Pakmor} {et~al.}(2016){Pakmor}, {Springel}, {Bauer}, {Mocz},
  {Munoz}, {Ohlmann}, {Schaal}, \& {Zhu}}]{2016MNRAS.455.1134P}
{Pakmor}, R., {Springel}, V., {Bauer}, A., {et~al.} 2016, \mnras, 455, 1134

\bibitem[{{Panithanpaisal} {et~al.}(2021){Panithanpaisal}, {Sanderson},
  {Wetzel}, {Cunningham}, {Bailin}, \&
  {Faucher-Gigu{\`e}re}}]{2021ApJ...920...10P}
{Panithanpaisal}, N., {Sanderson}, R.~E., {Wetzel}, A., {et~al.} 2021, \apj,
  920, 10

\bibitem[{{Perez} \& {Granger}(2007)}]{2007CSE.....9c..21P}
{Perez}, F. \& {Granger}, B.~E. 2007, Computing in Science and Engineering, 9,
  21

\bibitem[{{Pilkington} {et~al.}(2012){Pilkington}, {Few}, {Gibson}, {Calura},
  {Michel-Dansac}, {Thacker}, {Moll{\'a}}, {Matteucci}, {Rahimi}, {Kawata},
  {Kobayashi}, {Brook}, {Stinson}, {Couchman}, {Bailin}, \&
  {Wadsley}}]{2012A&A...540A..56P}
{Pilkington}, K., {Few}, C.~G., {Gibson}, B.~K., {et~al.} 2012, \aap, 540, A56

\bibitem[{{Planck Collaboration} {et~al.}(2014){Planck Collaboration}, {Ade},
  {Aghanim}, {Armitage-Caplan}, {Arnaud}, {Ashdown}, {Atrio-Barandela},
  {Aumont}, {Baccigalupi}, {Banday}, {Barreiro}, {Bartlett}, {Battaner},
  {Benabed}, {Beno{\^\i}t}, {Benoit-L{\'e}vy}, {Bernard}, {Bersanelli},
  {Bielewicz}, {Bobin}, {Bock}, {Bonaldi}, {Bond}, {Borrill}, {Bouchet},
  {Bridges}, {Bucher}, {Burigana}, {Butler}, {Calabrese}, {Cappellini},
  {Cardoso}, {Catalano}, {Challinor}, {Chamballu}, {Chary}, {Chen}, {Chiang},
  {Chiang}, {Christensen}, {Church}, {Clements}, {Colombi}, {Colombo},
  {Couchot}, {Coulais}, {Crill}, {Curto}, {Cuttaia}, {Danese}, {Davies},
  {Davis}, {de Bernardis}, {de Rosa}, {de Zotti}, {Delabrouille}, {Delouis},
  {D{\'e}sert}, {Dickinson}, {Diego}, {Dolag}, {Dole}, {Donzelli}, {Dor{\'e}},
  {Douspis}, {Dunkley}, {Dupac}, {Efstathiou}, {Elsner}, {En{\ss}lin},
  {Eriksen}, {Finelli}, {Forni}, {Frailis}, {Fraisse}, {Franceschi}, {Gaier},
  {Galeotta}, {Galli}, {Ganga}, {Giard}, {Giardino}, {Giraud-H{\'e}raud},
  {Gjerl{\o}w}, {Gonz{\'a}lez-Nuevo}, {G{\'o}rski}, {Gratton}, {Gregorio},
  {Gruppuso}, {Gudmundsson}, {Haissinski}, {Hamann}, {Hansen}, {Hanson},
  {Harrison}, {Henrot-Versill{\'e}}, {Hern{\'a}ndez-Monteagudo}, {Herranz},
  {Hildebrandt}, {Hivon}, {Hobson}, {Holmes}, {Hornstrup}, {Hou}, {Hovest},
  {Huffenberger}, {Jaffe}, {Jaffe}, {Jewell}, {Jones}, {Juvela},
  {Keih{\"a}nen}, {Keskitalo}, {Kisner}, {Kneissl}, {Knoche}, {Knox}, {Kunz},
  {Kurki-Suonio}, {Lagache}, {L{\"a}hteenm{\"a}ki}, {Lamarre}, {Lasenby},
  {Lattanzi}, {Laureijs}, {Lawrence}, {Leach}, {Leahy}, {Leonardi},
  {Le{\'o}n-Tavares}, {Lesgourgues}, {Lewis}, {Liguori}, {Lilje},
  {Linden-V{\o}rnle}, {L{\'o}pez-Caniego}, {Lubin}, {Mac{\'\i}as-P{\'e}rez},
  {Maffei}, {Maino}, {Mandolesi}, {Maris}, {Marshall}, {Martin},
  {Mart{\'\i}nez-Gonz{\'a}lez}, {Masi}, {Massardi}, {Matarrese}, {Matthai},
  {Mazzotta}, {Meinhold}, {Melchiorri}, {Melin}, {Mendes}, {Menegoni},
  {Mennella}, {Migliaccio}, {Millea}, {Mitra}, {Miville-Desch{\^e}nes},
  {Moneti}, {Montier}, {Morgante}, {Mortlock}, {Moss}, {Munshi}, {Murphy},
  {Naselsky}, {Nati}, {Natoli}, {Netterfield}, {N{\o}rgaard-Nielsen},
  {Noviello}, {Novikov}, {Novikov}, {O'Dwyer}, {Osborne}, {Oxborrow}, {Paci},
  {Pagano}, {Pajot}, {Paladini}, {Paoletti}, {Partridge}, {Pasian},
  {Patanchon}, {Pearson}, {Pearson}, {Peiris}, {Perdereau}, {Perotto},
  {Perrotta}, {Pettorino}, {Piacentini}, {Piat}, {Pierpaoli}, {Pietrobon},
  {Plaszczynski}, {Platania}, {Pointecouteau}, {Polenta}, {Ponthieu}, {Popa},
  {Poutanen}, {Pratt}, {Pr{\'e}zeau}, {Prunet}, {Puget}, {Rachen}, {Reach},
  {Rebolo}, {Reinecke}, {Remazeilles}, {Renault}, {Ricciardi}, {Riller},
  {Ristorcelli}, {Rocha}, {Rosset}, {Roudier}, {Rowan-Robinson},
  {Rubi{\~n}o-Mart{\'\i}n}, {Rusholme}, {Sandri}, {Santos}, {Savelainen},
  {Savini}, {Scott}, {Seiffert}, {Shellard}, {Spencer}, {Starck}, {Stolyarov},
  {Stompor}, {Sudiwala}, {Sunyaev}, {Sureau}, {Sutton}, {Suur-Uski}, {Sygnet},
  {Tauber}, {Tavagnacco}, {Terenzi}, {Toffolatti}, {Tomasi}, {Tristram},
  {Tucci}, {Tuovinen}, {T{\"u}rler}, {Umana}, {Valenziano}, {Valiviita}, {Van
  Tent}, {Vielva}, {Villa}, {Vittorio}, {Wade}, {Wandelt}, {Wehus}, {White},
  {White}, {Wilkinson}, {Yvon}, {Zacchei}, \& {Zonca}}]{2014A&A...571A..16P}
{Planck Collaboration}, {Ade}, P.~A.~R., {Aghanim}, N., {et~al.} 2014, \aap,
  571, A16

\bibitem[{{Queiroz} {et~al.}(2021){Queiroz}, {Chiappini}, {Perez-Villegas},
  {Khalatyan}, {Anders}, {Barbuy}, {Santiago}, {Steinmetz}, {Cunha},
  {Schultheis}, {Majewski}, {Minchev}, {Minniti}, {Beaton}, {Cohen}, {da
  Costa}, {Fern{\'a}ndez-Trincado}, {Garcia-Hern{\'a}ndez}, {Geisler},
  {Hasselquist}, {Lane}, {Nitschelm}, {Rojas-Arriagada}, {Roman-Lopes},
  {Smith}, \& {Zasowski}}]{2021A&A...656A.156Q}
{Queiroz}, A.~B.~A., {Chiappini}, C., {Perez-Villegas}, A., {et~al.} 2021,
  \aap, 656, A156

\bibitem[{{Reichert} {et~al.}(2020){Reichert}, {Hansen}, {Hanke},
  {Sk{\'u}lad{\'o}ttir}, {Arcones}, \& {Grebel}}]{2020A&A...641A.127R}
{Reichert}, M., {Hansen}, C.~J., {Hanke}, M., {et~al.} 2020, \aap, 641, A127

\bibitem[{{Renaud} {et~al.}(2021){Renaud}, {Agertz}, {Andersson}, {Read},
  {Ryde}, {Bensby}, {Rey}, \& {Feuillet}}]{2021MNRAS.503.5868R}
{Renaud}, F., {Agertz}, O., {Andersson}, E.~P., {et~al.} 2021, \mnras, 503,
  5868

\bibitem[{{Roederer} {et~al.}(2018){Roederer}, {Hattori}, \&
  {Valluri}}]{2018AJ....156..179R}
{Roederer}, I.~U., {Hattori}, K., \& {Valluri}, M. 2018, \aj, 156, 179

\bibitem[{{Rupke} {et~al.}(2010){Rupke}, {Kewley}, \&
  {Barnes}}]{2010ApJ...710L.156R}
{Rupke}, D. S.~N., {Kewley}, L.~J., \& {Barnes}, J.~E. 2010, \apjl, 710, L156

\bibitem[{{Sanders} {et~al.}(2021){Sanders}, {Belokurov}, \&
  {Man}}]{2021MNRAS.506.4321S}
{Sanders}, J.~L., {Belokurov}, V., \& {Man}, K. T.~F. 2021, \mnras, 506, 4321

\bibitem[{{Sanders} \& {Binney}(2013)}]{2013MNRAS.433.1813S}
{Sanders}, J.~L. \& {Binney}, J. 2013, \mnras, 433, 1813

\bibitem[{{Shetrone} {et~al.}(2003){Shetrone}, {Venn}, {Tolstoy}, {Primas},
  {Hill}, \& {Kaufer}}]{2003AJ....125..684S}
{Shetrone}, M., {Venn}, K.~A., {Tolstoy}, E., {et~al.} 2003, \aj, 125, 684

\bibitem[{{Shipp} {et~al.}(2018){Shipp}, {Drlica-Wagner}, {Balbinot},
  {Ferguson}, {Erkal}, {Li}, {Bechtol}, {Belokurov}, {Buncher}, {Carollo},
  {Carrasco Kind}, {Kuehn}, {Marshall}, {Pace}, {Rykoff}, {Sevilla-Noarbe},
  {Sheldon}, {Strigari}, {Vivas}, {Yanny}, {Zenteno}, {Abbott}, {Abdalla},
  {Allam}, {Avila}, {Bertin}, {Brooks}, {Burke}, {Carretero}, {Castander},
  {Cawthon}, {Crocce}, {Cunha}, {D'Andrea}, {da Costa}, {Davis}, {De Vicente},
  {Desai}, {Diehl}, {Doel}, {Evrard}, {Flaugher}, {Fosalba}, {Frieman},
  {Garc{\'\i}a-Bellido}, {Gaztanaga}, {Gerdes}, {Gruen}, {Gruendl}, {Gschwend},
  {Gutierrez}, {Hartley}, {Honscheid}, {Hoyle}, {James}, {Johnson}, {Krause},
  {Kuropatkin}, {Lahav}, {Lin}, {Maia}, {March}, {Martini}, {Menanteau},
  {Miller}, {Miquel}, {Nichol}, {Plazas}, {Romer}, {Sako}, {Sanchez},
  {Santiago}, {Scarpine}, {Schindler}, {Schubnell}, {Smith}, {Smith},
  {Sobreira}, {Suchyta}, {Swanson}, {Tarle}, {Thomas}, {Tucker}, {Walker},
  {Wechsler}, \& {DES Collaboration}}]{2018ApJ...862..114S}
{Shipp}, N., {Drlica-Wagner}, A., {Balbinot}, E., {et~al.} 2018, \apj, 862, 114

\bibitem[{{Simpson} {et~al.}(2019){Simpson}, {Gargiulo}, {G{\'o}mez}, {Grand},
  {Maffione}, {Cooper}, {Deason}, {Frenk}, {Helly}, {Marinacci}, \&
  {Pakmor}}]{2019MNRAS.490L..32S}
{Simpson}, C.~M., {Gargiulo}, I., {G{\'o}mez}, F.~A., {et~al.} 2019, \mnras,
  490, L32

\bibitem[{{Smith} {et~al.}(2009){Smith}, {Evans}, {Belokurov}, {Hewett},
  {Bramich}, {Gilmore}, {Irwin}, {Vidrih}, \& {Zucker}}]{2009MNRAS.399.1223S}
{Smith}, M.~C., {Evans}, N.~W., {Belokurov}, V., {et~al.} 2009, \mnras, 399,
  1223

\bibitem[{{Springel}(2005)}]{2005MNRAS.364.1105S}
{Springel}, V. 2005, \mnras, 364, 1105

\bibitem[{{Springel} {et~al.}(2005){Springel}, {White}, {Jenkins}, {Frenk},
  {Yoshida}, {Gao}, {Navarro}, {Thacker}, {Croton}, {Helly}, {Peacock}, {Cole},
  {Thomas}, {Couchman}, {Evrard}, {Colberg}, \& {Pearce}}]{2005Natur.435..629S}
{Springel}, V., {White}, S. D.~M., {Jenkins}, A., {et~al.} 2005, \nat, 435, 629

\bibitem[{{Steinmetz} {et~al.}(2020){Steinmetz}, {Matijevi{\v{c}}}, {Enke},
  {Zwitter}, {Guiglion}, {McMillan}, {Kordopatis}, {Valentini}, {Chiappini},
  {Casagrande}, {Wojno}, {Anguiano}, {Bienaym{\'e}}, {Bijaoui}, {Binney},
  {Burton}, {Cass}, {de Laverny}, {Fiegert}, {Freeman}, {Fulbright}, {Gibson},
  {Gilmore}, {Grebel}, {Helmi}, {Kunder}, {Munari}, {Navarro}, {Parker},
  {Ruchti}, {Recio-Blanco}, {Reid}, {Seabroke}, {Siviero}, {Siebert}, {Stupar},
  {Watson}, {Williams}, {Wyse}, {Anders}, {Antoja}, {Birko}, {Bland-Hawthorn},
  {Bossini}, {Garc{\'\i}a}, {Carrillo}, {Chaplin}, {Elsworth}, {Famaey},
  {Gerhard}, {Jofre}, {Just}, {Mathur}, {Miglio}, {Minchev}, {Monari},
  {Mosser}, {Ritter}, {Rodrigues}, {Scholz}, {Sharma}, {Sysoliatina}, \& {RAVE
  Collaboration}}]{2020AJ....160...82S}
{Steinmetz}, M., {Matijevi{\v{c}}}, G., {Enke}, H., {et~al.} 2020, \aj, 160, 82

\bibitem[{{Taylor}(2005)}]{2005ASPC..347...29T}
{Taylor}, M.~B. 2005, in Astronomical Society of the Pacific Conference Series,
  Vol. 347, Astronomical Data Analysis Software and Systems XIV, ed.
  P.~{Shopbell}, M.~{Britton}, \& R.~{Ebert}, 29

\bibitem[{{Tissera} {et~al.}(2014){Tissera}, {Beers}, {Carollo}, \&
  {Scannapieco}}]{2014MNRAS.439.3128T}
{Tissera}, P.~B., {Beers}, T.~C., {Carollo}, D., \& {Scannapieco}, C. 2014,
  \mnras, 439, 3128

\bibitem[{{Tissera} {et~al.}(2018){Tissera}, {Machado}, {Carollo}, {Minniti},
  {Beers}, {Zoccali}, \& {Meza}}]{2018MNRAS.473.1656T}
{Tissera}, P.~B., {Machado}, R. E.~G., {Carollo}, D., {et~al.} 2018, \mnras,
  473, 1656

\bibitem[{{Tolstoy} {et~al.}(2009){Tolstoy}, {Hill}, \&
  {Tosi}}]{2009ARA&A..47..371T}
{Tolstoy}, E., {Hill}, V., \& {Tosi}, M. 2009, \araa, 47, 371

\bibitem[{{Tremonti} {et~al.}(2004){Tremonti}, {Heckman}, {Kauffmann},
  {Brinchmann}, {Charlot}, {White}, {Seibert}, {Peng}, {Schlegel}, {Uomoto},
  {Fukugita}, \& {Brinkmann}}]{2004ApJ...613..898T}
{Tremonti}, C.~A., {Heckman}, T.~M., {Kauffmann}, G., {et~al.} 2004, \apj, 613,
  898

\bibitem[{{van der Walt} {et~al.}(2011){van der Walt}, {Colbert}, \&
  {Varoquaux}}]{2011CSE....13b..22V}
{van der Walt}, S., {Colbert}, S.~C., \& {Varoquaux}, G. 2011, Computing in
  Science and Engineering, 13, 22

\bibitem[{{van Dokkum} {et~al.}(2019){van Dokkum}, {Gilhuly}, {Bonaca},
  {Merritt}, {Danieli}, {Lokhorst}, {Abraham}, {Conroy}, \&
  {Greco}}]{2019ApJ...883L..32V}
{van Dokkum}, P., {Gilhuly}, C., {Bonaca}, A., {et~al.} 2019, \apjl, 883, L32

\bibitem[{{Vasiliev}(2019)}]{2019MNRAS.482.1525V}
{Vasiliev}, E. 2019, \mnras, 482, 1525

\bibitem[{{Vincenzo} {et~al.}(2019){Vincenzo}, {Spitoni}, {Calura},
  {Matteucci}, {Silva Aguirre}, {Miglio}, \& {Cescutti}}]{2019MNRAS.487L..47V}
{Vincenzo}, F., {Spitoni}, E., {Calura}, F., {et~al.} 2019, \mnras, 487, L47

\bibitem[{{Virtanen} {et~al.}(2020){Virtanen}, {Gommers}, {Oliphant},
  {Haberland}, {Reddy}, {Cournapeau}, {Burovski}, {Peterson}, {Weckesser},
  {Bright}, {van der Walt}, {Brett}, {Wilson}, {Jarrod Millman}, {Mayorov},
  {Nelson}, {Jones}, {Kern}, {Larson}, {Carey}, {Polat}, {Feng}, {Moore}, {Vand
  erPlas}, {Laxalde}, {Perktold}, {Cimrman}, {Henriksen}, {Quintero}, {Harris},
  {Archibald}, {Ribeiro}, {Pedregosa}, {van Mulbregt}, \&
  {Contributors}}]{2020SciPy-NMeth}
{Virtanen}, P., {Gommers}, R., {Oliphant}, T.~E., {et~al.} 2020, Nature
  Methods, 17, 261

\bibitem[{{Vogelsberger} {et~al.}(2008){Vogelsberger}, {White}, {Helmi}, \&
  {Springel}}]{2008MNRAS.385..236V}
{Vogelsberger}, M., {White}, S. D.~M., {Helmi}, A., \& {Springel}, V. 2008,
  \mnras, 385, 236

\bibitem[{{W}es {M}c{K}inney(2010)}]{mckinney-proc-scipy-2010}
{W}es {M}c{K}inney. 2010, in {P}roceedings of the 9th {P}ython in {S}cience
  {C}onference, ed. {S}t\'efan van~der {W}alt \& {J}arrod {M}illman, 56--61

\bibitem[{{White} \& {Frenk}(1991)}]{1991ApJ...379...52W}
{White}, S. D.~M. \& {Frenk}, C.~S. 1991, \apj, 379, 52

\end{thebibliography}

\end{document}